\documentclass[preprint,sort&compress,3p,times]{elsarticle}
\usepackage{caption}
\usepackage[caption=false]{subfig}
\usepackage{amsfonts}
\usepackage{amsmath, amsthm, amssymb}
\usepackage{mathtools}
\usepackage{graphicx}
\usepackage{float}
\usepackage{breqn}
\usepackage{hyperref}
\usepackage{bm}
\usepackage{url}
\usepackage{placeins}
\usepackage{multirow}
\newcommand\numberthis{\addtocounter{equation}{1}\tag{\theequation}}

\numberwithin{equation}{section}

\begin{document}
\begin{frontmatter}
\title{Full-Wave Algorithm to Model Effects of Bedding Slopes on the Response of Subsurface Electromagnetic Geophysical Sensors near Unconformities}

\renewcommand{\thefootnote}{\fnsymbol{footnote}}

\author[ad1]{Kamalesh Sainath\corref{cor1}}
\ead{sainath.1@osu.edu}
\author[ad1]{Fernando L. Teixeira}
\ead{teixeira@ece.osu.edu}
\cortext[cor1]{Corresponding author}
\address[ad1]{The Ohio State University, ElectroScience Laboratory, 1330 Kinnear Road, Columbus, Ohio, USA 43212}
\begin{abstract}
We propose a full-wave pseudo-analytical numerical electromagnetic (EM) algorithm to model subsurface induction sensors, traversing planar-layered geological formations of arbitrary EM material anisotropy and loss, which are used, for example, in the exploration of hydrocarbon reserves. Unlike past pseudo-analytical planar-layered modeling algorithms that impose parallelism between the formation's bed junctions however, our method involves judicious employment of Transformation Optics techniques to address challenges related to modeling \emph{relative} slope (i.e., tilting) between said junctions (including arbitrary azimuth orientation of each junction). The algorithm exhibits this flexibility, both with respect to loss and anisotropy in the formation layers as well as junction tilting, via employing special planar slabs that coat each ``flattened" (i.e., originally tilted) planar interface, locally redirecting the incident wave within the coating slabs to cause wave fronts to interact with the flattened interfaces \emph{as if} they were still tilted with a specific, user-defined orientation. Moreover, since the coating layers are homogeneous rather than exhibiting continuous material variation, a minimal number of these layers must be inserted and hence reduces added simulation time and computational expense. As said coating layers are not reflectionless however, they do induce artificial field scattering that corrupts legitimate field signatures due to the (effective) interface tilting. Numerical results, for two half-spaces separated by a tilted interface, quantify error trends versus effective interface tilting, material properties, transmitter/receiver spacing, sensor position, coating slab thickness, and transmitter and receiver orientation, helping understand the spurious scattering's effect on reliable (effective) tilting this algorithm can model. Under the effective tilting constraints suggested by the results of said error study, we finally exhibit responses of sensors traversing three-layered media, where we vary the anisotropy, loss, and \emph{relative} tilting of the formations and explore the sensitivity of the sensor's complex-valued measurements to both the magnitude of effective relative interface tilting (polar rotation) as well as azimuthal orientation of the effectively tilted interfaces.
\end{abstract}

\end{frontmatter}

\section{\label{Intro}Introduction}

Long-standing and sustained interest has been directed towards the numerical evaluation of electromagnetic (EM) fields produced by sensors embedded in complex, layered-medium environments~\cite{sommerfeld}. In particular, within the context of geophysical exploration (of hydrocarbon reserves, for example), there exists great interest to computationally model the response of induction tools that can remotely sense the electrical and structural properties of complex geological formations (and consequently, their hydrocarbon productivity)~\cite{anderson1,teixeira5}. Indeed, high-fidelity, rapid, and geometry-robust computational forward-modeling aids fundamental understanding of how factors such as formation's global inhomogeneity structure, conductive anisotropy in formation bed layers, induction tool geometry, exploration borehole geometry, and drilling fluid type (among other factors) affect the sensor's responses. This knowledge informs both effective and robust geophysical parameter retrieval algorithms (inverse problem), as well as sound data interpretation techniques~\cite{zhdanov,anderson1}.  Developing forward-modeling algorithms which not only deliver rapid, accuracy-controllable results, but also simulate the effects of a greater number of dominant, geophysical features without markedly increased computational burden, represents a high priority in subsurface geophysical exploration~\cite{sainath3,sainath6}. 

In the interest of obtaining a good trade-off between the forward modeler's solution speed while still satisfactorily modeling the EM behavior of the environment's dominant geophysical features, a layered-medium approximation of the geophysical formation often proves very useful. Indeed cylindrical layering, planar layering, and a combination of the two (for example, to model the cylindrical exploratory borehole and invasion zone embedded within a stack of planar formation beds) are arguably three of the most widely used layering approximations in subsurface geophysics~\cite{sofia,wei,anderson1,zhdanov,wang,tao,moran1,rabinovich1,sato1,teixeira2,teixeira7,teixeira13,teixeira18,doll,sainath,sainath2,sainath3,moon,moon2}, for both onshore and offshore geophysical exploration modeling~\cite{macgregor1,key1,key2,weiss1,weidelt1,constable1,constable2,constable3,constable4,um1,sainath4}. The prevalence of layered-medium approximations stems in large part, at least from a computational modeling standpoint, due to the typical availability of closed-form eigenfunction expansions to compute the EM field~\cite{chew}[Ch. 2-3]. These full-wave techniques are quite attractive since they can \emph{robustly} deliver rapid solutions with high, user-controlled accuracy under widely varying conditions with respect to anisotropy and loss in the formation's layers, orientation and position of the electric or (equivalent) magnetic current-based sensors (viz., electric loop antennas), and source frequency~\cite{sainath,sainath4}. The robustness to physical parameters is highly desirable in geophysics applications since geological structures are known to exhibit a wide range of inhomogeneity profiles with respect to conductivity, anisotropy, and geometrical layering~\cite{tao,anderson1,sainath6}. For example, with respect to formation conductivity properties, diverse geological structures can embody macro-scale conductive anisotropy in the induction frequency regime, such as (possibly deviated) sand-shale micro-laminate deposits, clean-sand micro-laminate deposits, and either natural or drilling-induced fractures. In the sub-2MHz frequency regime, the electrical conduction current transport characteristics of such structures indeed are often mathematically described by a uniaxial or biaxial conductivity tensor exhibiting directional electrical conductivities whose value range can span in excess of four orders of magnitude~\cite{rabinovich1,anderson1,sofia}. 

When employing planar and cylindrical layer approximations one almost always assumes that the interfaces are parallel, i.e. exhibit common central axes (say, along $z$) in the case of cylindrical layers~\cite{moon,moon2}, or interfaces that are all parallel to a common plane in the case of planar layers~\cite{tao,teixeira5}. However, it may be more appropriate in many cases to admit layered media with material property variation along the direction(s) conventionally presumed homogeneous. For example, in cylindrically-layered medium problems involving deviated drilling, gravitational effects may induce a downward diffusion of the drilling fluid that leads to a cylindrical invasion zone angled relative to the cylindrical exploratory borehole~\cite{sainath6}. Similarly, formations that locally (i.e., in the proximity of the EM sensor) appear as a ``stack" of beds with {\it tilted} (sloped) planar interfaces can appear (for example) due to temporal discontinuities in the formation's geological record. These temporal discontinuities in turn can manifest as commensurately abrupt spatial discontinuities, known as unconformities (especially, angular unconformities)~\cite{gabrielsen1,karim1,qu1}; see Figure \ref{Fig1a} below for a schematic illustration. Indeed, the effects of unconformities and other complex formation properties (such as fractures) have garnered increasing attention over the past ten years~\cite{filomena1,volkan1,omeragic4}, particularly in light of the relatively recent availability of induction sensor systems offering a rich diversity of measurement information with respect to sensor radiation frequency, transmitter and receiver orientation (``directional" diversity), and transmitter/receiver separation~\cite{walton1,omeragic1,omeragic2,omeragic3,omeragic4}. 

A natural question arises as to which numerical technique is best suited to modeling these more complex geometries involving tilted layers. In principle, one could resort to brute-force techniques such as finite difference and finite element methods~\cite{teixeira2,teixeira5,teixeira9,omeragic2,1048989}. The potential for low-frequency instability (e.g., when modeling geophysical sensors operating down to the magnetotelluric frequency range (fraction of a Hertz)), high computational cost (unacceptable especially when many repeated forward solutions are required to solve the inverse problem), and accuracy limitations due to mesh truncation issues (say, via perfectly matched layers or other approaches~\cite{teixeiraJEWA,jflee}) associated with the lack of transverse symmetry in the tilted-layer domain~\cite{sjohnson}, render these numerical methods less suitable for developing fast forward-modeler engines for tilted-layer problems. Another potential approach involves asymptotic solutions which traces the progress of incident rays and their specular reflections within subsurface formations~\cite{wait}. However, this approach is limited to ``sufficiently  high" frequencies and hence may often be unsuitable for modeling induction-regime sensors operating in zones where highly resistive and highly conductive (not to mention anisotropic) layers may coexist~\cite{felsen}. Yet a third possible approach, called the ``Tilt Operator" method, which assumes lossless media and negligible EM near-fields to avoid spurious exponential field growths (arising from violation of ``primitive" causality [i.e., cause preceding effect], which is inherent in this method), is another possibility~\cite{zhang1,zhang2}. Akin to the other mentioned high-frequency approach however~\cite{wait}, the Tilt Operator method is not appropriate for our more general class of problems with respect to sensor and geological formation characteristics.

Turning our attention henceforth to tilted planar-layered media, we propose a pseudo-analytical method based on EM plane wave eigenfunction expansions that manifest mathematically as two-dimensional (2-D) Fourier integrals. This is in contrast to faster, but more restrictive (with respect to allowed media) 1-D Fourier-Bessel (``Sommerfeld") and Fourier-Hankel integral transforms that express EM fields in planar-layered media as integral expansions of EM conical wave eigenfunctions~\cite{chew}[Ch. 2]. Our choice rests upon robust error control capabilities and speed performance of the 2-D integral transform with respect to source radiation frequency, source distribution, and material properties~\cite{sainath3,sainath4}. The use of eigenfunction expansions for modeling EM behavior of non-parallel layers is enabled here by the use of Transformation Optics (T.O.) techniques~\cite{pendry1,pendry6,pendry7,teixeiraJEWA,leonhardt,Ozgun2010921} to effectively replace the original problem (with tilted interfaces) by an equivalent problem with strictly parallel interfaces, where additional ``interface-flattening" layers with anisotropic response are inserted into the geometry to mimic the effect of the original tilted geometry (c.f. Fig. \ref{Fig1b}).
We remark that exploitation of T.O. techniques to facilitate numerical field computation based on brute-force algorithms\textemdash rather than the eigenfunction-based spectral approach considered here\textemdash was recently proposed in~\cite{Ozgun20141616,Ozgun2015165}.
We moreover emphasize that the 2-D Fourier integral is capable of modeling propagation and scattering behavior of these interface-flattening media, which possess azimuthal \emph{non-symmetric} material tensors; the two stated 1-D integral transforms, which are inherently restricted to modeling azimuthal-\emph{symmetric} media, by contrast lack this numerical modeling capability. 

Before proceeding, we note that the coating slabs are spatially homogeneous in the employed Cartesian coordinate system, and hence we require only one planar layer to represent each coating slab, and that too to represent each slab's spatial material profile \emph{exactly}. This homogeneity characteristic is important from a computational efficiency standpoint, as it means that for each ``flattened" interface only two coating layers' (source-independent) EM eigenfunctions, Fresnel reflection and transmission matrices, etc. need to be computed~\cite{sainath}. Second, as we will be fundamentally approximating the transverse translation-\emph{variant} geometry as a transverse translation-\emph{invariant} one, modeling spurious scattering from the ``apexes" and more complex intersection junctions of the tilted beds is out of the question. As we concern ourselves with subsurface geophysical media, which typically present inherent conductivity (typically on the order of at least $10^{-3}$S/m to 2S/m~\cite{anderson1}), scattering from these intersections should typically be negligible so long as the sensors are not in the immediate neighborhood of said intersections (rarely the case, for the small tilting explored herein). Third, the interface-flattening slabs are \emph{not} impedance-matched to the respective ambient medium layers into which they are inserted. Indeed, one of the objectives of this paper is to quantify the impedance mismatch of the artificial slabs, which we will find practically constrains (i.e., for a given desired computation accuracy) the amount of interface tilt that can be reliably modeled. 

This paper is organized as follows. In Section \ref{Form} we overview the 2-D plane wave expansion algorithm, derive the material blueprints for the planar ``interface-flattening" coating slabs, and show how to systematically incorporate these into the computational model. Sections \ref{Err}-\ref{Err2} show the error analysis to quantify how the accuracy of the results varies with effective interface tilting, material profile, transmitter/receiver spacing, sensor position, coating slab thickness, and complex-valued measurement component (both its real and imaginary parts). In Section \ref{Val} we apply the algorithm to predicting EM multi-component induction tool responses when the tool traverses (effectively) tilted formation beds for different interface tilt orientations as well as central bed anisotropic conductivity profiles. The formation anisotropies studied will span the full gamut: All the way from isotropic to (``Transverse-Isotropic") non-deviated uniaxial, (``cross-bedded") deviated uniaxial, and full biaxial anisotropy. We adopt the exp($-i\omega t$) convention, as well as assume all EM media are spatially non-dispersive, time-invariant, and are representable by \emph{diagonalizable} anisotropic $3\times 3$ material tensors.\footnote{Diagonalizability of the material tensors, which physically corresponds to a medium having well-defined electrical and magnetic responses for any direction of applied electric and magnetic field, is required for completeness of the plane wave basis. All naturally-occurring media, as well as the introduced interface-flattening slabs applied to geophysical media, are characterized by diagonalizable material tensors.}

\section{\label{Form}Formulation}
\subsection{Background: Electromagnetic Plane Wave Eigenfunction Expansions}
In deriving the planar multi-layered medium eigenfunction expansion expressions, first assume a homogeneous formation whose dielectric (i.e., excluding conductivity), relative magnetic permeability, and electric conductivity constitutive anisotropic material tensors write as $\boldsymbol{\bar{\epsilon}}_r$, $\boldsymbol{\bar{\mu}}_r$, and $\boldsymbol{\bar{\sigma}}$. Specifically, the assumed material tensors are those of the layer (i.e., in the anticipated multi-layered case), labeled $M$, within which the transmitter resides. Maxwell's Equations in the frequency domain, upon impressing causative electric current $\bm{\mathcal{J}}(\bold{r})$ and/or (equivalent) magnetic current $\bm{\mathcal{M}}(\bold{r})$, yields the electric field vector wave equation (duality in Maxwell's Equations yields the magnetic field vector wave equation)~\cite{sainath,chew}:\footnote{$\epsilon_0$, $c$, and $\mu_0=1/(\epsilon_0c^2)$ represent vacuum electric permittivity, vacuum speed of light, and vacuum magnetic permeability, respectively. Furthermore, $\omega=2\pi f$ is the angular temporal radiation frequency, $k_0=\omega/c$ is the vacuum wave number, and $\eta_0=\sqrt{\mu_0/\epsilon_0}$ is the intrinsic vacuum plane wave impedance~\cite{balanis1,chew}.}
\begin{equation}\label{WaveOp} \bm{\mathcal{\bar{A}}}(\cdot)=\nabla \times \boldsymbol{\bar{\mu}}_r^{-1} \cdot \nabla \times - k_0^2 \left(\boldsymbol{\bar{\epsilon}}_r+i\boldsymbol{\bar{\sigma}}/\omega\right) \cdot ,\ \bm{\mathcal{\bar{A}}} \left(\bm{\mathcal{E}}\right) = ik_0\eta_0\bm{\mathcal{J}}-\nabla \times \boldsymbol{\bar{\mu}}_r^{-1} \cdot \bm{\mathcal{M}} \end{equation}
Now define the three-dimensional spatial Fourier Transform (FT) pair for some generic vector field $\bm{\mathcal{L}}$ (e.g., the magnetic field or current source vector)~\cite{sainath}:
\begin{equation}\label{FT3D} \ \bold{\tilde{L}}(\bold{k}) \ = \ \iiint\limits_{-\infty}^{+\infty} \bm{\mathcal{L}}(\bold{r}) \, \mathrm{e}^{-i\bold{k} \cdot \bold{r}} \, \mathrm{d}x \, \mathrm{d}y \, \mathrm{d}z,\ \bm{\mathcal{L}}(\bold{r}) \ = \ \left(\frac{1}{2\pi}\right)^3\iiint\limits_{-\infty}^{+\infty} \bold{\tilde{L}}(\bold{k}) \, \mathrm{e}^{i\bold{k} \cdot \bold{r}} \, \mathrm{d}k_x \,  \mathrm{d}k_y \, \mathrm{d}k_z \end{equation}
where $\bold{r}=(x,y,z)$ is the position vector and $\bold{k}=(k_x,k_y,k_z)$ is the wave vector. Expanding the left and right hand sides, of the second equation in Eqn. \eqref{WaveOp}, in their respective wave number domain 3-D integral representations and matching the Fourier-domain integrands on both sides, one can multiply the inverse of $\bold{\tilde{\bar{A}}}$ (the FT of $\bm{\mathcal{\bar{A}}}$) to the left of both integrands. Admitting a single Hertzian/infinitesimal-point transmitter current source located at $\bold{r}'=(x',y',z')$, and denoting the receiver location $\bold{r}$, one can then procure the ``direct" (i.e., homogeneous medium) radiated time-harmonic electric field $\bm{\mathcal{E}}_d(\bold{r})$~\cite{sainath}. Indeed, performing ``analytically" (i.e., via contour integration and residue calculus techniques) the $k_z$ integral leads to the following expression: 
\begin{equation}\label{EHM2a} \bm{\mathcal{E}}_d(\bold{r})=\frac{i}{(2\pi)^{2}} \iint \limits_{-\infty}^{+\infty}
\left[u(z-z')\sum_{n=1}^2{\tilde{a}_{M,n}^D\bold{\tilde{e}}_{M,n}\mathrm{e}^{i\tilde{k}_{M,nz}\Delta z}}+ u(z'-z)\sum_{n=3}^4{\tilde{a}_{M,n}^D\bold{\tilde{e}}_{M,n}\mathrm{e}^{i\tilde{k}_{M,nz}\Delta z}}
\right] \mathrm{e}^{ik_x\Delta x+ik_y\Delta y} \, \mathrm{d}k_x \, \mathrm{d}k_y \end{equation}
where $\Delta x=x-x'$, $\Delta y=y-y'$, $\Delta z=z-z'$, $u(\cdot)$ denotes the Heaviside step function, and $\{\bold{\tilde{e}}_{p,n},\tilde{k}_{p,nz},\tilde{a}_{p,n}^D \}$ stand for the electric field polarization state vector, longitudinal wave number component, and direct field polarization amplitude of the $p$th formation bed's $n$th plane wave polarization ($1 \leq p \leq N$, $1 \leq n \leq 4$), respectively. Now introducing additional formation beds will induce a modification, via reflection and transmission mechanisms interfering with the direct field, to the total observed electric field. We mathematically codify this interference phenomenon by deriving the (transmitter layer [$M$] and receiver layer [$L$]-dependent) time-harmonic scattered electric field $\bm{\mathcal{E}}_s(\bold{r})$~\cite{sainath}:
\begin{equation}\label{Espace1}   \bm{\mathcal{E}}_s(\bold{r})  =
\frac{i}{(2\pi)^{2}} \iint\limits_{-\infty}^{+\infty}\left[ (1-\delta_{L,N})\sum_{n=1}^2{\tilde{a}^s_{L,n}\bold{\tilde{e}}_{L,n}\mathrm{e}^{i\tilde{k}_{L,nz}z}}+
(1-\delta_{L,1})\sum_{n=3}^4{\tilde{a}^s_{L,n}\bold{\tilde{e}}_{L,n}\mathrm{e}^{i\tilde{k}_{L,nz}z}}\right] \times \\ \mathrm{e}^{ik_x\Delta x+ik_y\Delta y} \, \mathrm{d}k_x \, \mathrm{d}k_y\end{equation}
 where $\tilde{a}^s_{p,n}$ is the scattered field polarization amplitude of the $n$th polarization in layer $p$, and $\delta_{P1,P2}$ denotes the Kronecker Delta function. 

\subsection{\label{Form2}Tilted Layer Modeling}

\begin{figure}[H]
\centering
\subfloat[\label{Fig1a}]{\includegraphics[height=1.9in]{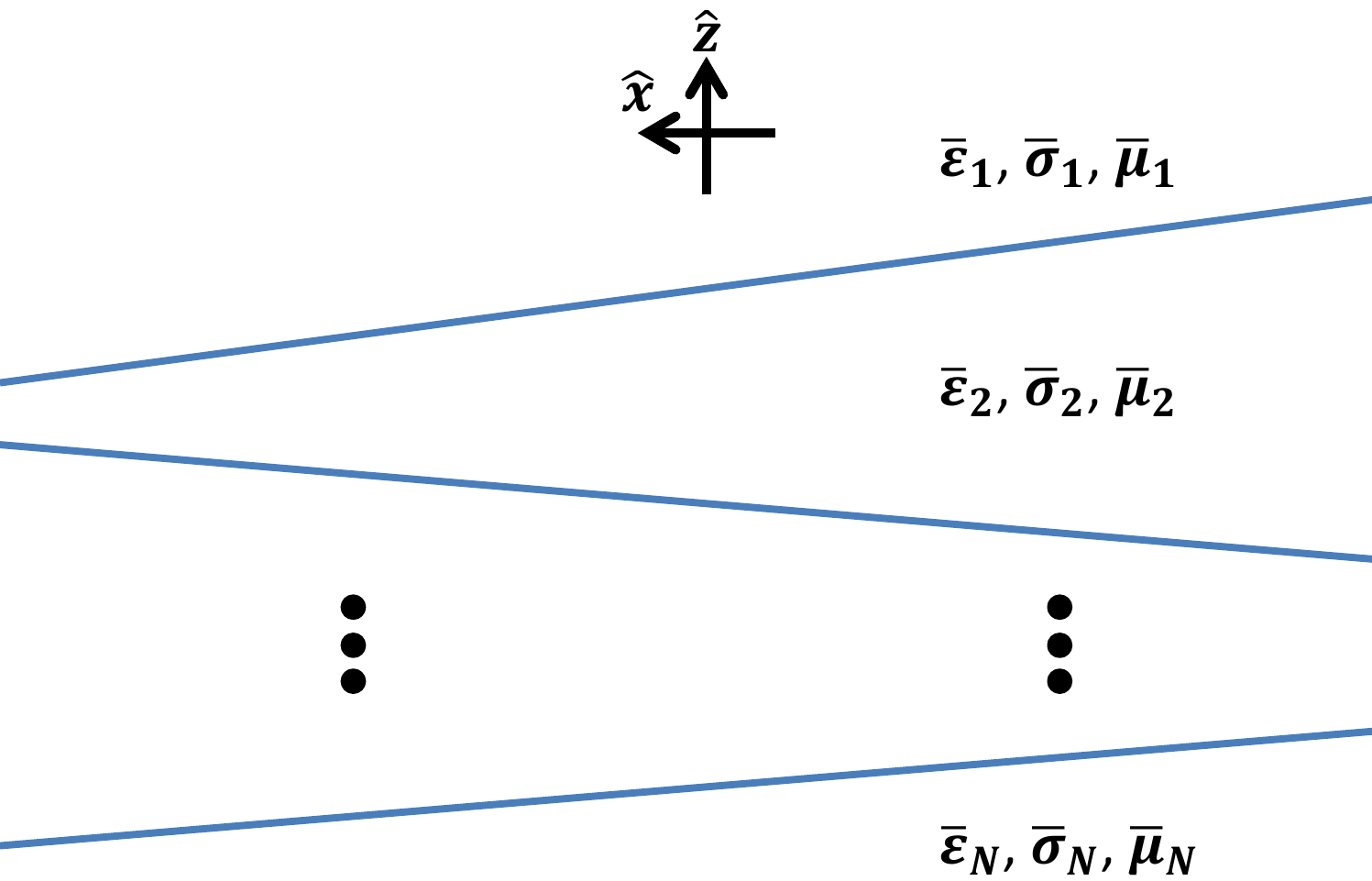}}
\subfloat[\label{Fig1b}]{\includegraphics[height=1.9in]{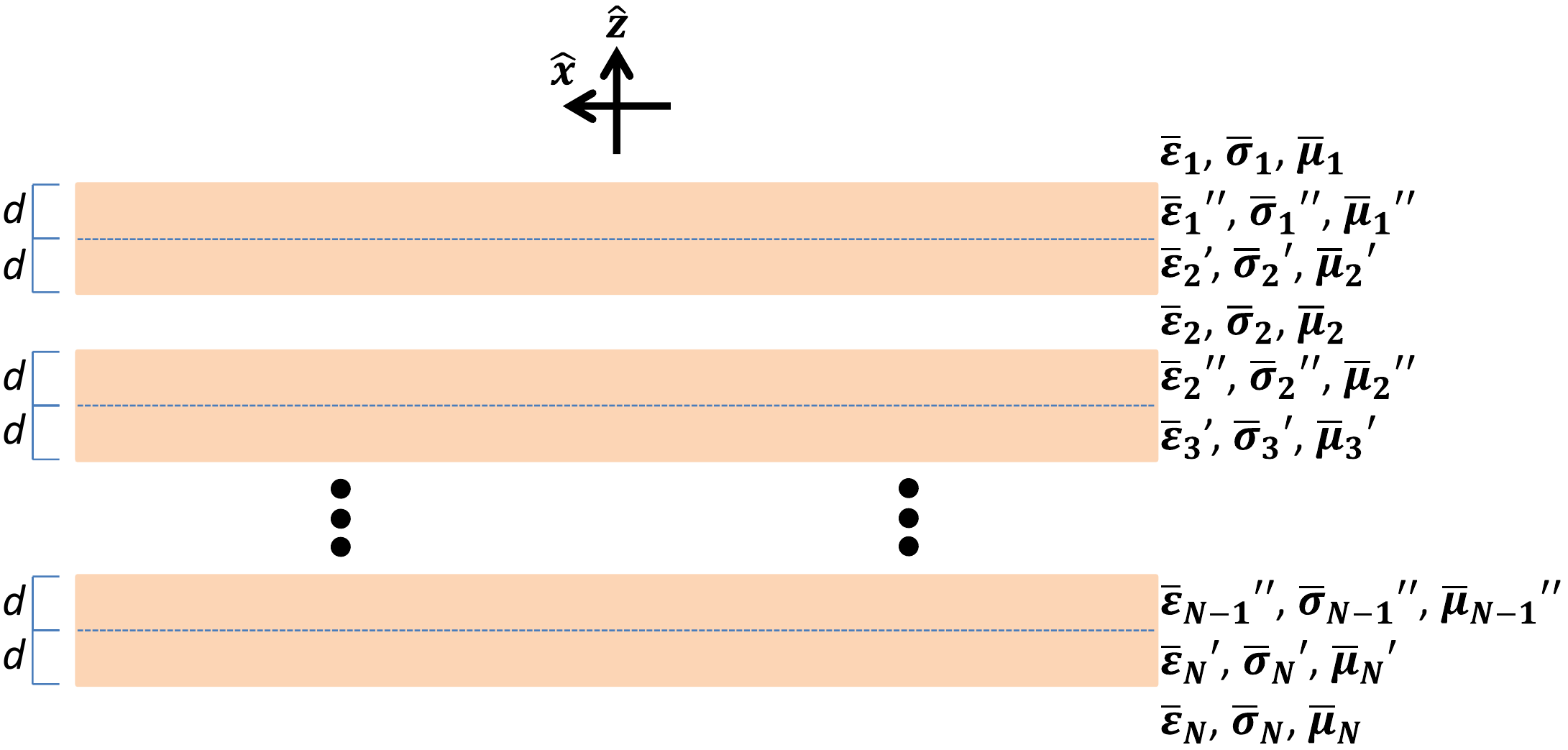}}
\caption{\label{TiltProb} \small (Color online) Figure \ref{Fig1a} shows the original problem with tilted planar interfaces in an $N$-layer geological formation possessing the EM material tensors $\{\boldsymbol{\bar{\epsilon}}_p,\boldsymbol{\bar{\mu}}_p,\boldsymbol{\bar{\sigma}}_p \}$. Figure \ref{Fig1b} shows the transformed, equivalent problem obtained through employing special ``interface-flattening" media (c.f. Eqn. \eqref{eq1}) that coat the underside ($\{\boldsymbol{\bar{\epsilon}}_{m+1}',\boldsymbol{\bar{\mu}}_{m+1}',\boldsymbol{\bar{\sigma}}_{m+1}' \}$) and over-side ($\{\boldsymbol{\bar{\epsilon}}_m^{\prime\prime},\boldsymbol{\bar{\mu}}_m^{\prime\prime},\boldsymbol{\bar{\sigma}}_m^{\prime\prime}\}$) of the $m$th interface. $d$ represents the thickness of each T.O. slab in meters. For simplicity of illustration, all interfaces here are tilted within the $xz$ plane (i.e., interface-tilting azimuth orientation angles $\{\beta_m'\}=0^{\circ}$).}
\end{figure}

Admit an $N$-layer medium where the $m$th planar interface ($m$=1,2,...,$N-1$) is characterized as follows. First, its upward-pointing area normal vector $\bold{\hat{z}}_m'$ is rotated by polar angle $-90^{\circ} \leq \alpha_m' \leq 90^{\circ}$ relative to the $z$ axis,\footnote{Albeit as becomes apparent below, this ``polar" angle corresponds to rotation in the direction \emph{opposite} to that ascribed to the spherical coordinate system.} and azimuth angle $0^{\circ} \leq \beta_m' \leq 180^{\circ}$ relative to the $x$ axis. Second, the $m$th interface's ``depth" $z_m'$ is defined at the Cartesian coordinate system's transverse origin $(x,y)=(0,0)$. See Figure \ref{TiltProb} for a schematic illustration of the environment geometry's parametrization.
To make the $m$th planar interface parallel to the $xy$ plane yet retain its tilted-interface scattering characteristics, as the first of two steps we abstractly define, within the two slab regions ($[z_m'-d] \leq z < z_m'$ and $z_m' \leq z < [z_m'+d]$) bounding the $m$th interface, a ``coordinate stretching" transformation. Namely this transformation relates Cartesian coordinates ($x,y,z$), which parametrize the coordinate mesh of standard ``flat space", to new oblique coordinates ($\bar{x},\bar{y},\bar{z}$) that parametrize an imaginary ``deformed" space whose coordinate mesh deformation systematically induces in turn a well-defined distortion of the EM wave amplitude profile within said slab regions~\cite{pendry6,pendry5,pendry1}:\footnote{We remark in passing upon a strong similarity between the coordinate transform shown in Eqn. \eqref{transform1} versus the ``refractor" and ``beam shifter" coordinate transforms prescribed elsewhere~\cite{lin1,pendry5}. However, while the beam shifter transform (and the equivalent anisotropic medium it describes~\cite{pendry5}) is perfectly reflectionless due to \emph{continuously} transitioning the coordinate transformation back to the ambient medium (e.g., free space), our coordinate transformation is inherently \emph{discontinuous}. Indeed, note in Eqn. \eqref{transform1} that the mapping $\bar{z}$, which depends on $x$ and $y$ in addition to $z$, can not be made to continuously transition back to the (identity) coordinate transform $\bar{z}(z)=z$ implicitly present within the ambient medium. Alternatively stated, our defined anisotropic coating slabs have the exact same material properties as the beam shifter, but our slabs border the ambient medium at planes that, though parallel to each other, are orthogonal relative to the junction planes between the beam shifter and its ambient medium. See other references for the importance of the junction surface's orientation in ensuring a medium perfectly impedance-matched to the ambient medium~\cite{teixeira14,teixeira16}.}
\begin{equation}\label{transform1}
\bar{x} =x , \ \bar{y}=y , \ \bar{z}=z + a_m x + b_m y
\end{equation}
where $a_m=-\tan{\alpha_m'}\cos{\beta_m'}$ and $b_m=-\tan{\alpha_m'}\sin{\beta_m'}$. Indeed this coordinate transform will cause wave fronts to interact with the $m$th flattened interface \emph{as if} said interface were geometrically defined by the equation $z = z_m' - a_m x - b_m y$ rather than $z=z_m'$. As the second step in the interface-flattening procedure, we invoke a ``duality" (not to be confused with duality between the Ampere and Faraday Laws) between spatial coordinate transformations and doubly-anisotropic EM media which ``implement" the effects, of an effectively deformed spatial coordinate mesh (and hence effectively deformed spatial metric tensor), on EM waves propagating through flat space (see references deriving this ``duality"~\cite{pendry1,pendry6,pendry7,teixeiraJEWA,leonhardt}). Following one of two common, equivalent conventions~\cite{pendry1,teixeiraJEWA} leading seamlessly from coordinate transformation to equivalent anisotropic material properties, by defining the Jacobian coordinate transformation tensor~\cite{pendry1}:
\begin{equation}
\boldsymbol{\bar{\Lambda}}_m=\begin{bmatrix} \frac{\partial \bar{x}}{\partial x} & \frac{\partial \bar{y}}{\partial x} & \frac{\partial \bar{z}}{\partial x} \\ \frac{\partial \bar{x}}{\partial y} & \frac{\partial \bar{y}}{\partial y} & \frac{\partial \bar{z}}{\partial y} \\ \frac{\partial \bar{x}}{\partial z} & \frac{\partial \bar{y}}{\partial z} & \frac{\partial \bar{z}}{\partial z} \end{bmatrix}=\begin{bmatrix}1&0&a_m\\0&1&b_m\\ 0 & 0 & 1 \end{bmatrix} \label{eq0}
\end{equation}
within the region $(z_m'-d) \leq z < z_m'$ one has the interface-flattening material tensors $\{ \boldsymbol{\bar{\gamma}}_{m+1}'\}$ in place of the original formation's material parameters $\{ \boldsymbol{\bar{\gamma}}_{m+1}\}$ within layer $p=m+1$ ($\gamma=\epsilon,\mu,\sigma$). Similarly, within the region $z_m' \leq z < (z_m'+d)$ one has the interface-flattening material tensors $\{ \boldsymbol{\bar{\gamma}}_{m}^{\prime\prime}\}$ in place of the original formation's material parameters $\{ \boldsymbol{\bar{\gamma}}_{m}\}$ within layer $m$. How are the interface-flattening material tensors defined though? Quite simply, in fact, and this definition holds \emph{regardless} of the original formation layer's anisotropy and loss (``T" superscript denotes non-Hermitian transpose)~\cite{pendry1}:
\begin{equation}
\boldsymbol{\bar{\gamma}}_{m+1}'=\boldsymbol{\bar{\Lambda}}_{m}^{\mathrm{T}}\cdot \boldsymbol{\bar{\gamma}}_{m+1}\cdot\boldsymbol{\bar{\Lambda}}_{m},\ \boldsymbol{\bar{\gamma}}_{m}^{\prime \prime}=\boldsymbol{\bar{\Lambda}}_{m}^{\mathrm{T}}\cdot \boldsymbol{\bar{\gamma}}_{m}\cdot\boldsymbol{\bar{\Lambda}}_{m}
\numberthis \label{eq1}
\end{equation}
Note that if the $m$th interface lacks effective tilt then $\boldsymbol{\bar{\Lambda}}_m$ reduces to the identity matrix, which in turn leads to the two interface-flattening media reducing to the media of the respective formation layers from which they were derived using Eqn. \eqref{eq1}: $\boldsymbol{\bar{\gamma}}_{m+1}'=\boldsymbol{\bar{\gamma}}_{m+1}$ and $\boldsymbol{\bar{\gamma}}_{m}^{\prime\prime}=\boldsymbol{\bar{\gamma}}_{m}$ (as expected!). Now the new material profile, characterized by \emph{parallel} planar interfaces, appears for a simple three-layer, two-interface geometry as:
\begin{align}
\boldsymbol{\bar{\gamma}}_{1} &,\ (z_1'+d) \leq z < \infty \\
\boldsymbol{\bar{\gamma}}_{1}^{\prime\prime} &,\ z_1' \leq z < (z_1'+d) \\
\boldsymbol{\bar{\gamma}}_{2}' &, \ (z_1'-d) \leq z < z_1' \\
\boldsymbol{\bar{\gamma}}_{2} &, \ (z_2'+d) \leq z < (z_1'-d) \\
\boldsymbol{\bar{\gamma}}_{2}^{\prime\prime} &, \ z_2' \leq z < (z_2'+d) \\
\boldsymbol{\bar{\gamma}}_{3}' &, \ (z_2'-d) \leq z < z_2' \\
\boldsymbol{\bar{\gamma}}_{3} &, \ -\infty < z < (z_2'-d)
\end{align}
with an analogous material profile resultant for $N >3 $ layers.

There are two advisories worth mentioning. First, we recommend adaptively (i.e., depending on the transmitter and receiver positions) reducing the thickness $d$ of coating layer(s), within which receiver(s) and/or transmitter(s) may reside depending on their depth, just enough so that the receivers and transmitters are located once more within the formation layers. Why this recommendation? Although pseudo-analytical techniques are available to compute fields when the receiver and/or transmitter are located in such layers~\cite{sainath8}, the main reason is to eliminate spurious discontinuities of the normal ($z$ in our case) electric and magnetic field components manifest when the source (or, as can be anticipated from EM reciprocity, the receiver) traverse a boundary separating a true formation layer and a coating slab~\cite{teixeiraJEWA}. Second, the thickness $d$ of the coating slabs must also be adjusted to ensure the coating layer just beneath the $m$th interface does not cross over into the coating layer just above the $(m+1)$st interface. We account for these two points within our numerical results below. 

\section{Error Analysis}
\subsection{\label{Err}Overview}
To briefly recap: The proposed method relies upon insertion of specially-prescribed, doubly-anisotropic material slabs above and below each (effectively) tilted original interface to manipulate wave-fronts such that they interact with the coated interfaces \emph{as if} they were tilted. This technique allows one, in principle, to unequivocally and independently prescribe the arbitrary, effective polar and azimuthal tilting orientation of each interface. Moreover, in the limit of vanishingly small effective polar tilt for some interface (and irrespective of effective azimuth tilt), the material properties of the slab just above (below) this interface \emph{continuously} transitions back to the material properties of the ambient medium just above (below) said interface. However, as the material properties of each coating slab are (for non-zero effective tilt) \emph{not} perfectly impedance matched to the respective ambient medium into which it is inserted, spurious scattering will result that coherently interferes with the true field scattered from the effectively tilted interface. 
 
The spurious scattering corrupts the true responses (both co-polarized and cross-polarized ones) arising from tilt and hence needs to be quantified if we are to understand the practical limitations of the proposed tilted-layer algorithm, which is inherently \emph{perturbational} in nature with respect to the range of (effective) polar tilt that can be reliably modeled. It is also important to understand the error trends not just versus effective polar tilt, but also for different geophysical media (anisotropy and loss), transmitter/receiver spacing, sensor positions (relative to the interface), transmitter and receiver orientation, and thickness $d$ of the coating slabs. To simplify the geometry and admit a closed-form reference solution, we examine only a two-layered medium with an interface effectively tilted within the $xz$ plane by $\alpha=\alpha_1'$ degrees. Both the reference ($H_{wq}^{\mathrm{Ref}}$) and algorithm/transformed-domain ($H_{wq}^{\mathrm{TO}}$) field values are computed to fourteen Digits of Precision (DoP), with the relative error between their respective field values (either real or imaginary part) denoted $\epsilon$ ($-\mathrm{Log}_{10}\epsilon$ denotes DoP agreement): Either $\epsilon=|\mathrm{Re}[H_{wq}^{\mathrm{TO}}]-\mathrm{Re}[H_{wq}^{\mathrm{Ref}}]|/|\mathrm{Re}[H_{wq}^{\mathrm{Ref}}]|$ or $\epsilon=|\mathrm{Im}[H_{wq}^{\mathrm{TO}}]-\mathrm{Im}[H_{wq}^{\mathrm{Ref}}]|/|\mathrm{Im}[H_{wq}^{\mathrm{Ref}}]|$, where Re$[H_{wq}]=H_{wq}'$ and Im[$H_{wq}]=H_{wq}^{\prime \prime}$ denote the real and imaginary parts (resp.) of the $q$-oriented magnetic field component observed at the receiver due to the $w$-directed magnetic dipole transmitter $H_{wq}$ ($w,q=x,y,z$). Errors below $10^{-14}$ were artificially coerced in post-processing to $10^{-14}$.

In order to compute the reference field solution in this two-layer scenario, we first denote the transmitter-receiving spacing $L_s$, transmitter depth $z'$, and receiver depth $z=z'+L_s$ in the ``transformed" domain with a flat, parallel (to the $xy$ plane) interface (residing at $z_1'=0$) with two coating slabs.\footnote{Note: For all results in this paper, we assume a vertically-oriented sensor in the transformed domain.} In the equivalent domain, involving again a flat interface ($z_1'=0$) but with a rotated sensor, the new transmitter position ($x_T',y_T'=0,z_T'$) writes as $z_T'=z'\cos{\alpha}$ and $x_T'=z'\sin{\alpha}$, the new receiver position ($x_T,y_T=0,z_T$) writes as $z_T=(z'+L_s)\cos{\alpha}$ and $x_T=(z'+L_s)\sin{\alpha}$, the transmitting dipole orientations are physically rotated by $-\alpha$ degrees in the $xz$ plane, and the receiver antennas are also physically rotated by $-\alpha$ degrees in the $xz$ plane.\footnote{Equivalently, the observed fields at the receiver location are now observed relative to a cartesian coordinate system rotated by $-\alpha$ degrees.} Additionally, for the latter two material profiles (described, and denoted M3 and M4, below) involving anisotropic media, we rotate the anisotropic material tensors by $-\alpha$ degrees (isotropic tensors are invariant under rotation, by definition). Throughout this error study, when computing the reference and transformed-domain solutions the transmitter radiation frequency is held fixed at $f$=100kHz. Furthermore, both the reference and transformed-domain solutions are computed using the same numerical code, albeit with effective interface tilt set to zero for the reference solutions.

We show relative error $\mathrm{Log}_{10}\epsilon[H_{wq}']$ and $\mathrm{Log}_{10}\epsilon[H_{wq}^{\prime \prime}]$ versus the effective polar tilt angle $\alpha<0$ (degrees) and azimuth angle (fixed at $\beta'=0^{\circ}$). To better illustrate error trends (e.g., linear or quadratic) versus $\alpha$, we plot this Log-scale error on the vertical axis and Log$_{10}|\alpha|$ on the horizontal axis. To show error variations versus material profile and transmitter-receiver spacing, in each plot we exhibit relative errors for two different sensor spacings $L_s$ (=400mm [``S1"] and =1m [``S2"]) and four different conductivity profiles (each having two layers, with $\epsilon_r=\mu_r=1$, prior to inserting the two coating slabs): 
\begin{enumerate}
\item $\sigma_1=1$mS/m, $\sigma_2=2$mS/m ([``M1"]- Both layers isotropic, 2:1 conductivity contrast)
\item $\sigma_1=1$mS/m, $\sigma_2=20$mS/m ([``M2"]- Both layers isotropic, 20:1 conductivity contrast)
\item $\sigma_1=1$mS/m, $\sigma_{h2}=5$mS/m, $\sigma_{v2}=1$mS/m, $\alpha_2=60^{\circ}$, $\beta_2=0^{\circ}$ ([``M3"]- Deviated uniaxial bottom layer)
\item $\sigma_1=1$mS/m, $\sigma_{x2}=5$mS/m, $\sigma_{y2}=2.5$mS/m, $\sigma_{z2}=1$mS/m, $\alpha_2=\beta_2=0^{\circ}$ ([``M4"]- Non-deviated biaxial bottom layer)
\end{enumerate}
where the anisotropic tensor components $\{\sigma_{h2},\sigma_{v2},\sigma_{x2},\sigma_{y2},\sigma_{z2}\}$ and tensor spherical (polar and azimuth) deviation angles $\{\alpha_2,\beta_2\}$ define the principal conductivity values and orientation (resp.) of the conductivity tensor relative to the cartesian axes (discussed in Section \ref{Val} and~\cite{anderson1}). The curve labeled ``SXMY" refers to the sensor spacing denoted above in square brackets by [``$SX$"] ($X$=1,2) and the material scenario denoted above in square brackets by [``$MY$"] ($Y$=1,2,3,4). Fixing the flattened interface's depth at $z_1'=0$, for two coating slab thicknesses ($d=$2mm [``d1"] and $d=200$mm [``d2"]) we examine three mid-point sensor depths in the transformed domain: $D=2$m (transmitter and receiver in top layer [``O1"]), $D=0$m (transmitter in bottom layer, receiver in top layer [``O2"]), and $D=-2$m (transmitter and receiver in bottom layer [``O3"]). For the real and imaginary part of each field component we examine six sensor-location/$d$ permutations, which are denoted on each page with plot labels (``Scenario:O1d1", etc.). The transmitter depth ($z'$) and receiver depth ($z$) are computed using the sensor spacing $L_s$ and mid-point depth $D$: $z=D+L_s/2$ and $z'=D-L_s/2$.

A final remark before proceeding with error analysis, concerning our choices of examined $d$: One could in principle make the slabs arbitrarily thin or thick, which we have not tried (we only examined, in this error study, $d$=0.2m and $d$=2mm). The algorithm's present design does not dictate a specific ``optimal" value of $d$, unfortunately. What we \emph{can} say however is that making $d$ comparable to the local wavelength, on either side of the interface, is highly undesirable for at least two reasons. First, since the coating slabs are in fact reflective they can confine spurious guided-wave modes that may significantly corrupt computed sensor responses. Second, as indicated in Section 2.2, the thicker $d$ is one must adaptively reduce the thicknesses of slab(s) intersecting each other, as well as slab(s) containing transmitters or receivers, to eliminate artificial discontinuities in the normal field components. This adaptive thickness reduction, which becomes increasingly frequent for finer spatial sampling (i.e., versus sensor depth) of the sensor response profiles, introduces yet another level of arbitrariness into the algorithm. Namely, the desirable minimum space maintained between the receiver and coating layer/ambient medium interfaces (and likewise for the transmitter), as well as between any two ambient medium/coating slab junctions. Although both examined values of $d$ in this error study would not necessitate their adaptive thinning, due to the three specifically examined sensor locations, the thicker $d=$0.2m slabs \emph{would} require adaptive thinning in the Section \ref{Val} results due to the sensor's depth being varied (0.1m sampling period) throughout the studied three-layer formation profiles. 

For the cross-polarized field plots, we only show errors for $H_{xz}$ and $H_{zx}$ since the other cross-polarized field components ($H_{xy}$, $H_{yx}$, $H_{yz}$, $H_{zy}$) had zero magnitude to within numerical noise. Elaborating: Our chosen threshold for numerical noise is that either $|H_{wq}^{\mathrm{TO}}|\leq 10^{-12}$ and/or $|H_{wq}^{\mathrm{Ref}}|\leq 10^{-12}$ (-12 on Log$_{10}$ scale), which is based on the adaptive integration tolerance ($1.2\times 10^{-14}$) and an educated guess ($10^2$) of the maximum magnitude of field components that would experience cancellation during evaluation of the oscillatory Fourier double-integral (Eqns. \eqref{EHM2a}-\eqref{Espace1}).\footnote{Although rigorous justification for the threshold $10^{-12}$ is lacking, the conclusion of negligible $\{H_{xy},H_{yx},H_{yz},H_{zy}\}$ is physically reasonable. Indeed one can verify (using Eqns. \eqref{eq0}-\eqref{eq1}) that T.O. media, used to effect modeling of $xz$ plane tilting, only perturbs the ambient medium's response in the $xz$ plane (but not $y$ direction). Hence one can reason the T.O. media should not induce spurious scattering leading to non-zero $\{H_{xy},H_{yx},H_{yz},H_{zy}\}$ responses (i.e., the cross-pol responses involving a $y$-directed transmitter or receiver) if they were absent without the T.O. slabs.} In passing, we mention that to suppress numerical cancellation-based noise due to using finite-precision arithmetic, we ``symmetrically" integrate. That is to say, we integrate along the integration contour sub-sections $(a,b)$ and $(-b,-a)$, add these results, then integrate along contour sub-sections $(b,c)$ and $(-c,-b)$, add these results and update the accumulated contour integral result, and so on. 

\subsection{\label{Err2}Results and Discussion}

\begin{figure}[H]
\centering
\subfloat[\label{ReHxxO1D1}]{\includegraphics[width=3.25in]{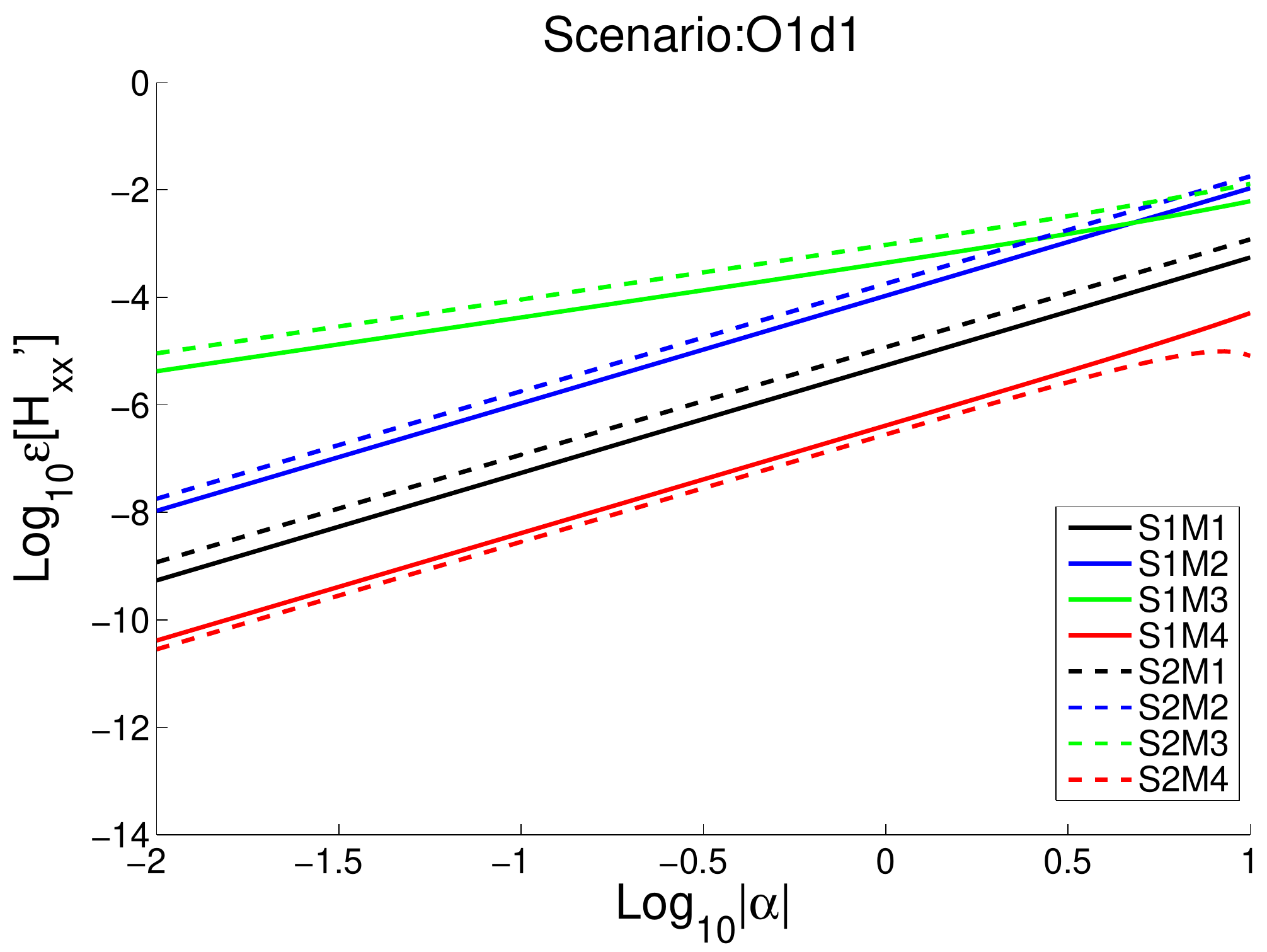}}
\subfloat[\label{ReHxxO1D2}]{\includegraphics[width=3.25in]{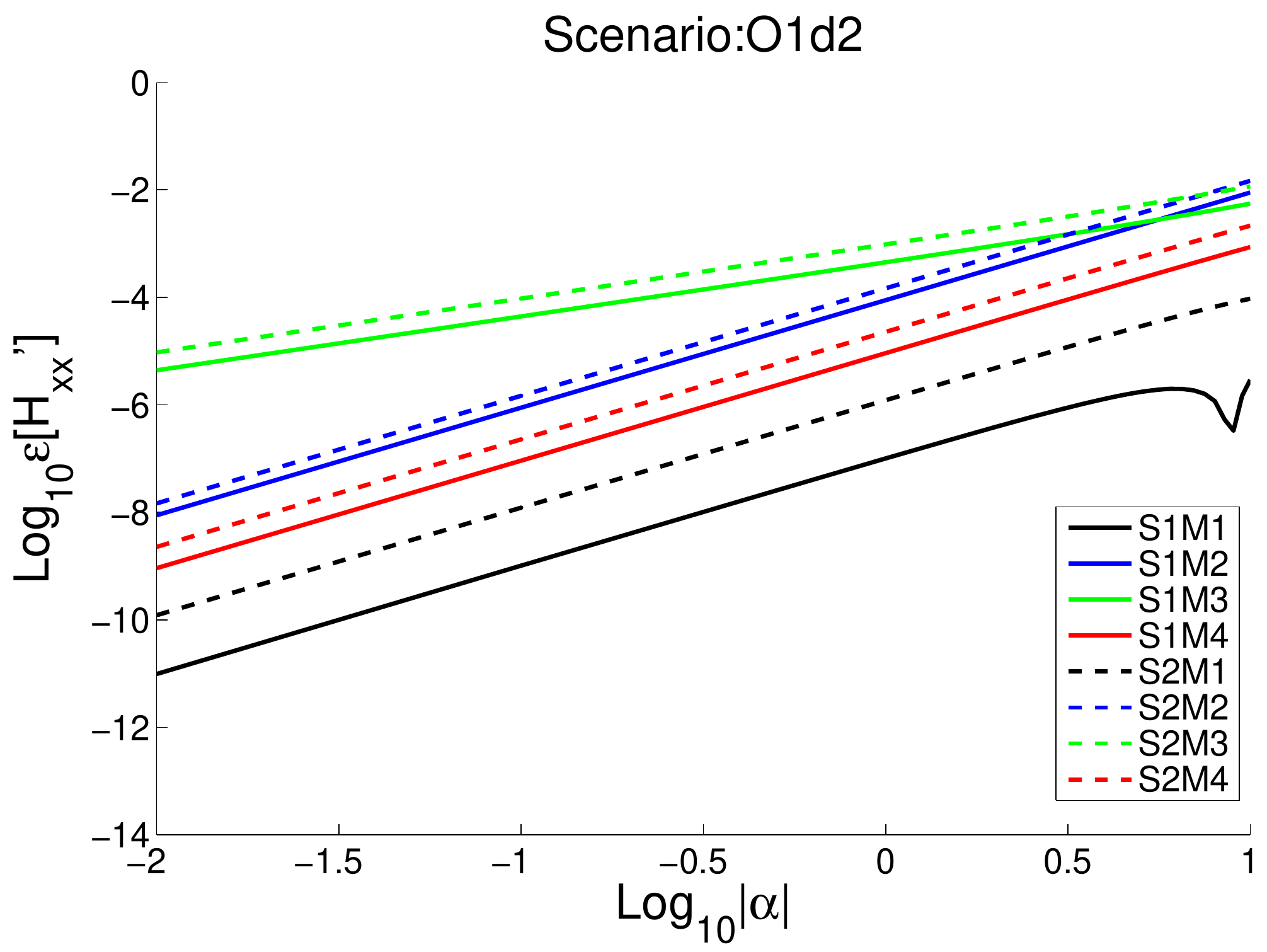}}

\subfloat[\label{ReHxxO2D1}]{\includegraphics[width=3.25in]{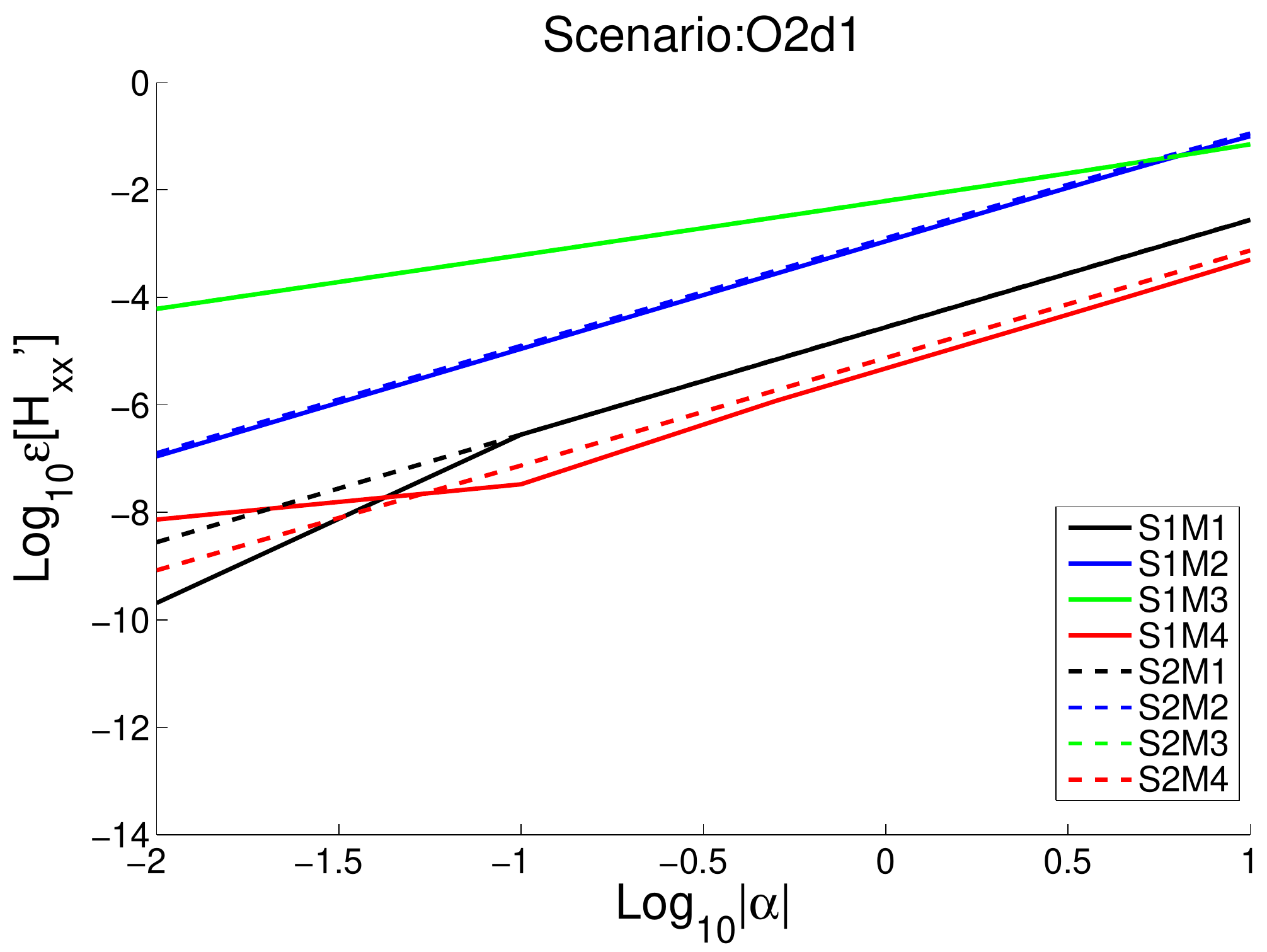}}
\subfloat[\label{ReHxxO2D2}]{\includegraphics[width=3.25in]{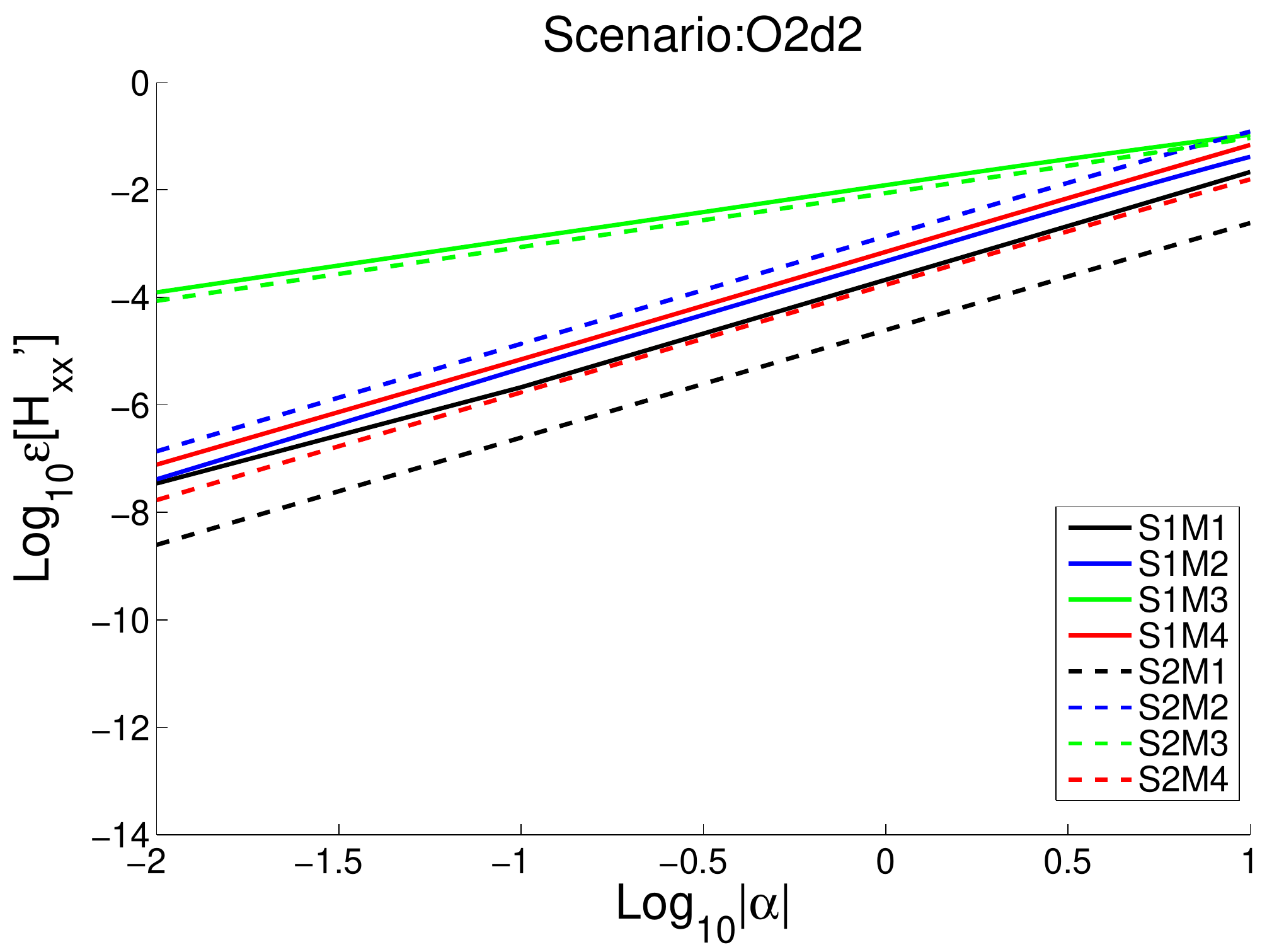}}

\subfloat[\label{ReHxxO3D1}]{\includegraphics[width=3.25in]{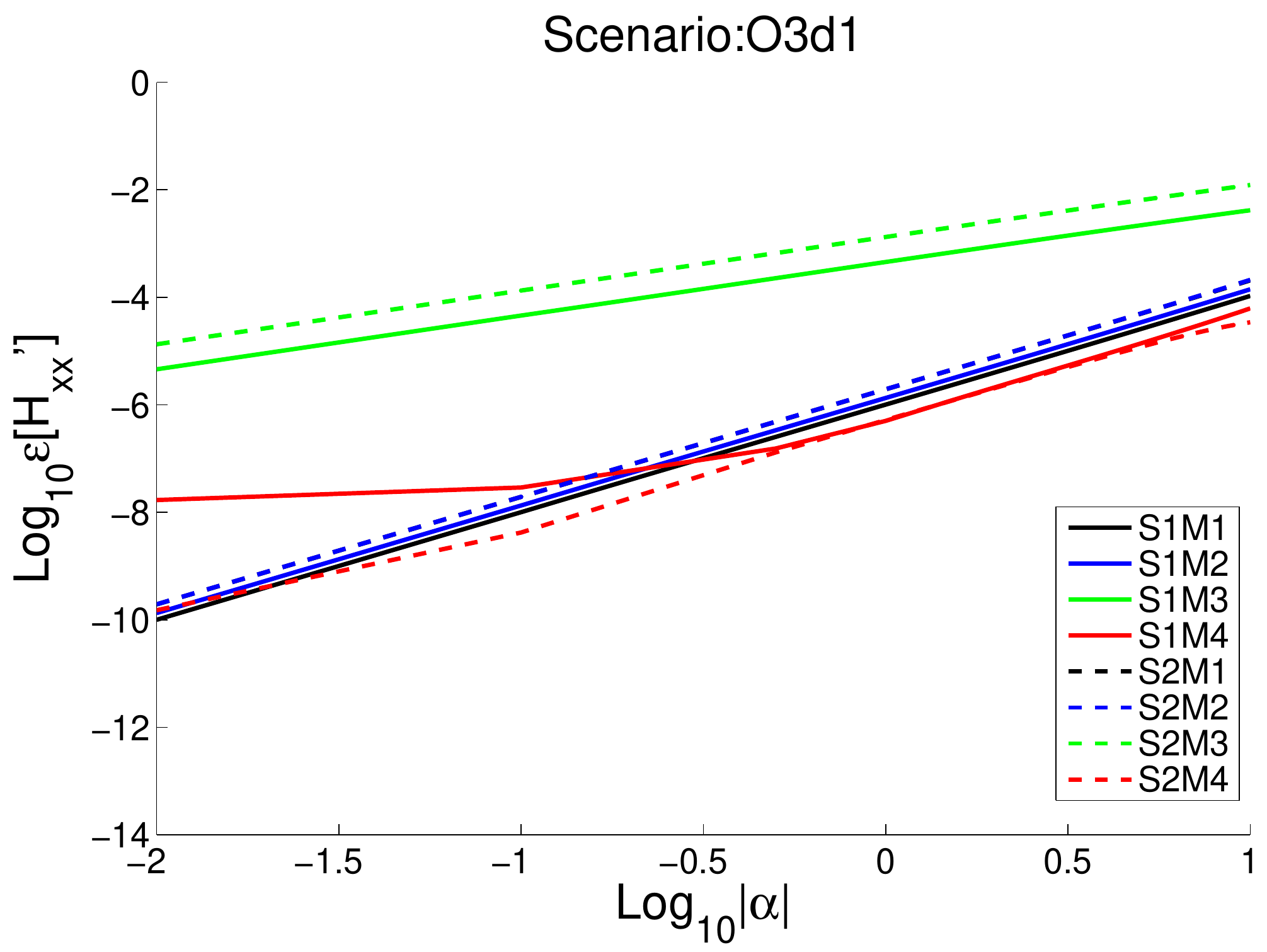}}
\subfloat[\label{ReHxxO3D2}]{\includegraphics[width=3.25in]{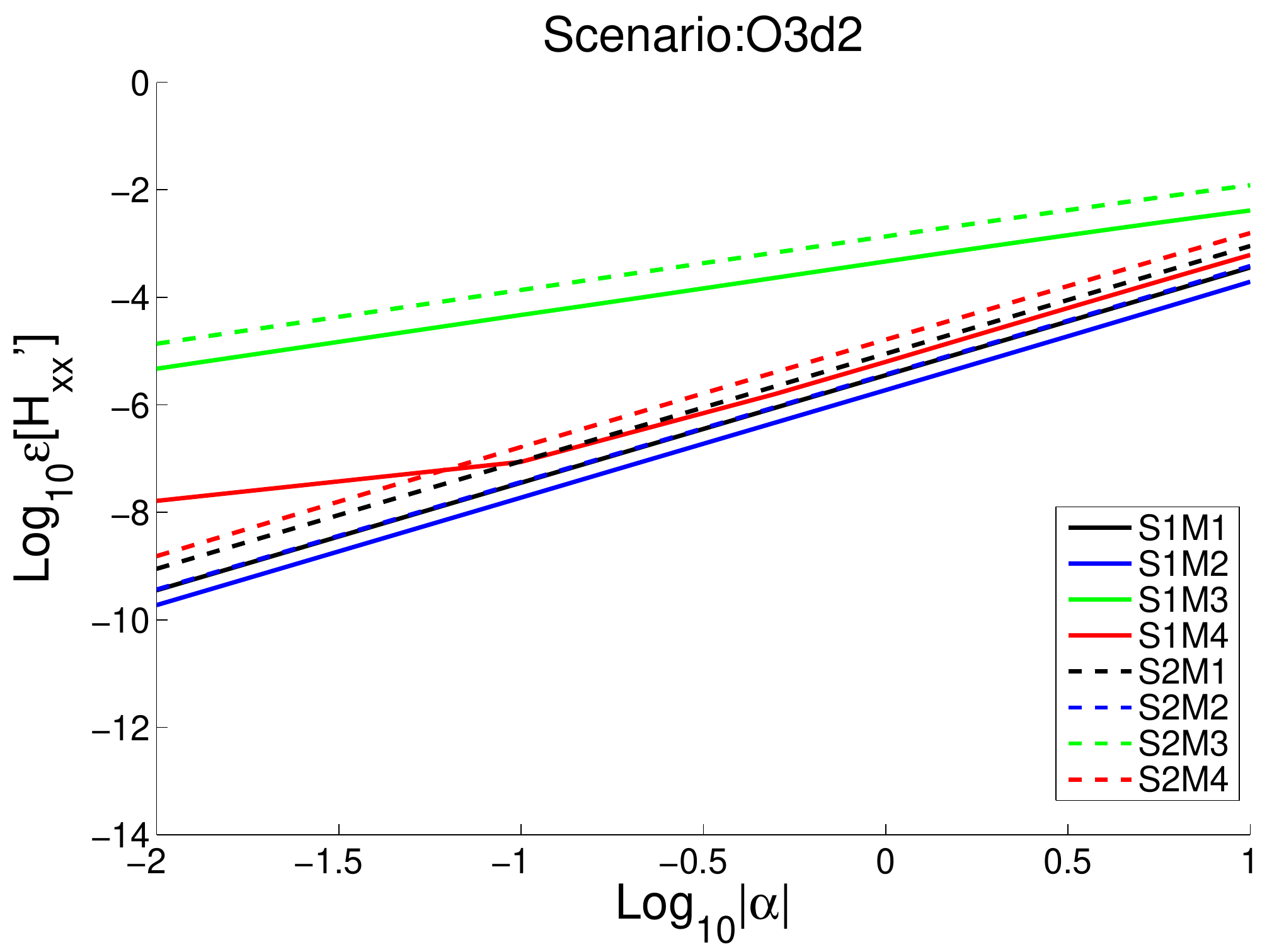}}
\caption{\small Relative error in computing $H_{xx}^{\prime}$=Re[$H_{xx}$].}
\label{ReHxx}
\end{figure}

\begin{figure}[H]
\centering
\subfloat[\label{ImHxxO1D1}]{\includegraphics[width=3.25in]{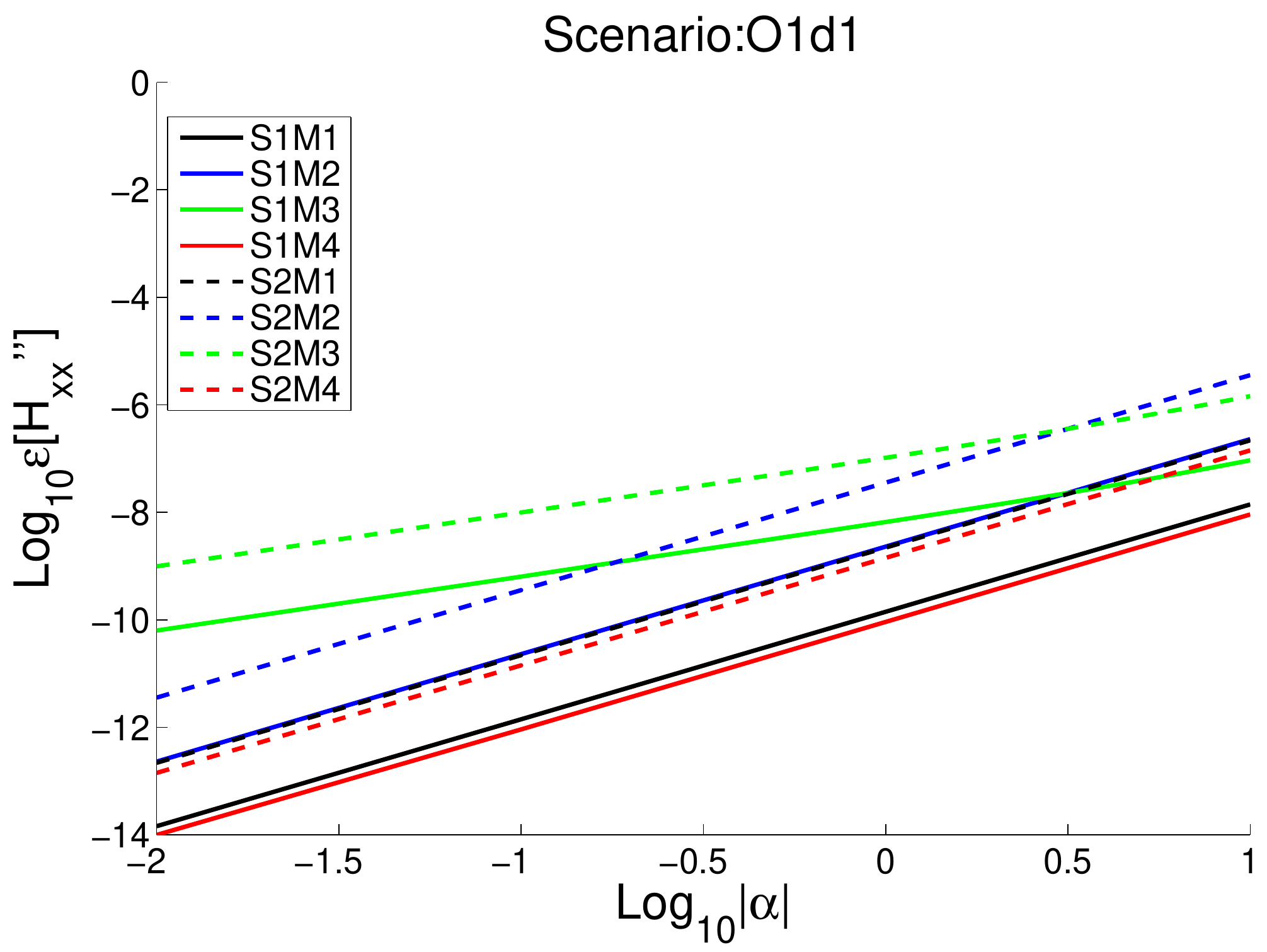}}
\subfloat[\label{ImHxxO1D2}]{\includegraphics[width=3.25in]{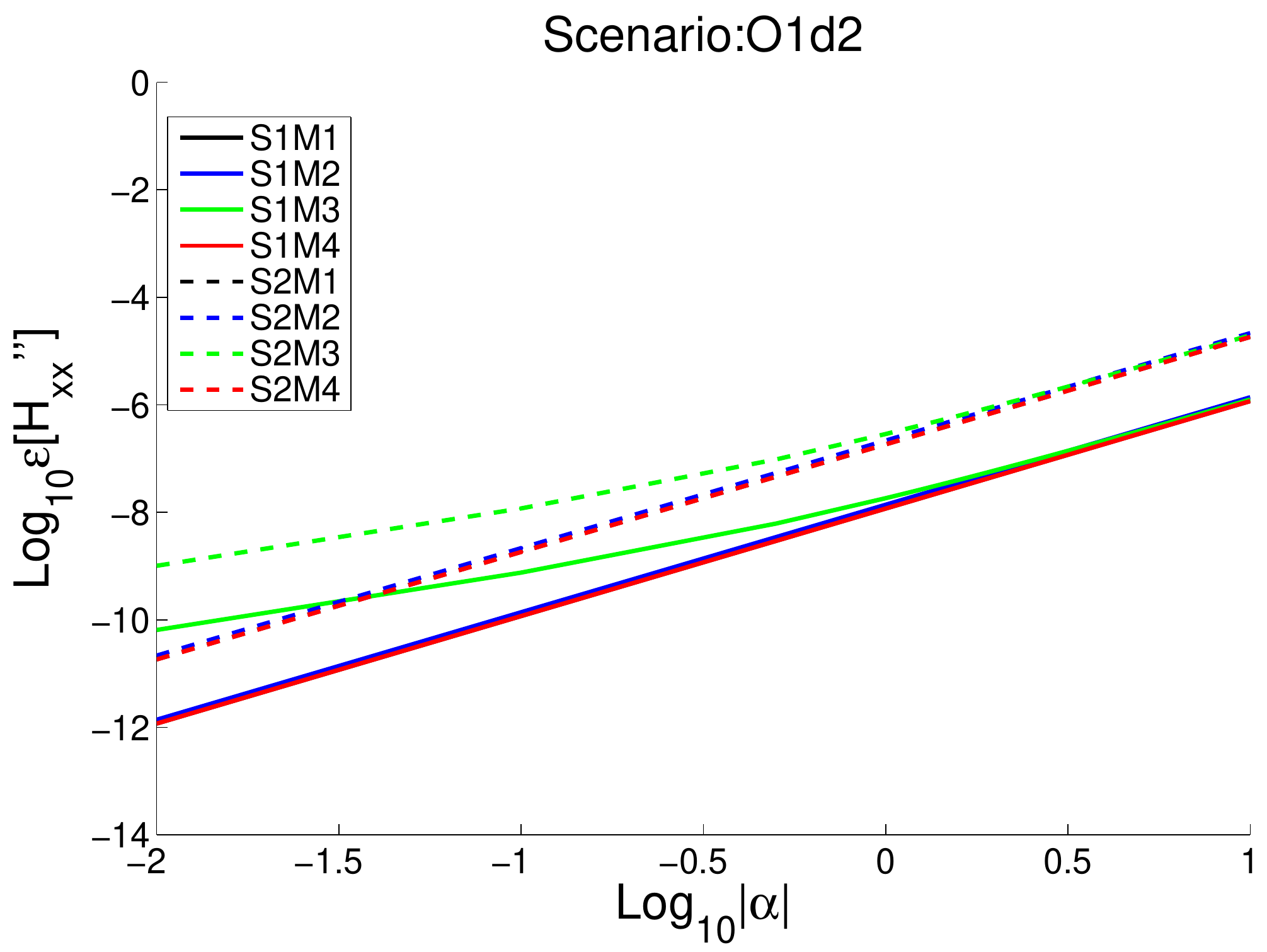}}

\subfloat[\label{ImHxxO2D1}]{\includegraphics[width=3.25in]{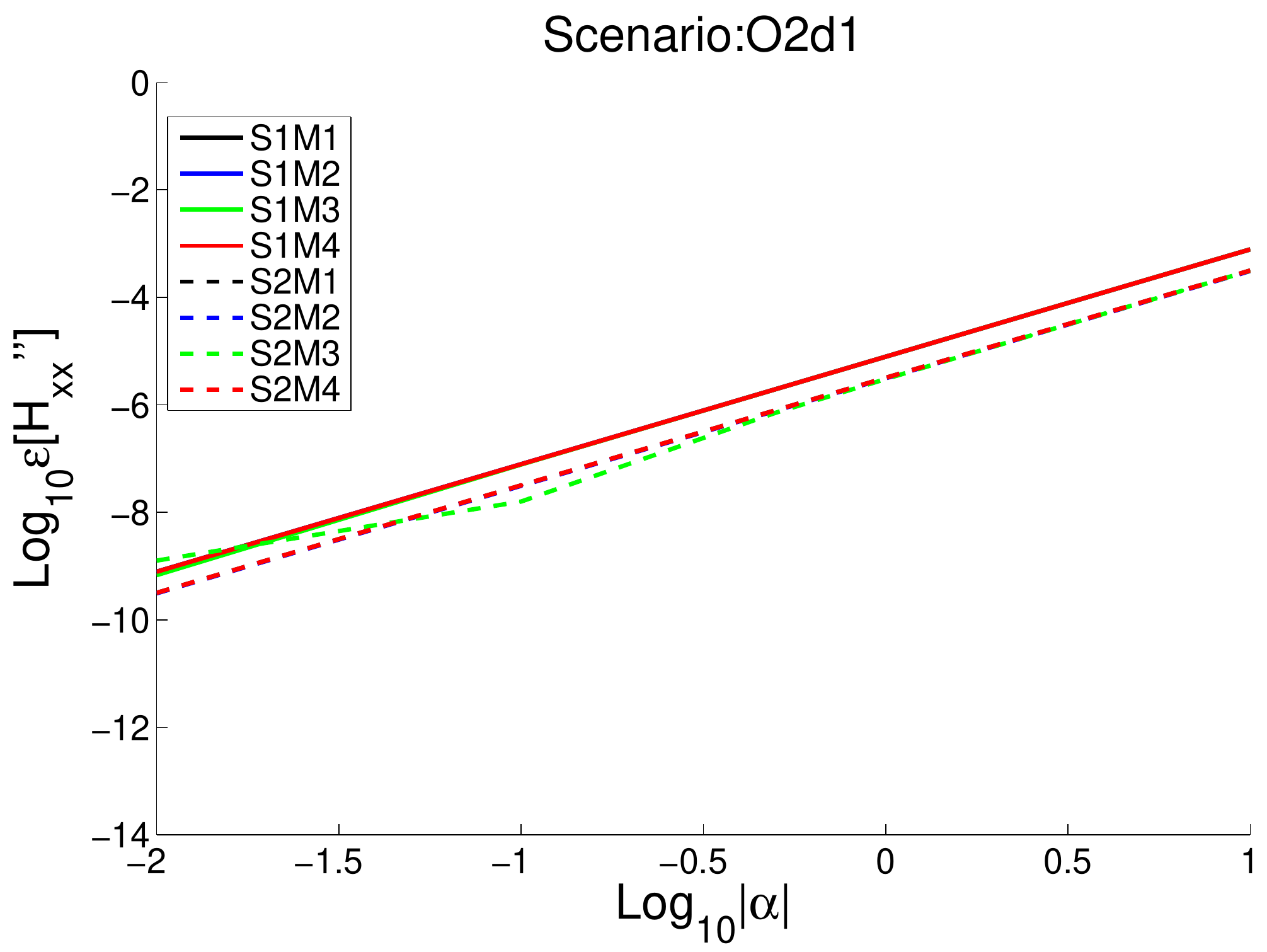}}
\subfloat[\label{ImHxxO2D2}]{\includegraphics[width=3.25in]{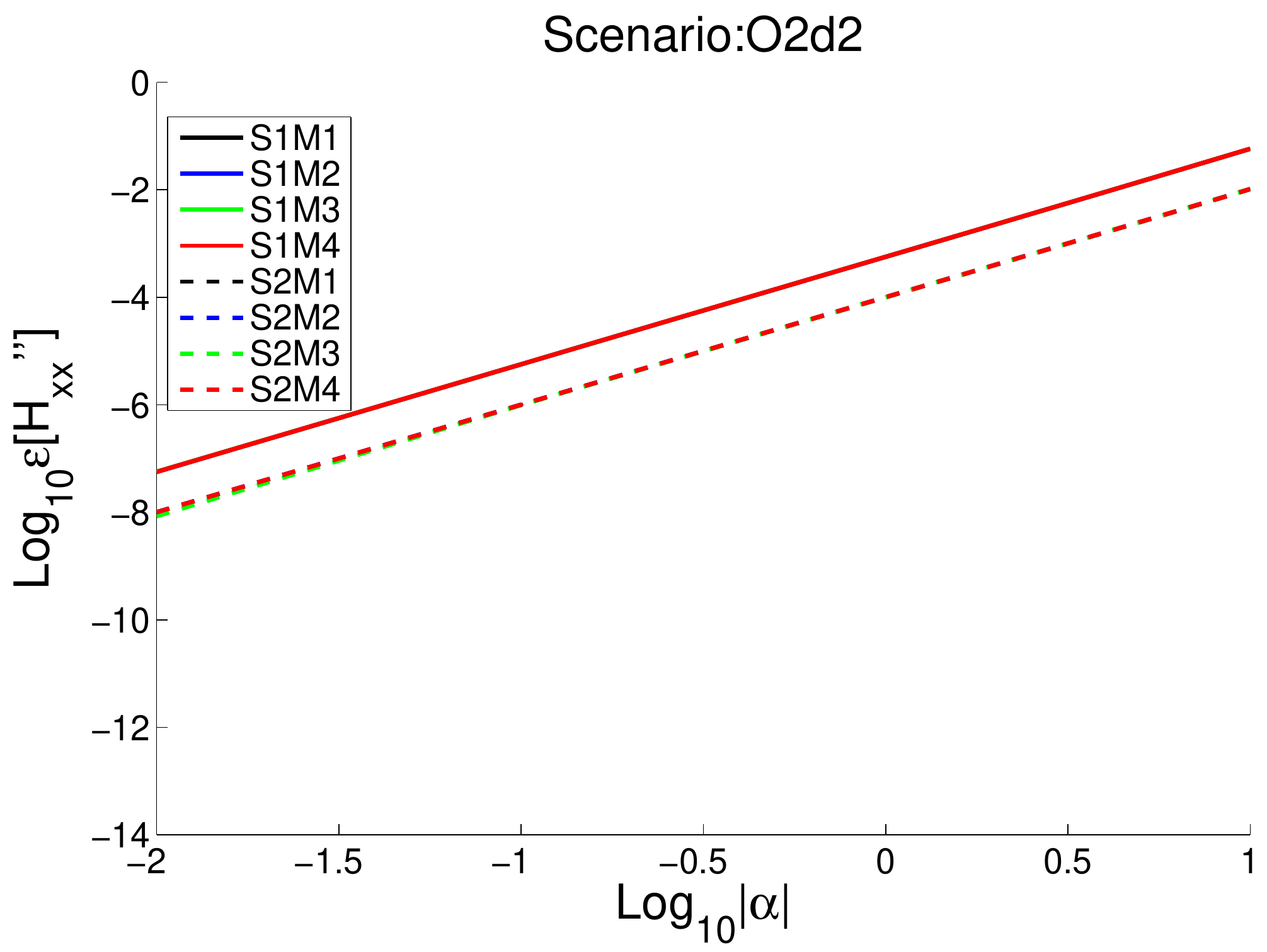}}

\subfloat[\label{ImHxxO3D1}]{\includegraphics[width=3.25in]{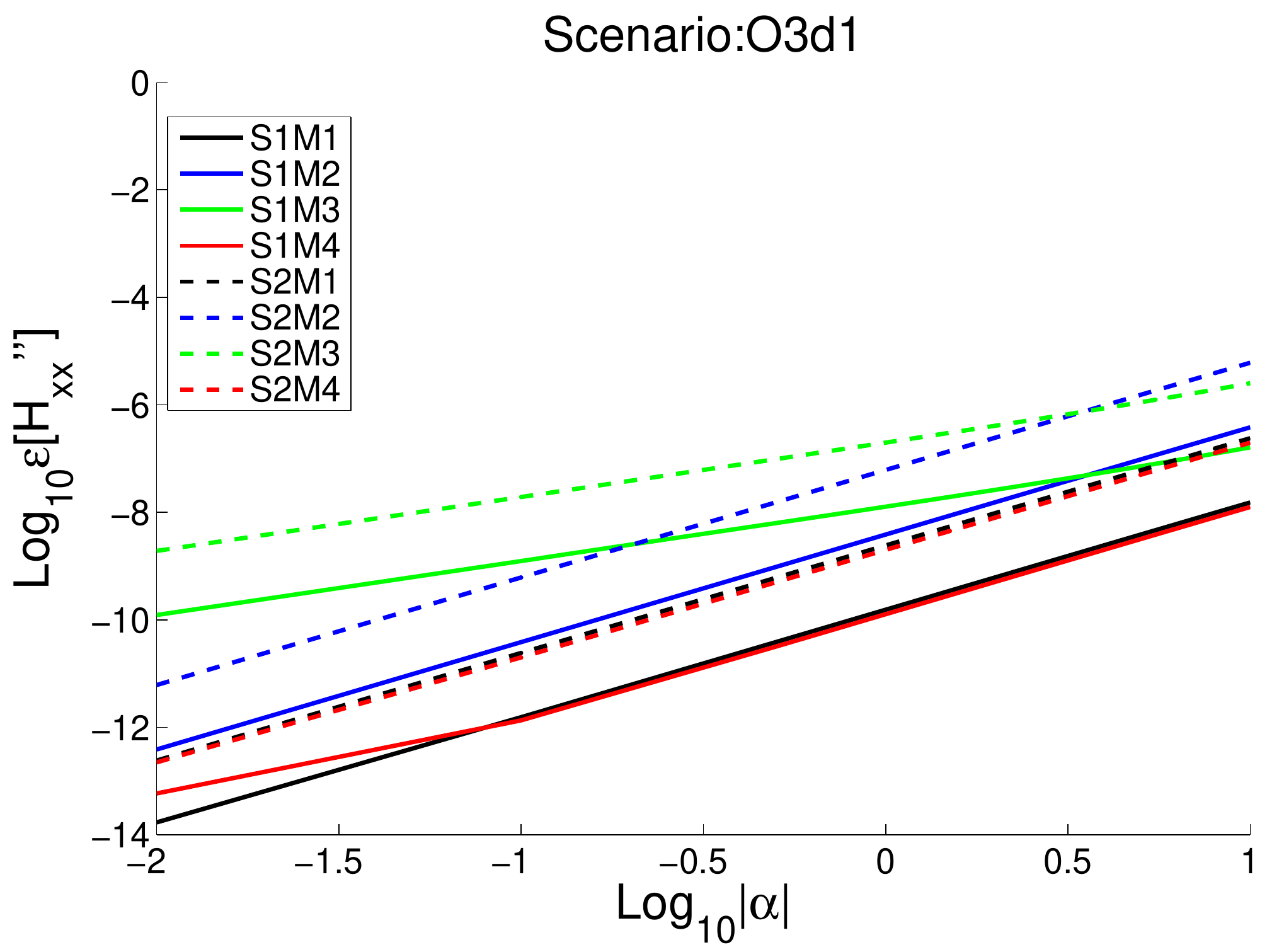}}
\subfloat[\label{ImHxxO3D2}]{\includegraphics[width=3.25in]{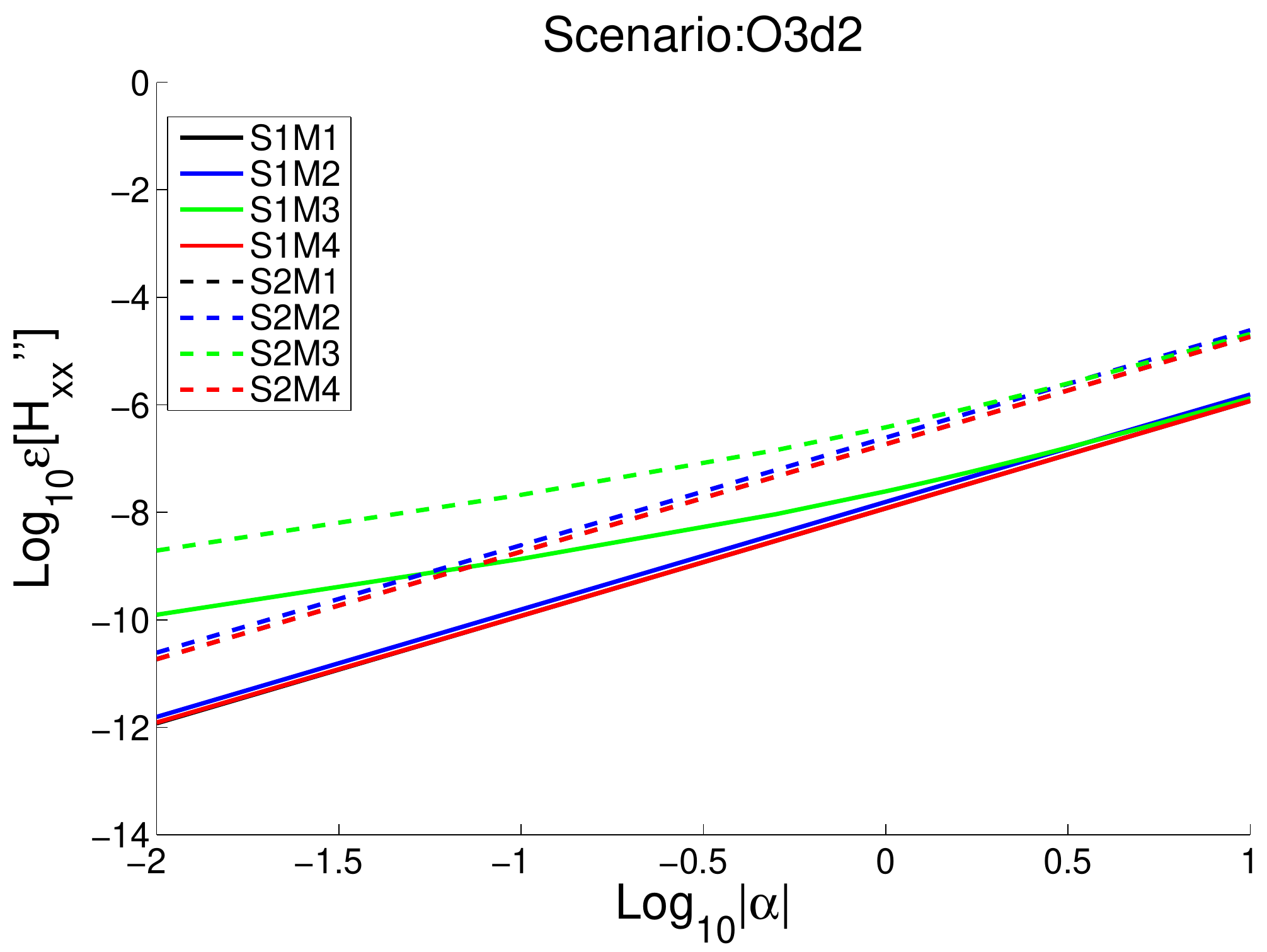}}
\caption{\small Relative error in computing $H_{xx}^{\prime \prime}$=Im[$H_{xx}$].}
\label{ImHxx}
\end{figure}

\begin{figure}[H]
\centering
\subfloat[\label{ReHyyO1D1}]{\includegraphics[width=3.25in]{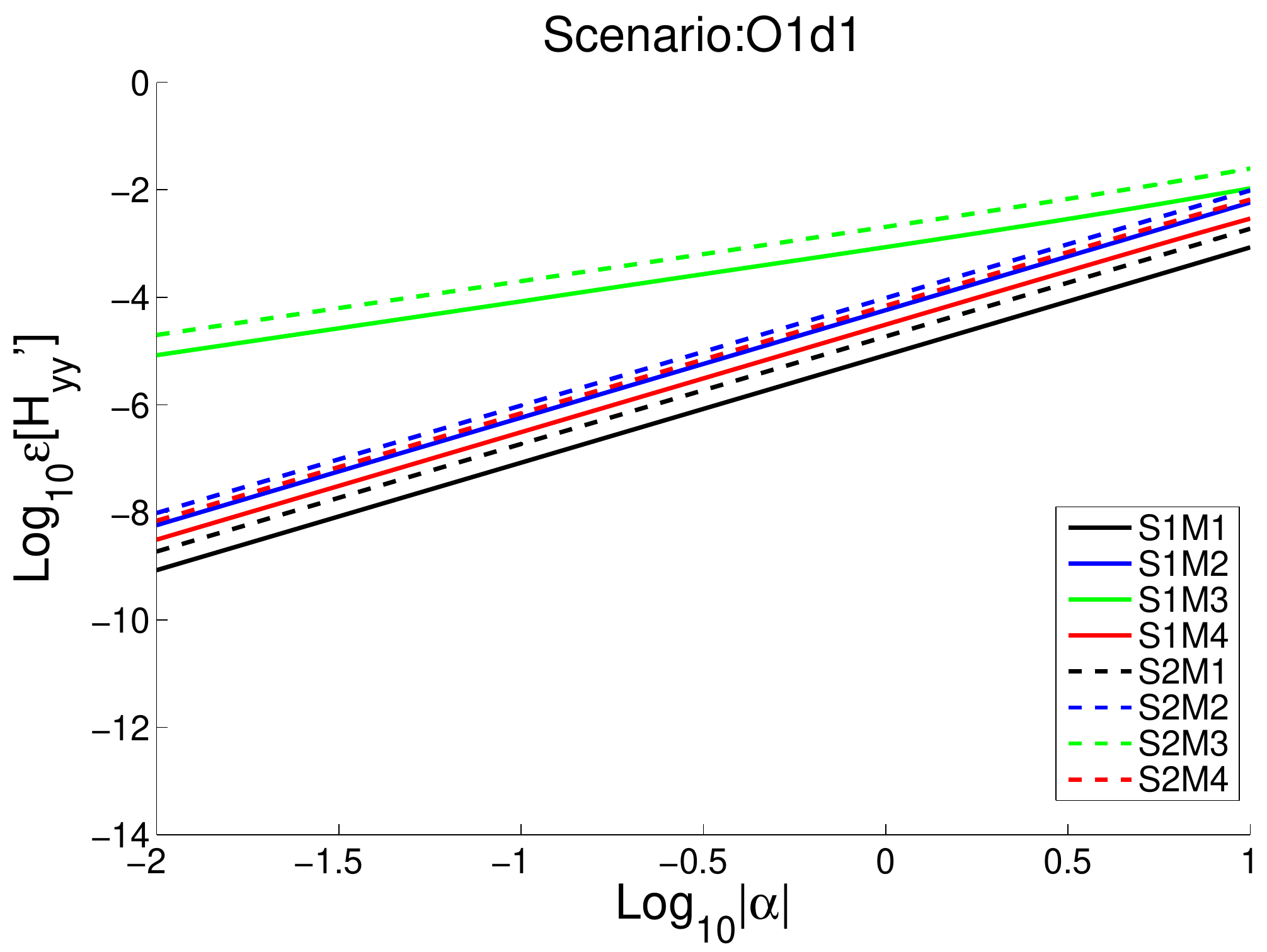}}
\subfloat[\label{ReHyyO1D2}]{\includegraphics[width=3.25in]{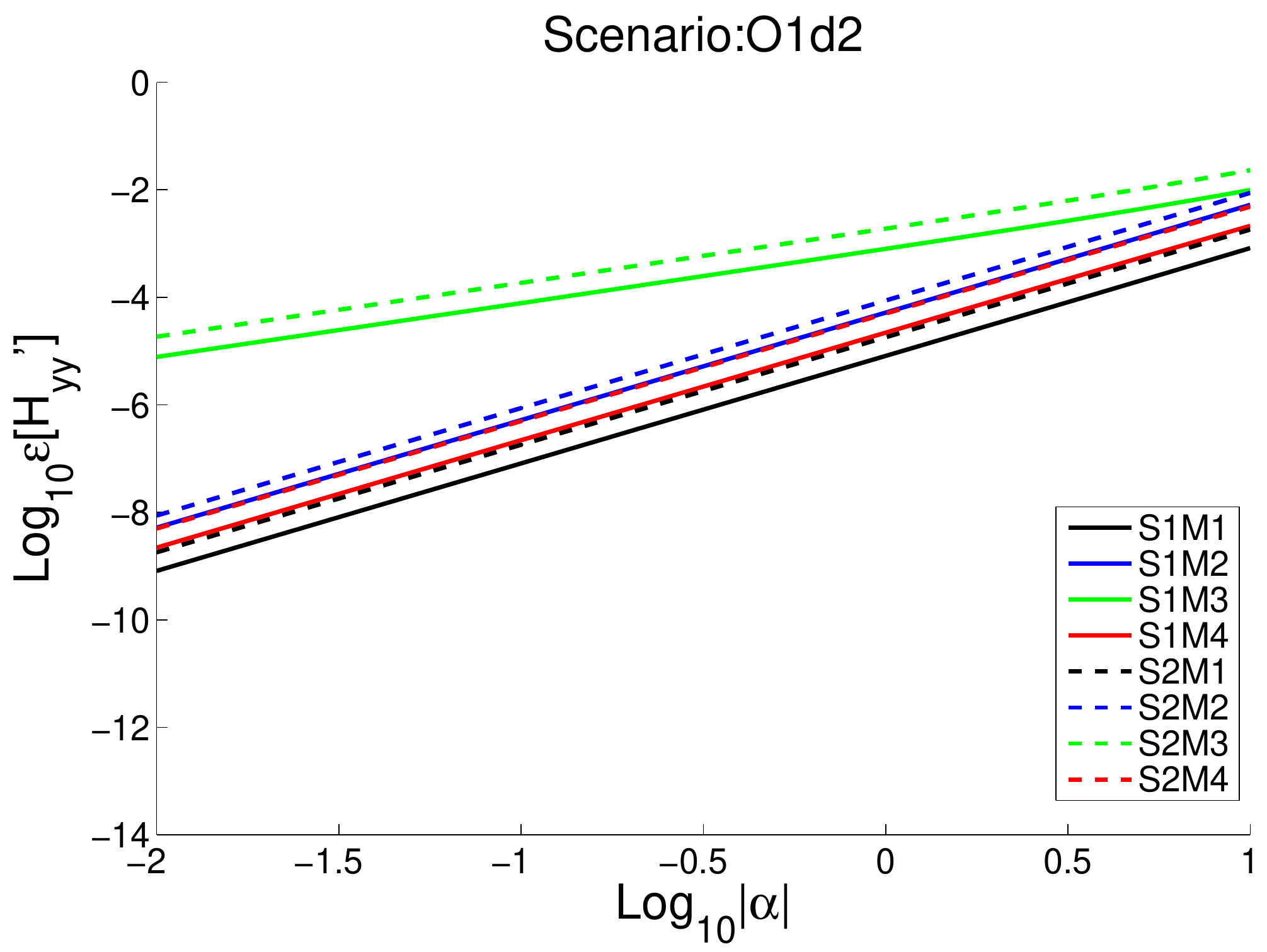}}

\subfloat[\label{ReHyyO2D1}]{\includegraphics[width=3.25in]{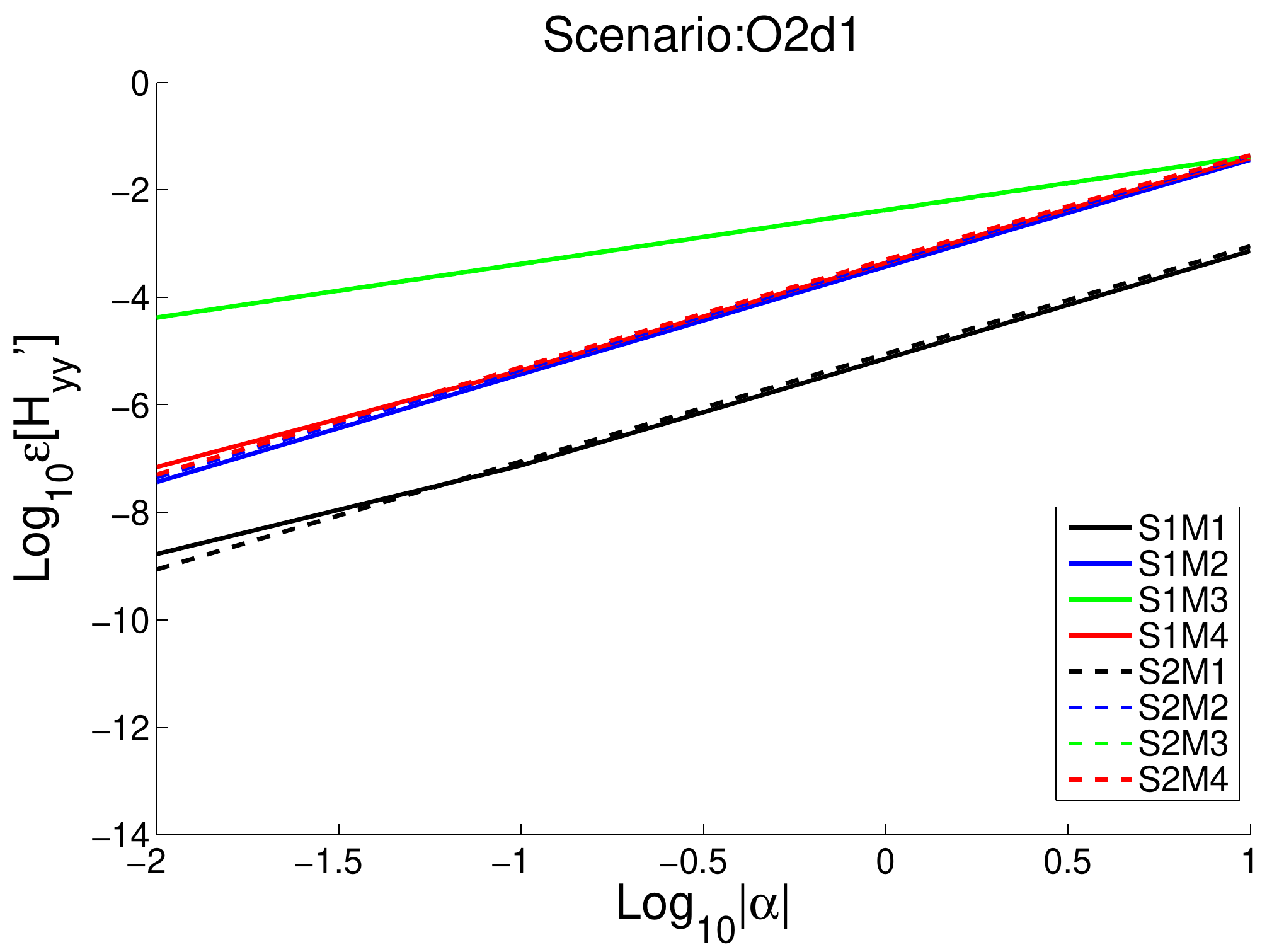}}
\subfloat[\label{ReHyyO2D2}]{\includegraphics[width=3.25in]{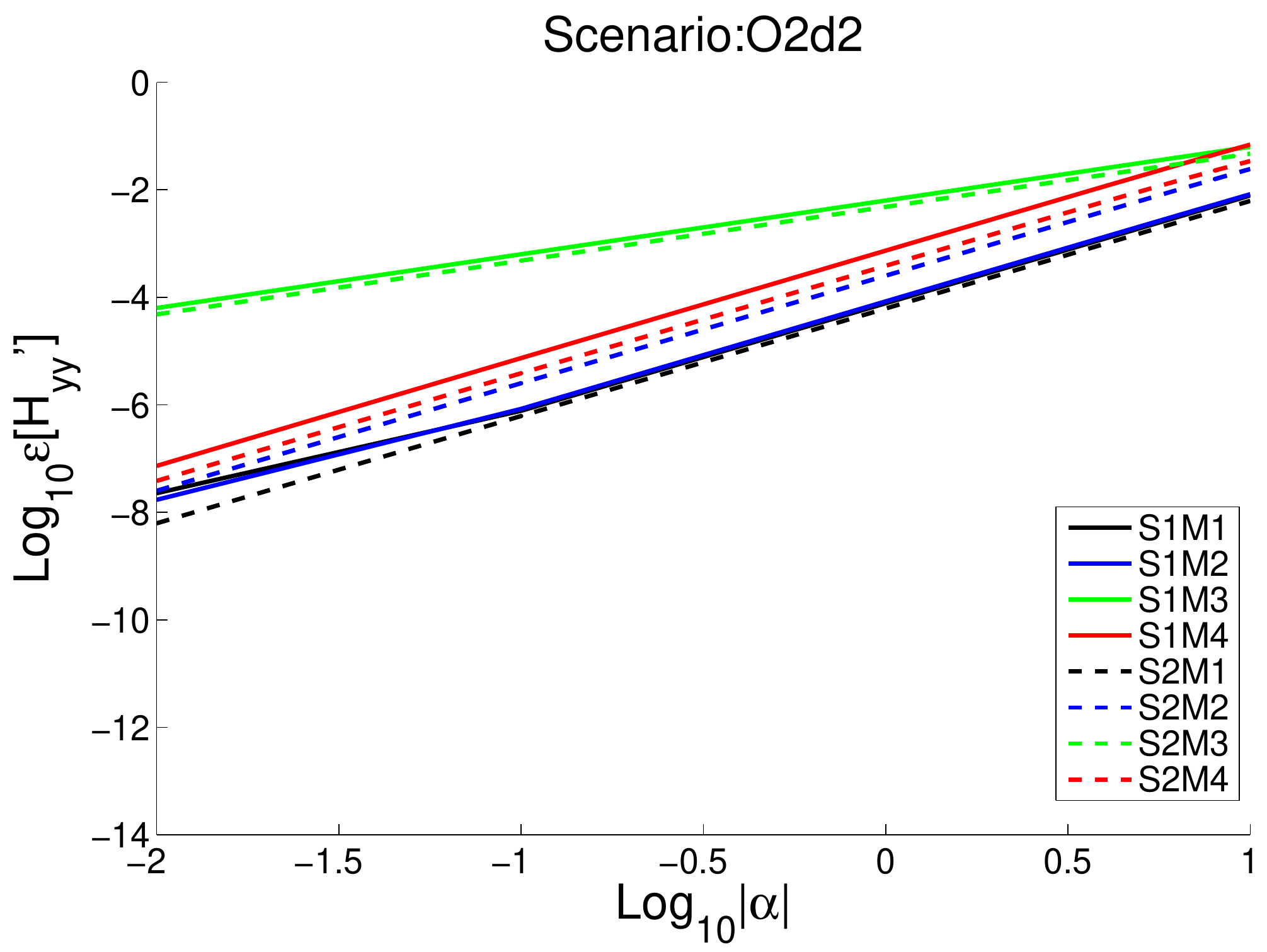}}

\subfloat[\label{ReHyyO3D1}]{\includegraphics[width=3.25in]{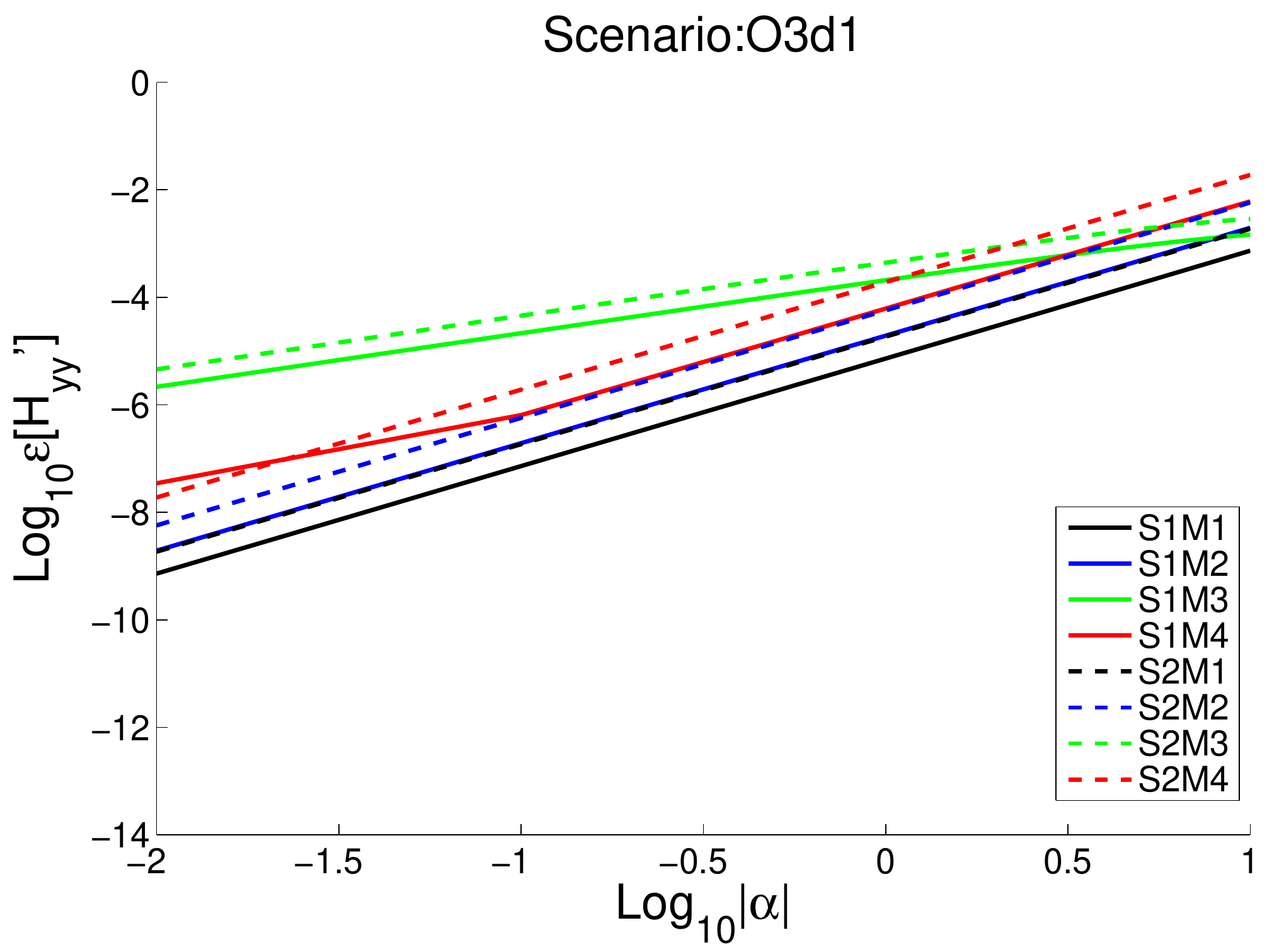}}
\subfloat[\label{ReHyyO3D2}]{\includegraphics[width=3.25in]{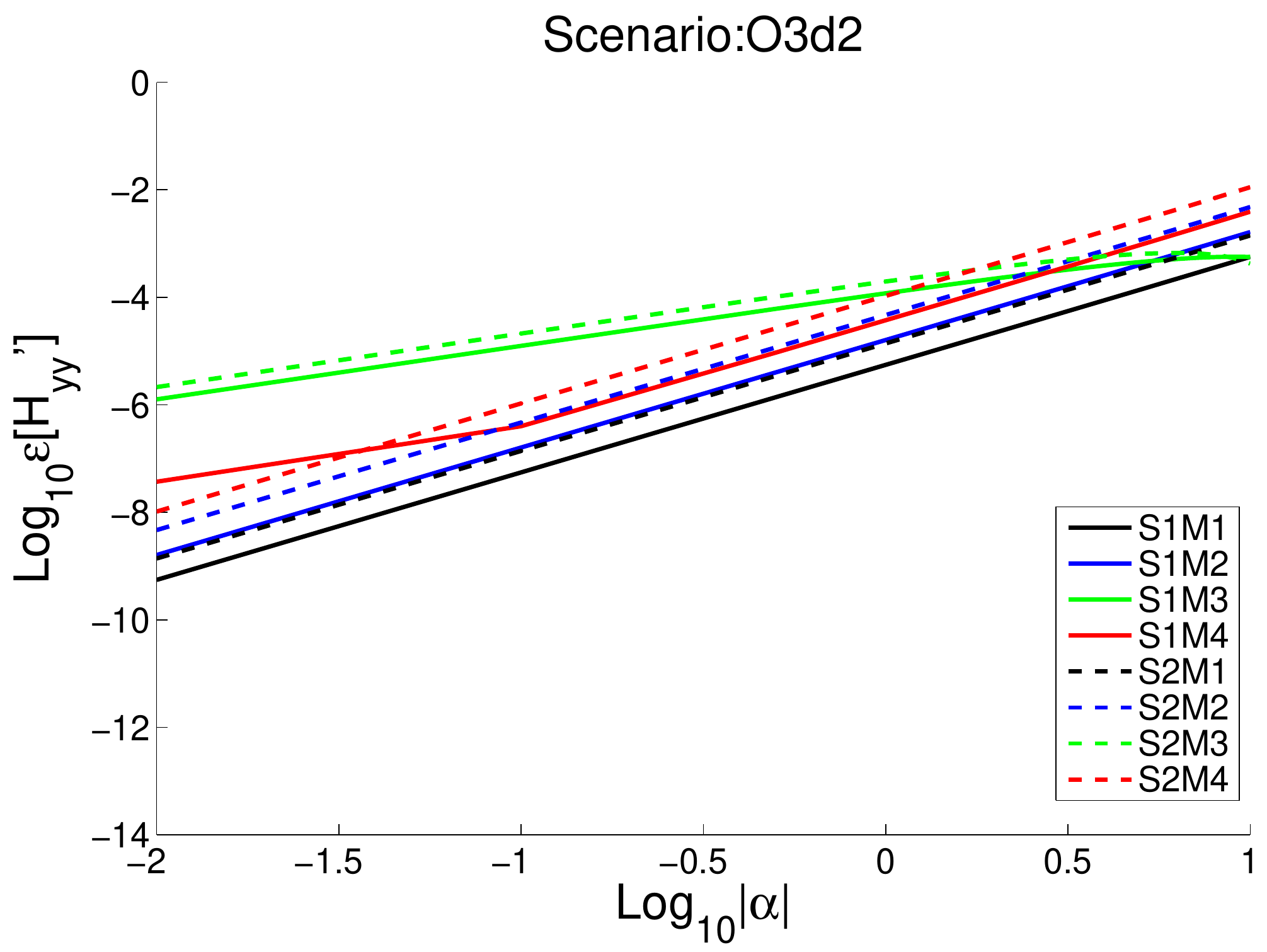}}
\caption{\small Relative error in computing $H_{yy}^{\prime}$=Re[$H_{yy}$].}
\label{ReHyy}
\end{figure}

\begin{figure}[H]
\centering
\subfloat[\label{ImHyyO1D1}]{\includegraphics[width=3.25in]{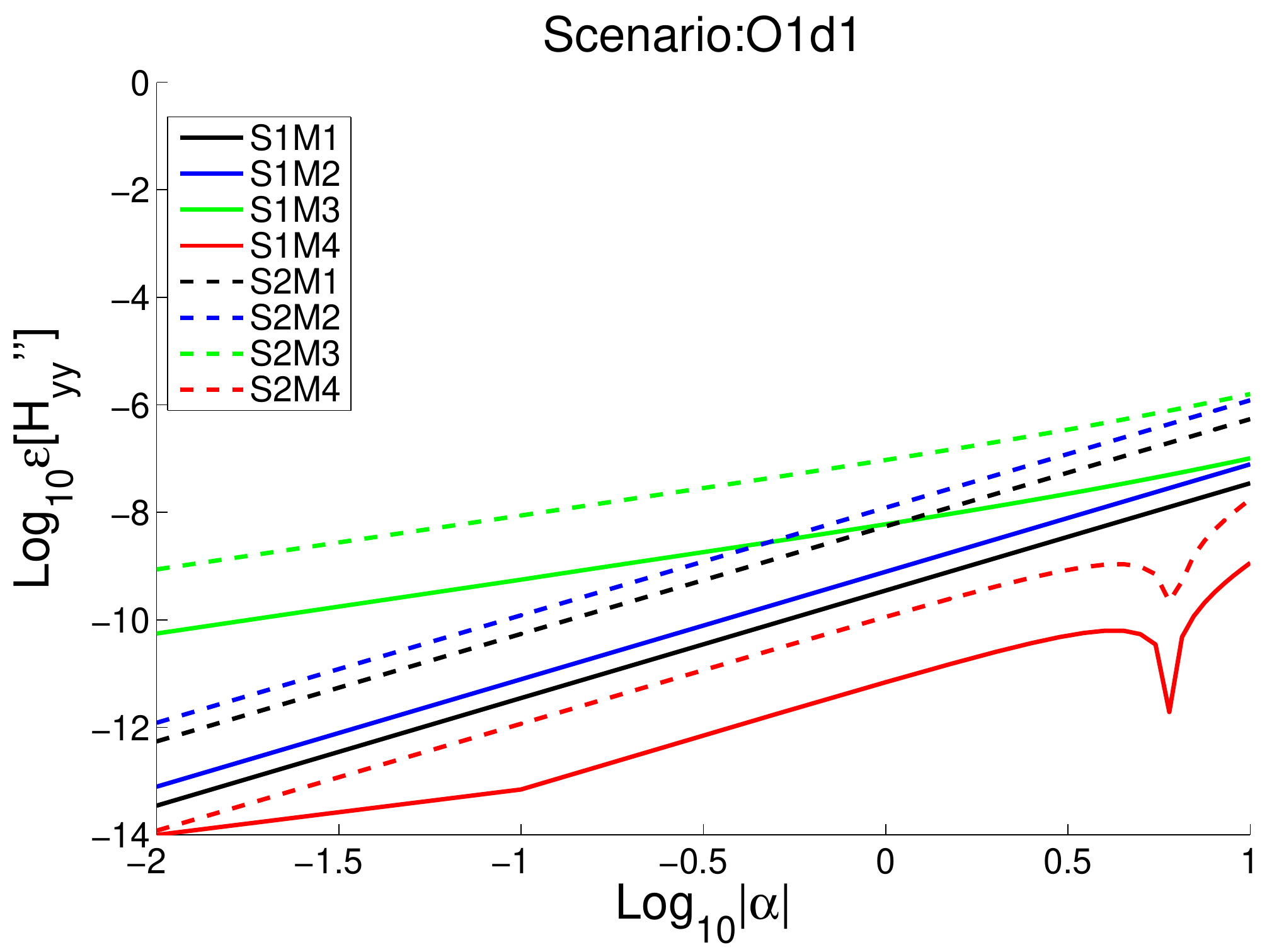}}
\subfloat[\label{ImHyyO1D2}]{\includegraphics[width=3.25in]{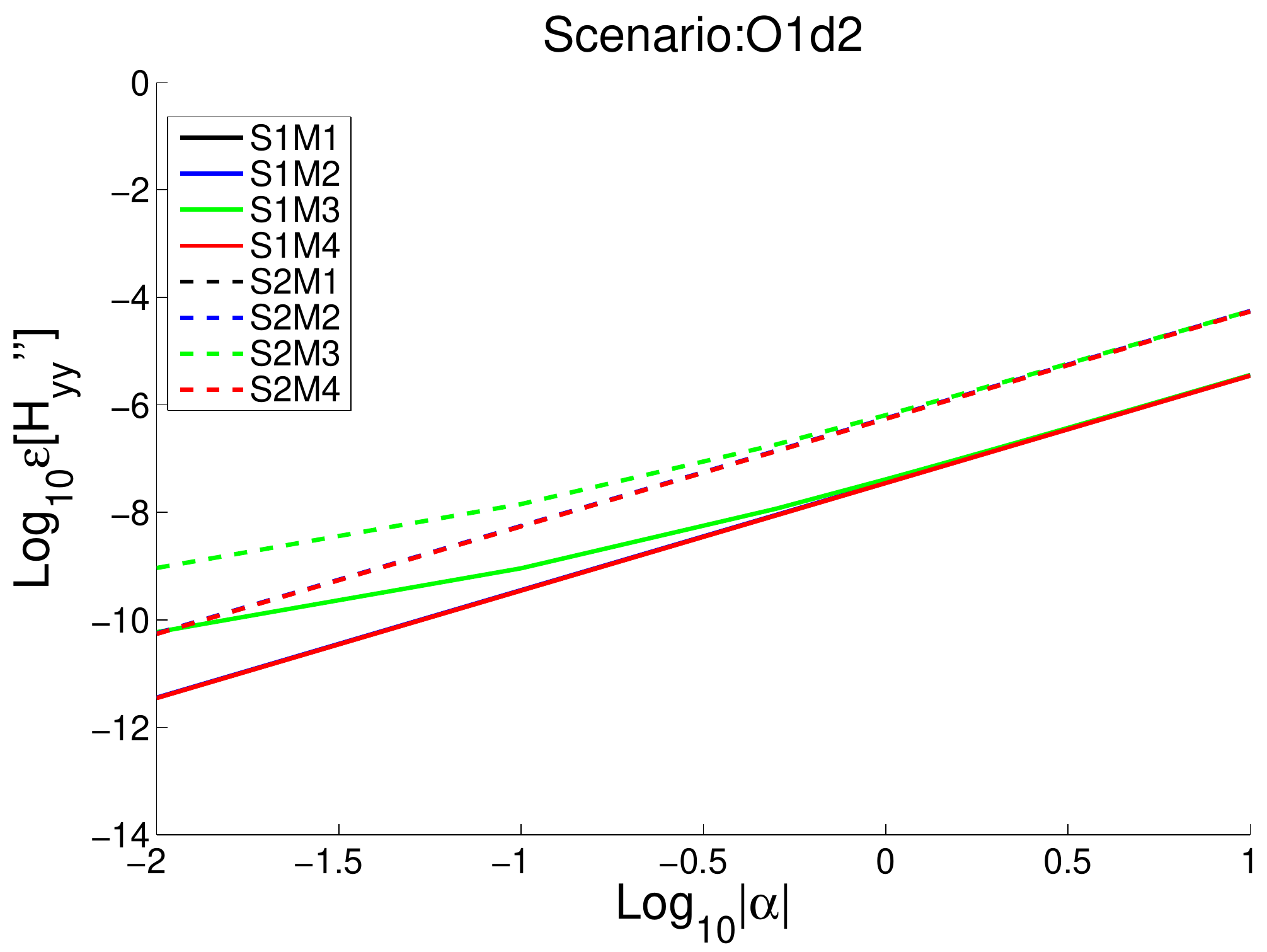}}

\subfloat[\label{ImHyyO2D1}]{\includegraphics[width=3.25in]{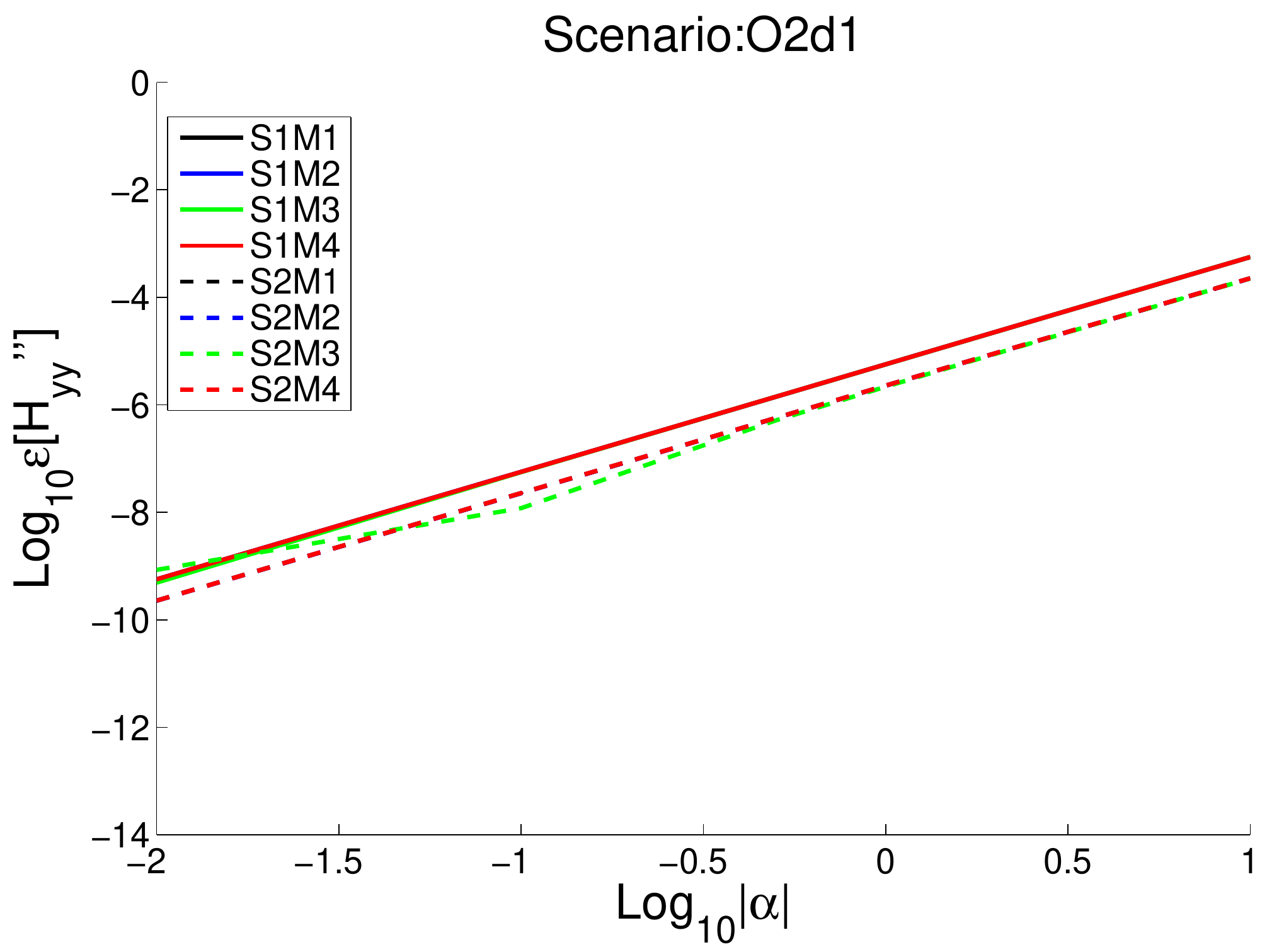}}
\subfloat[\label{ImHyyO2D2}]{\includegraphics[width=3.25in]{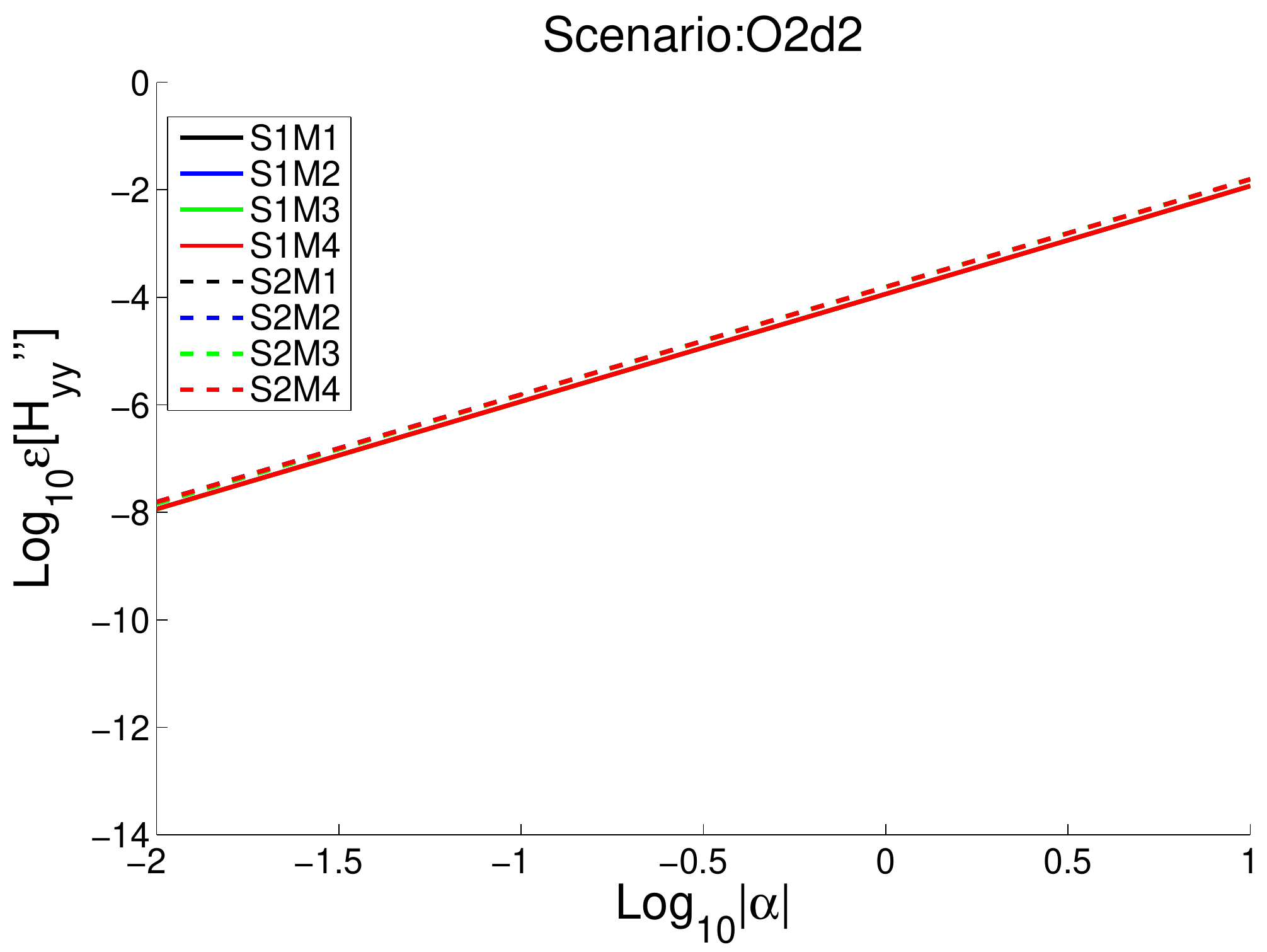}}

\subfloat[\label{ImHyyO3D1}]{\includegraphics[width=3.25in]{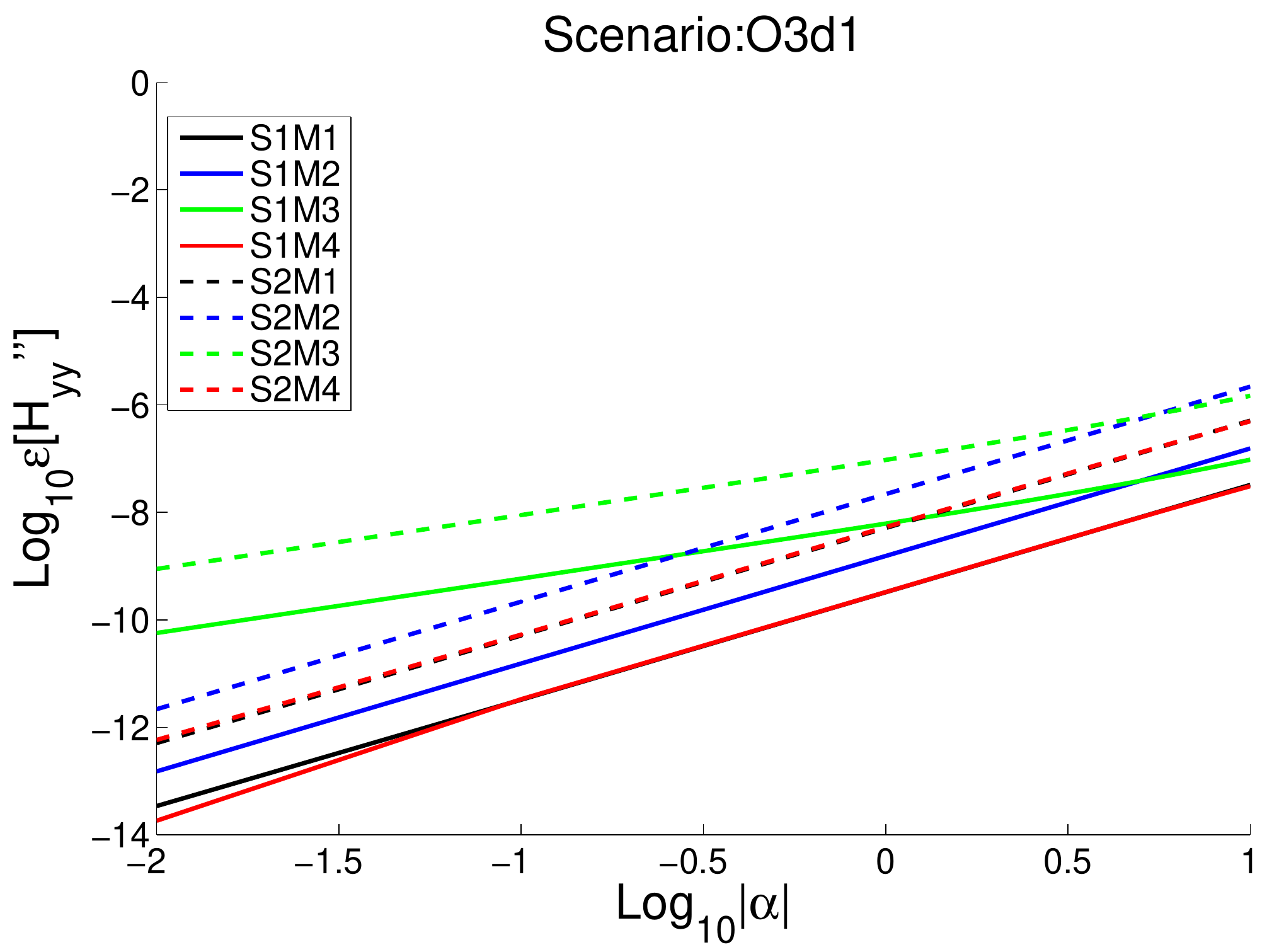}}
\subfloat[\label{ImHyyO3D2}]{\includegraphics[width=3.25in]{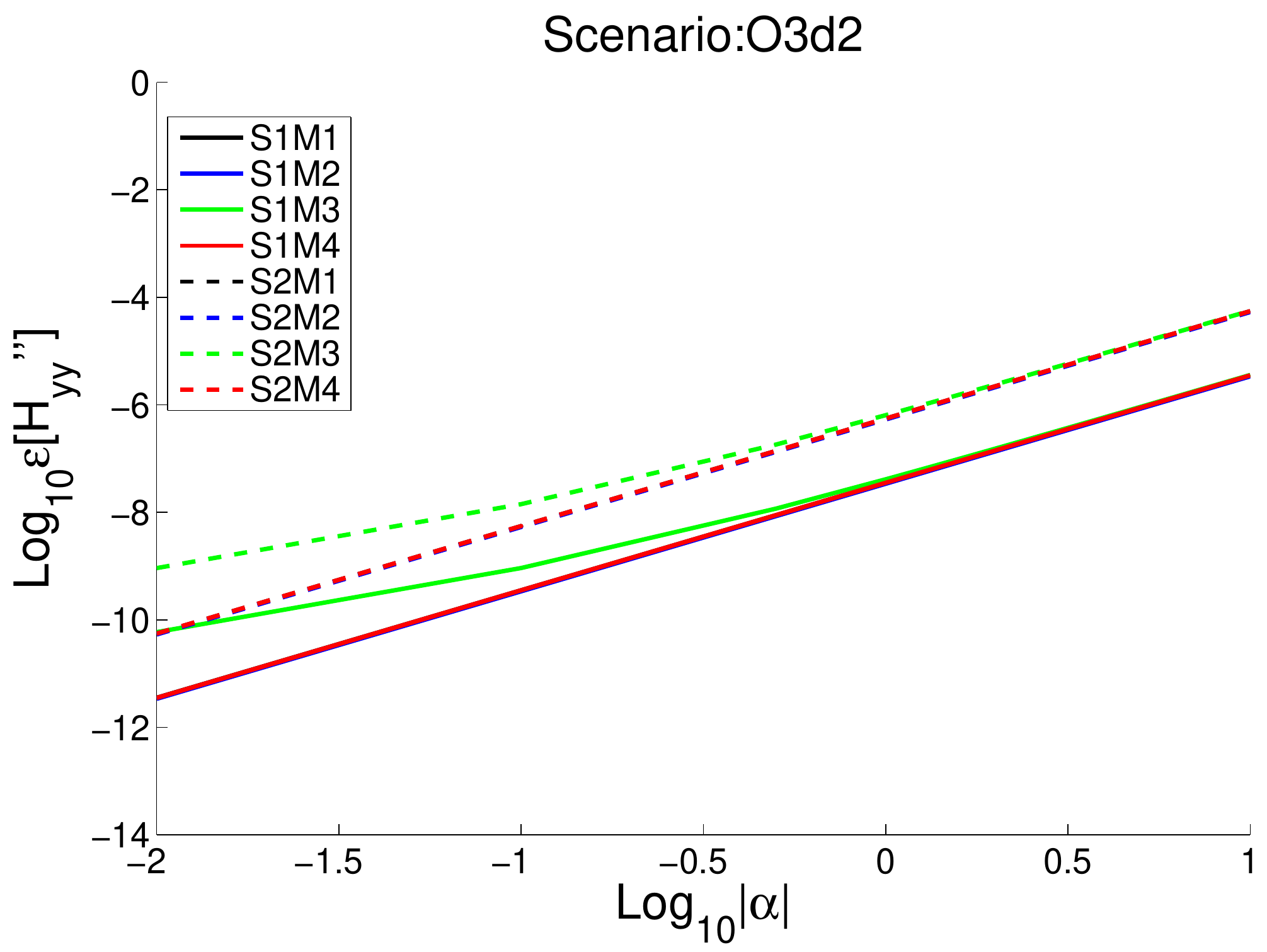}}
\caption{\small Relative error in computing $H_{yy}^{\prime \prime}$=Im[$H_{yy}$].}
\label{ImHyy}
\end{figure}

\begin{figure}[H]
\centering
\subfloat[\label{ReHzzO1D1}]{\includegraphics[width=3.25in]{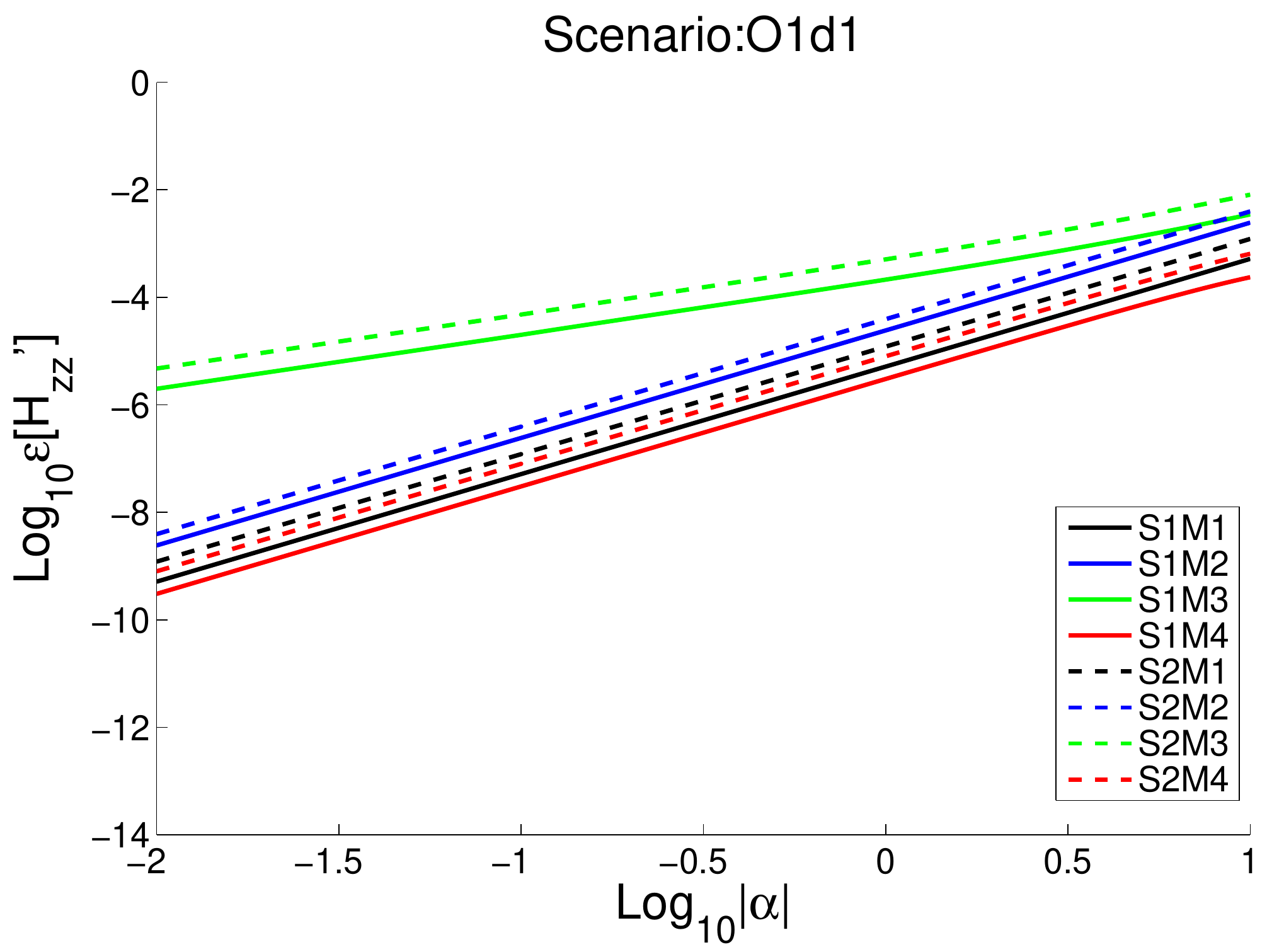}}
\subfloat[\label{ReHzzO1D2}]{\includegraphics[width=3.25in]{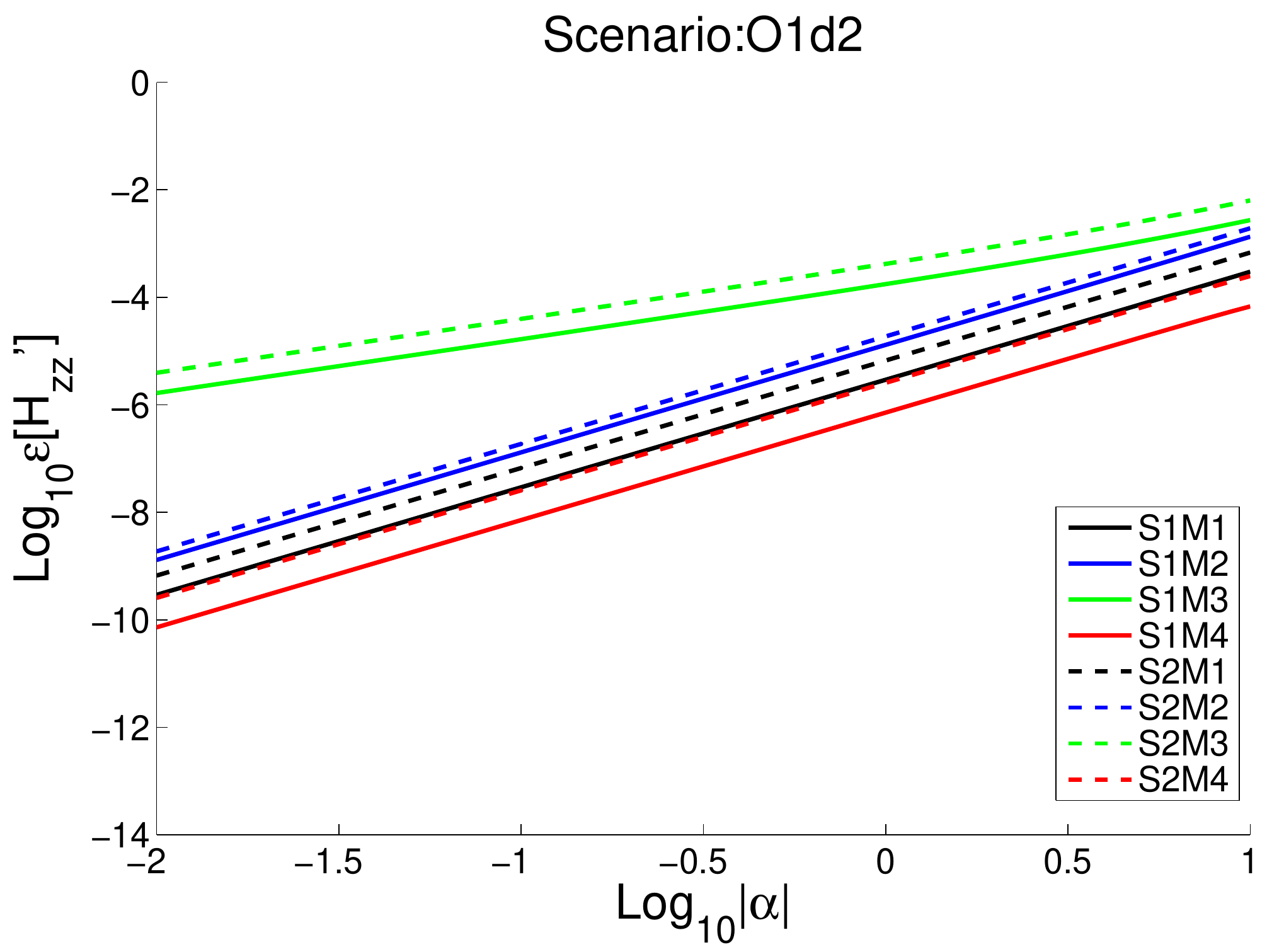}}

\subfloat[\label{ReHzzO2D1}]{\includegraphics[width=3.25in]{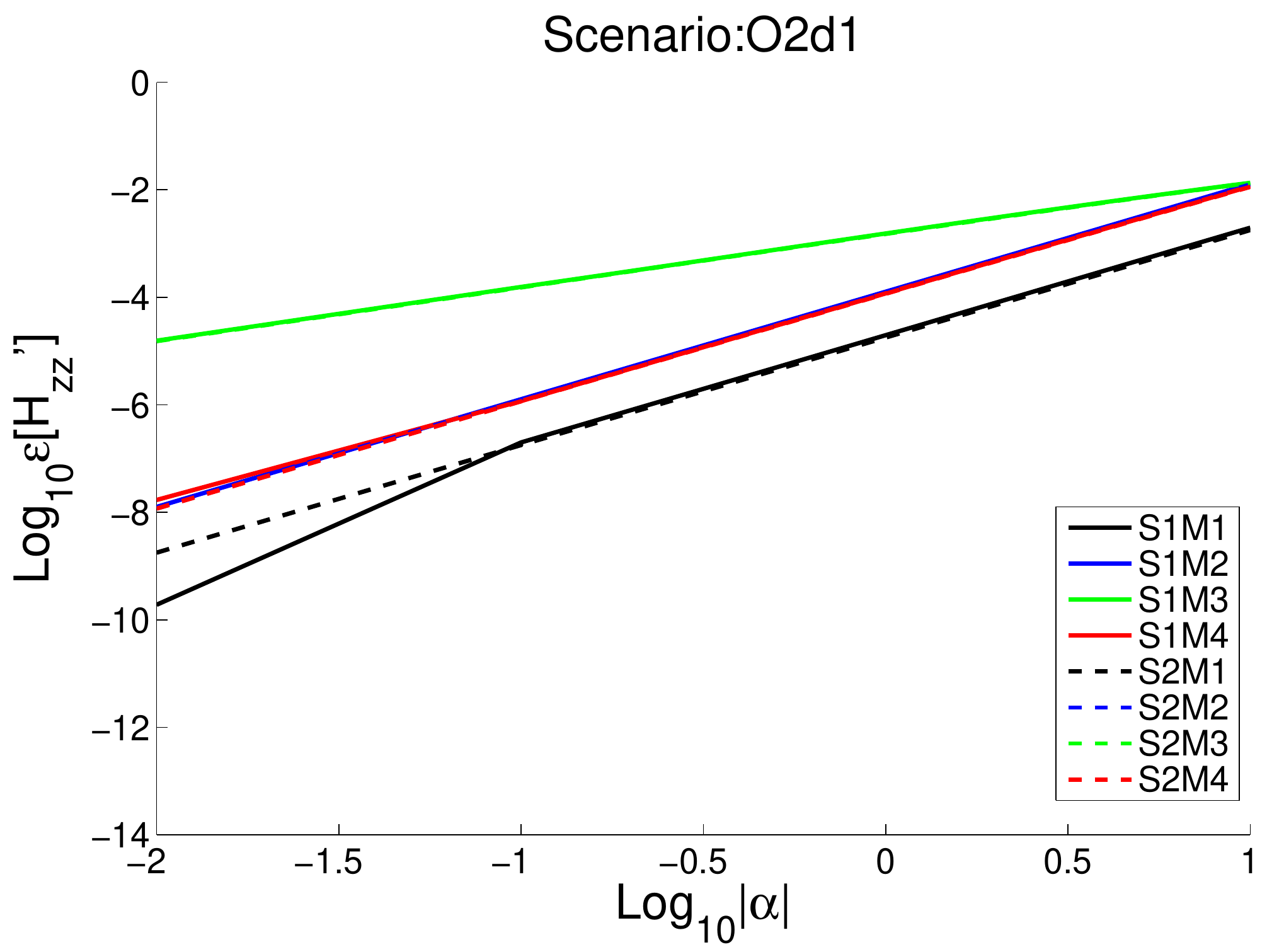}}
\subfloat[\label{ReHzzO2D2}]{\includegraphics[width=3.25in]{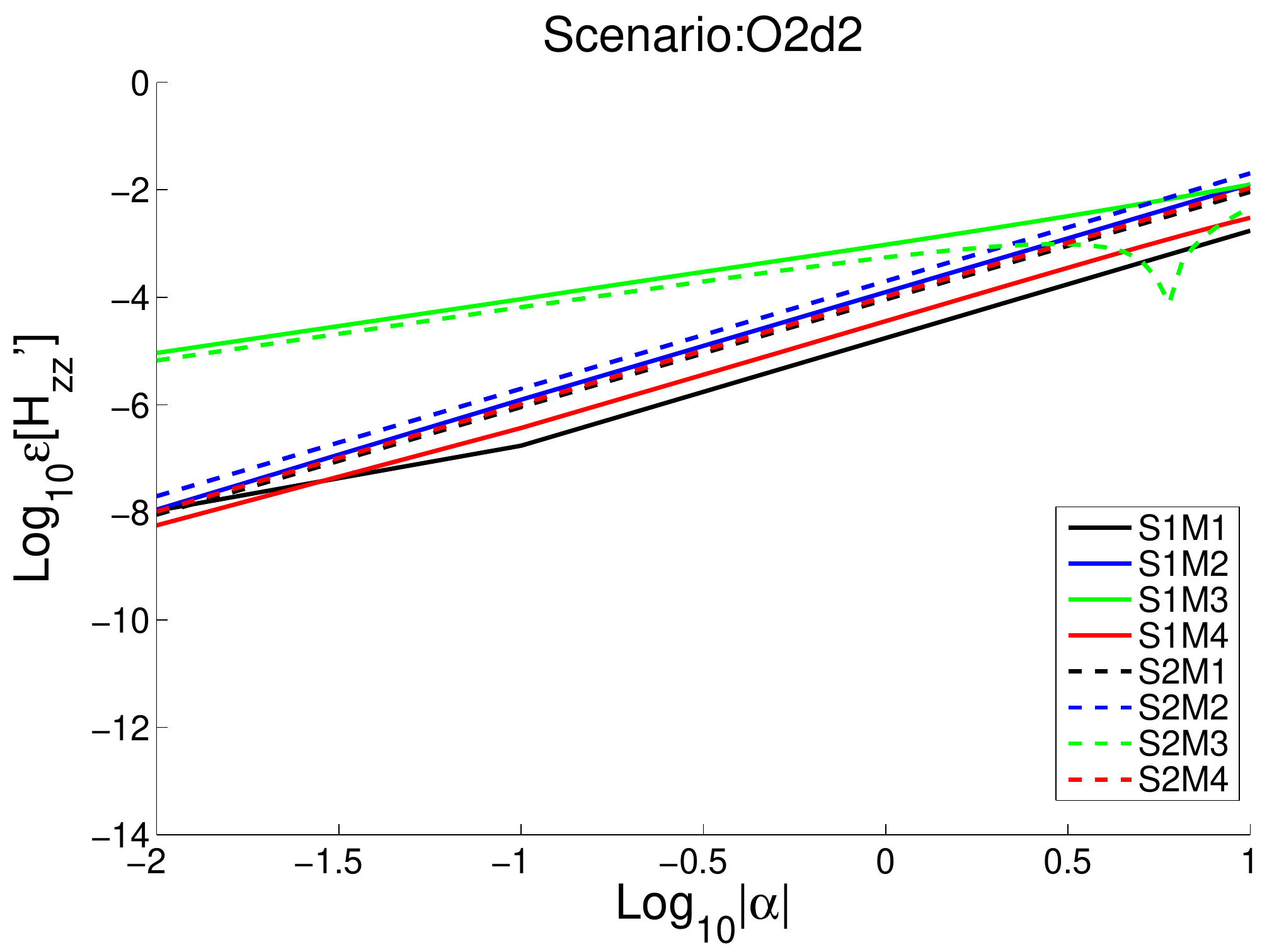}}

\subfloat[\label{ReHzzO3D1}]{\includegraphics[width=3.25in]{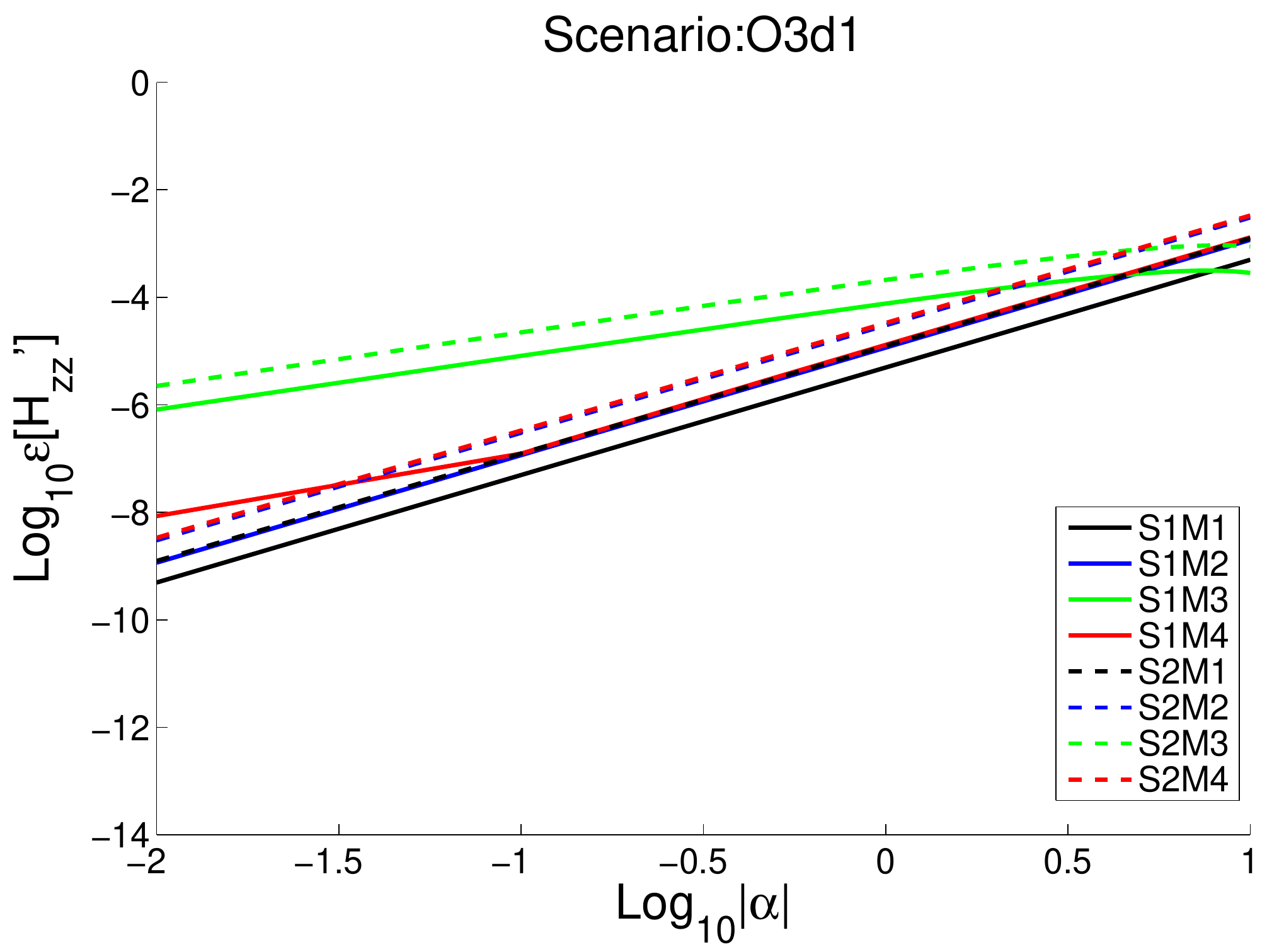}}
\subfloat[\label{ReHzzO3D2}]{\includegraphics[width=3.25in]{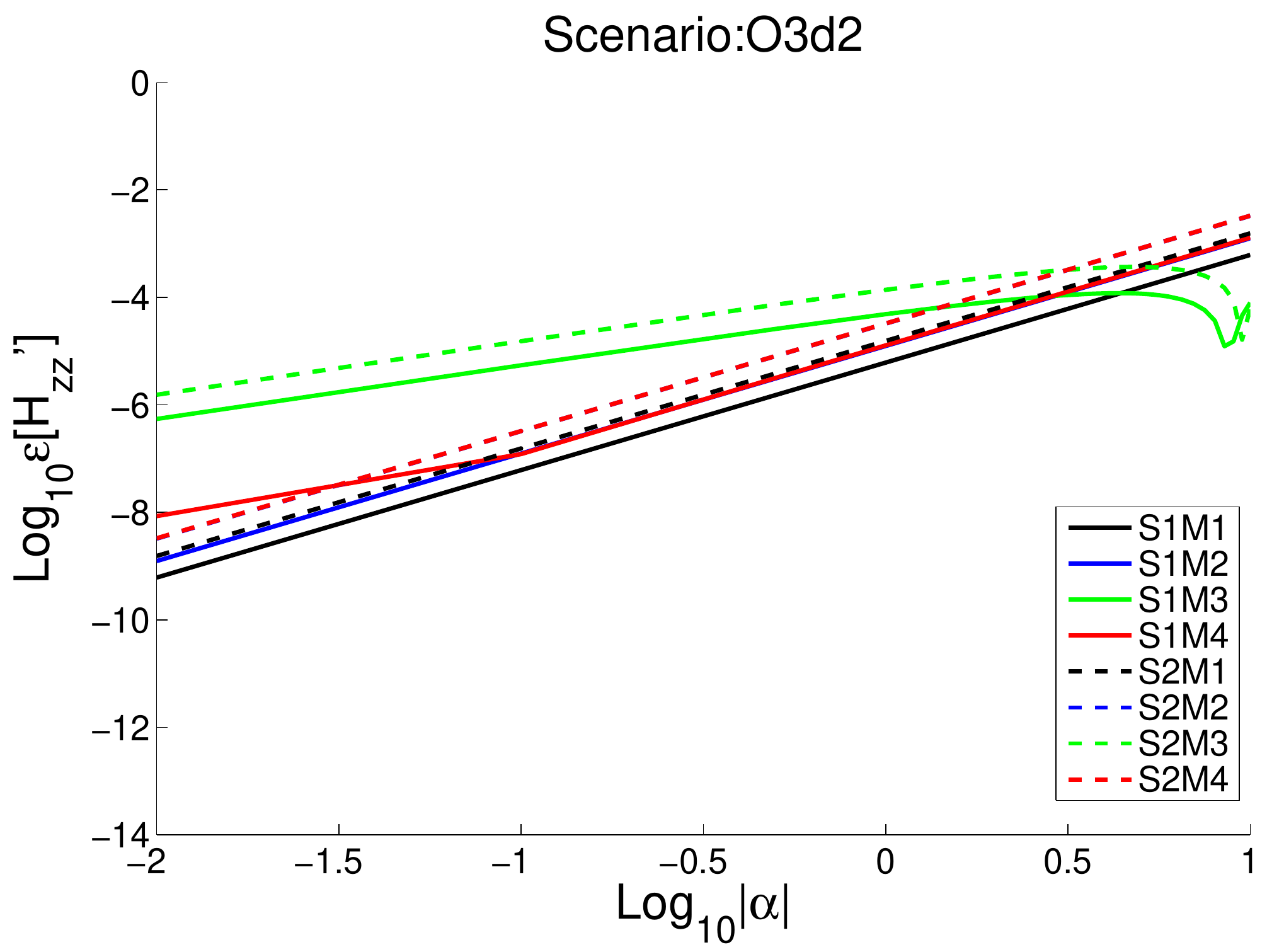}}
\caption{\small Relative error in computing $H_{zz}^{\prime}$=Re[$H_{zz}$].}
\label{ReHzz}
\end{figure}

\begin{figure}[H]
\centering
\subfloat[\label{ImHzzO1D1}]{\includegraphics[width=3.25in]{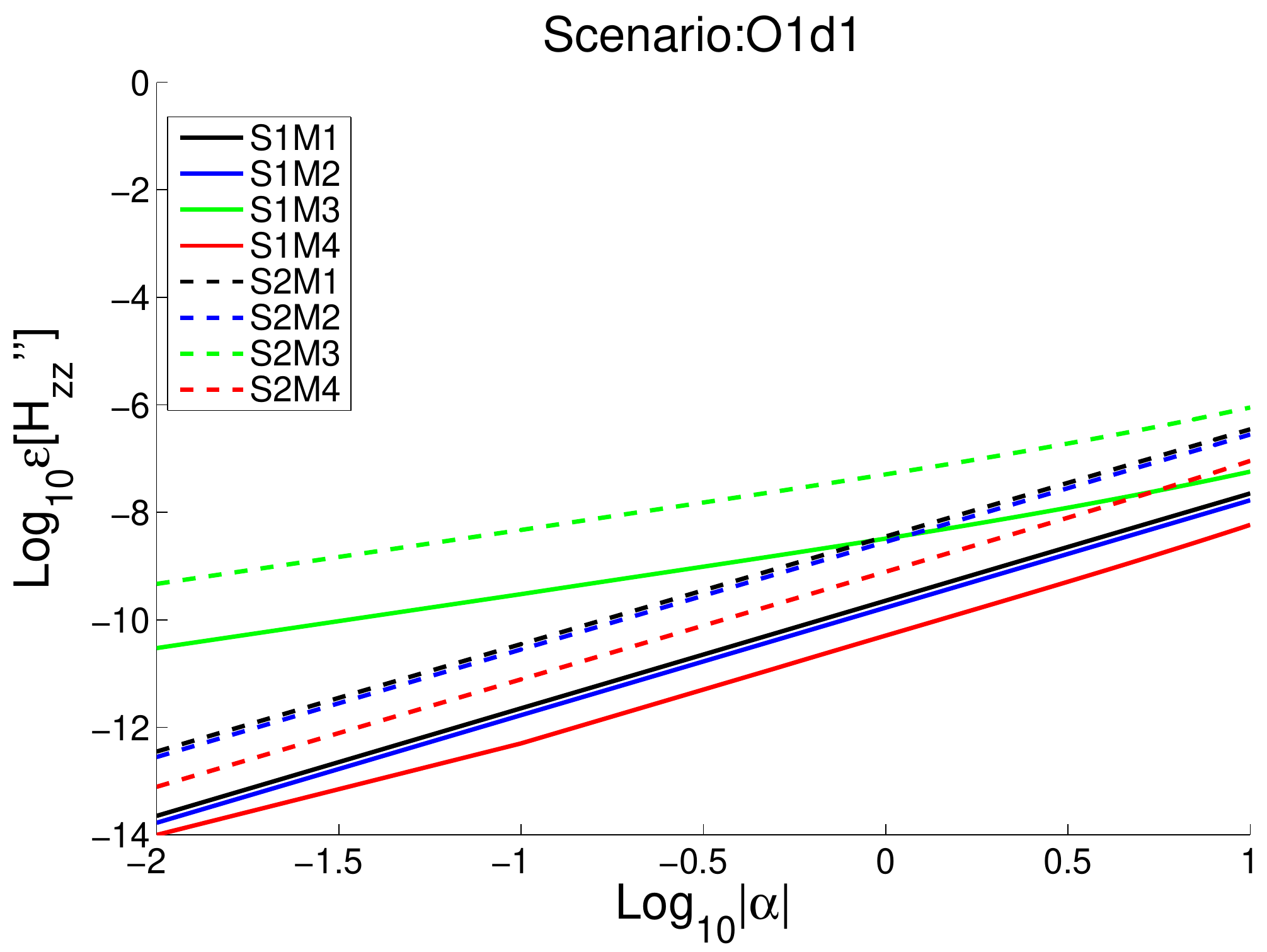}}
\subfloat[\label{ImHzzO1D2}]{\includegraphics[width=3.25in]{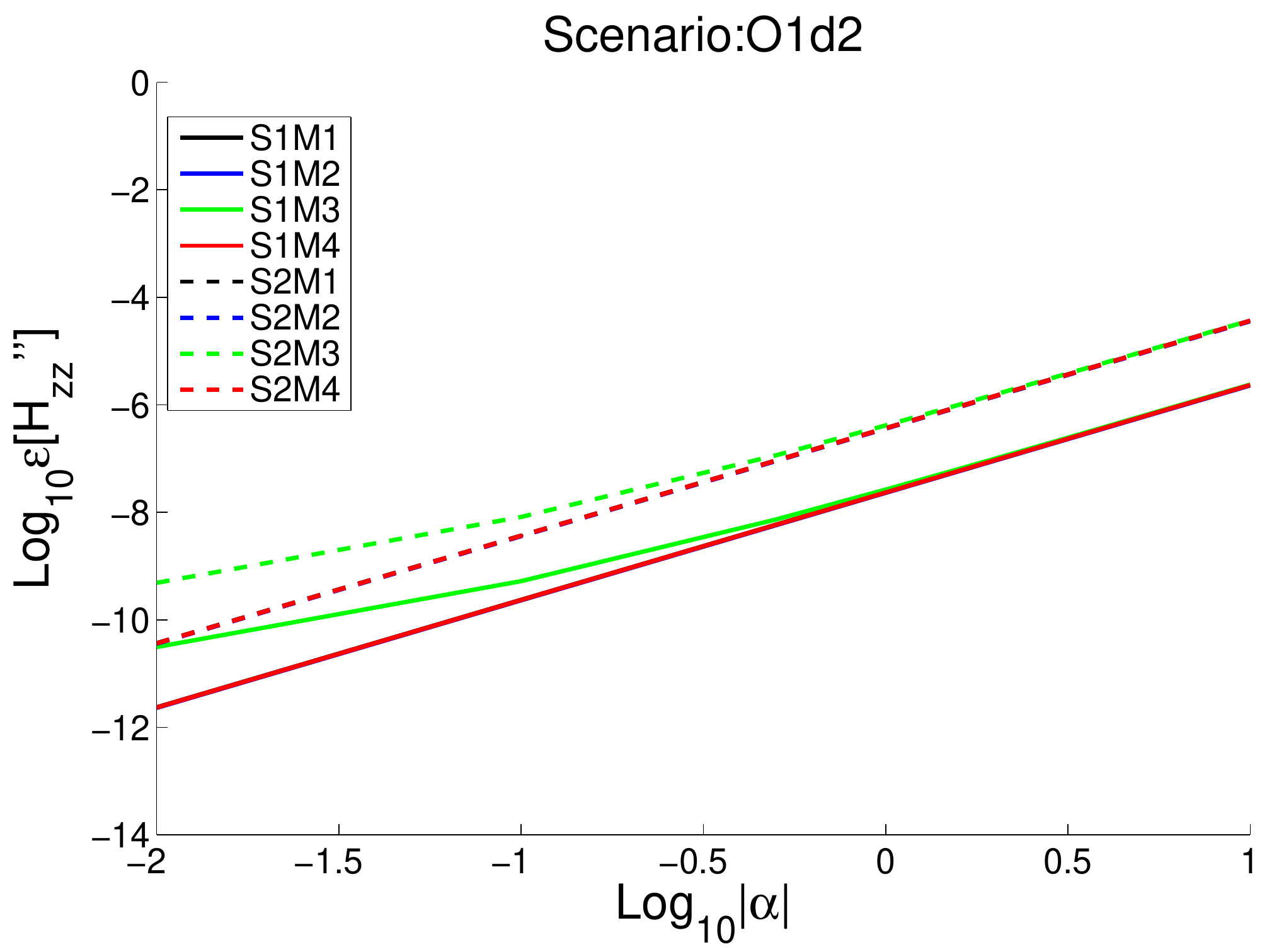}}

\subfloat[\label{ImHzzO2D1}]{\includegraphics[width=3.25in]{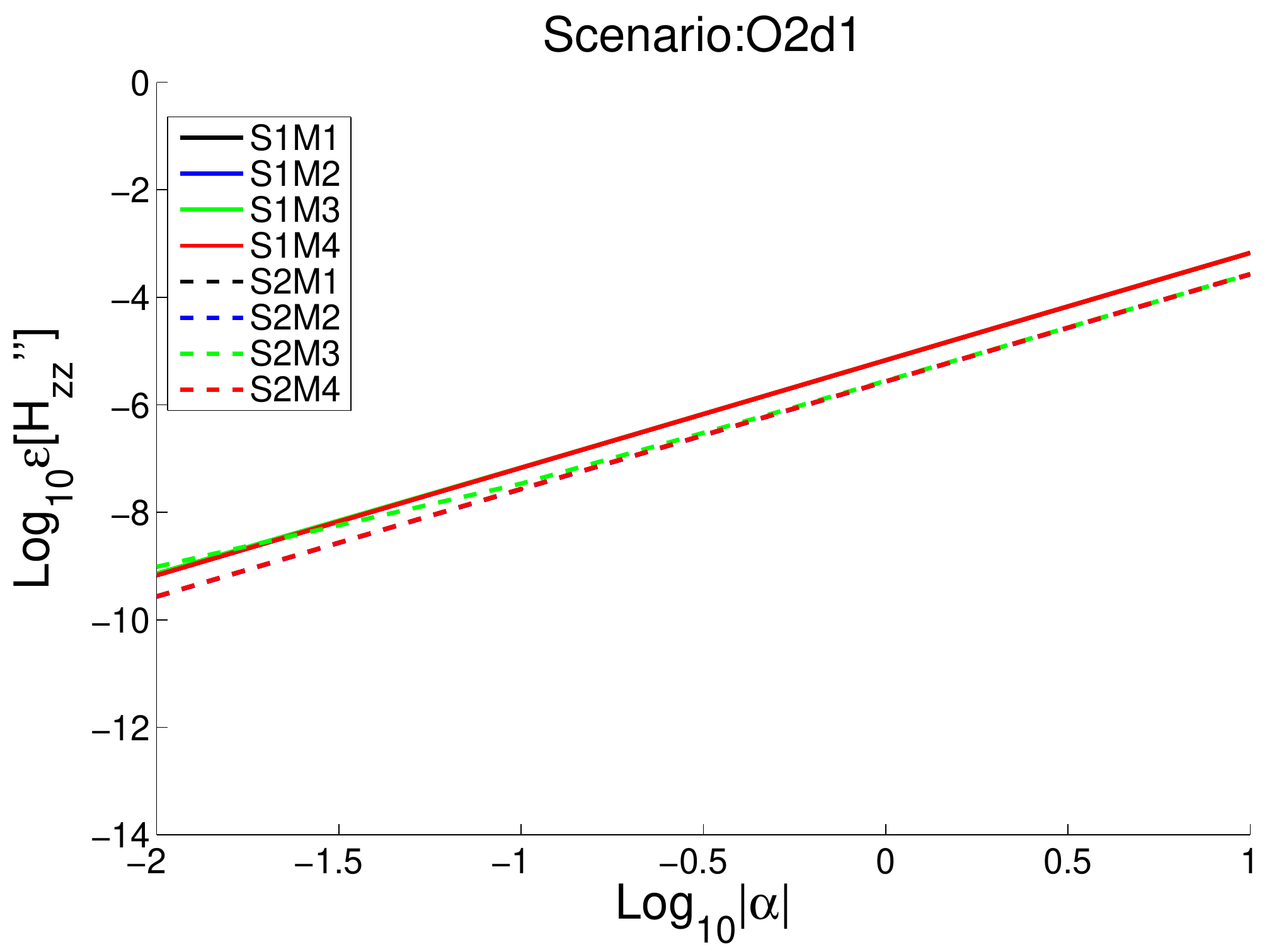}}
\subfloat[\label{ImHzzO2D2}]{\includegraphics[width=3.25in]{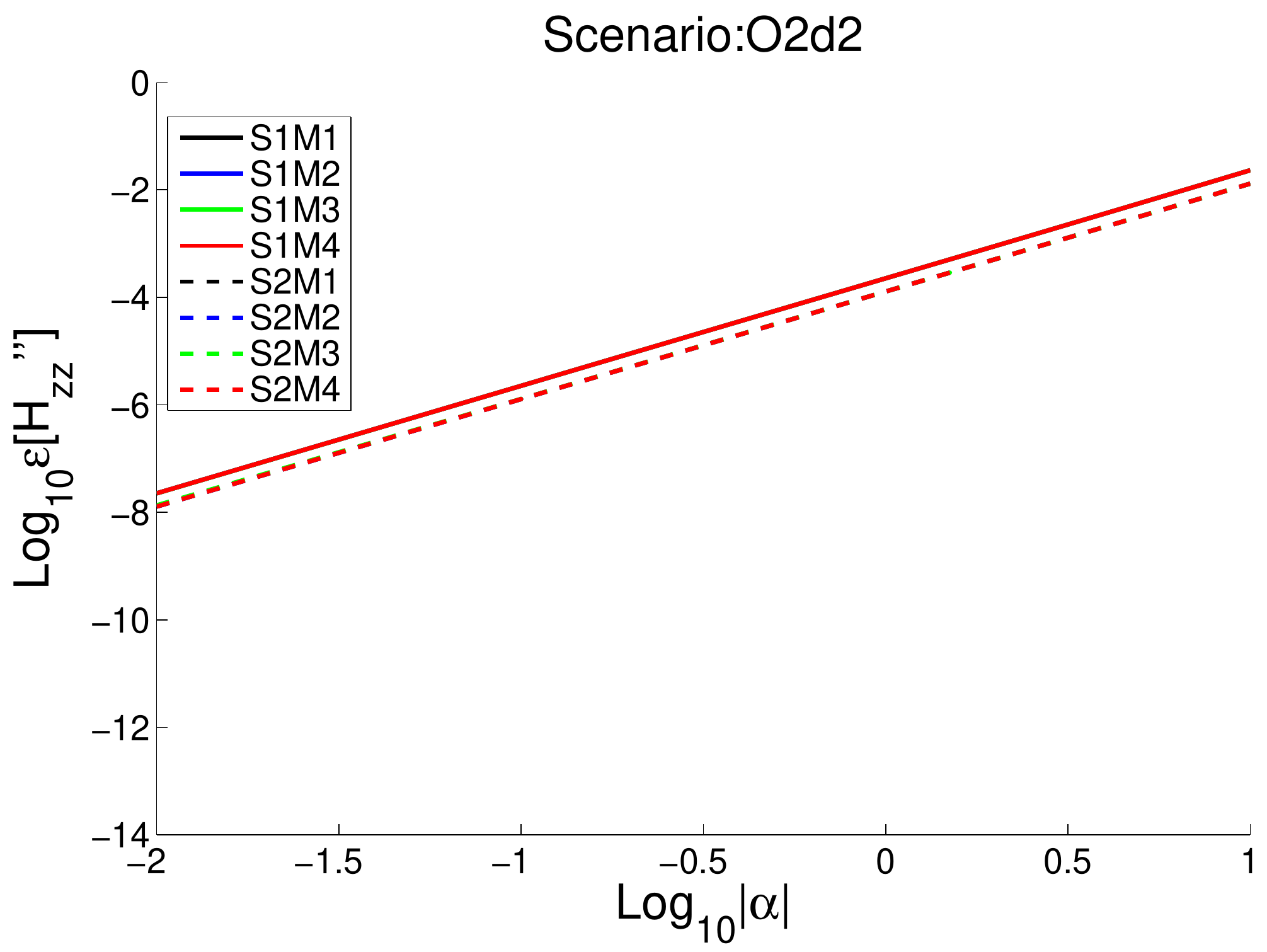}}

\subfloat[\label{ImHzzO3D1}]{\includegraphics[width=3.25in]{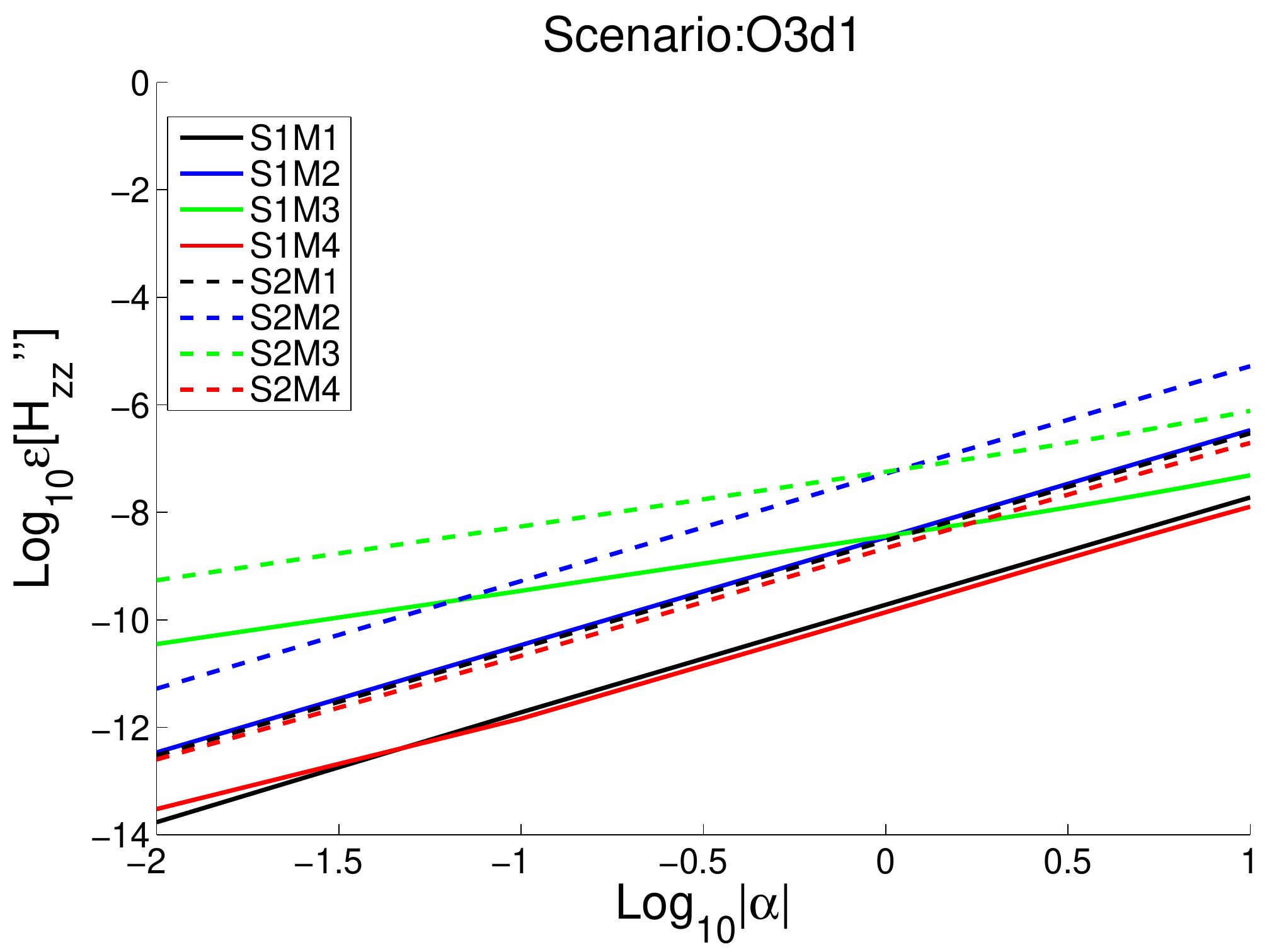}}
\subfloat[\label{ImHzzO3D2}]{\includegraphics[width=3.25in]{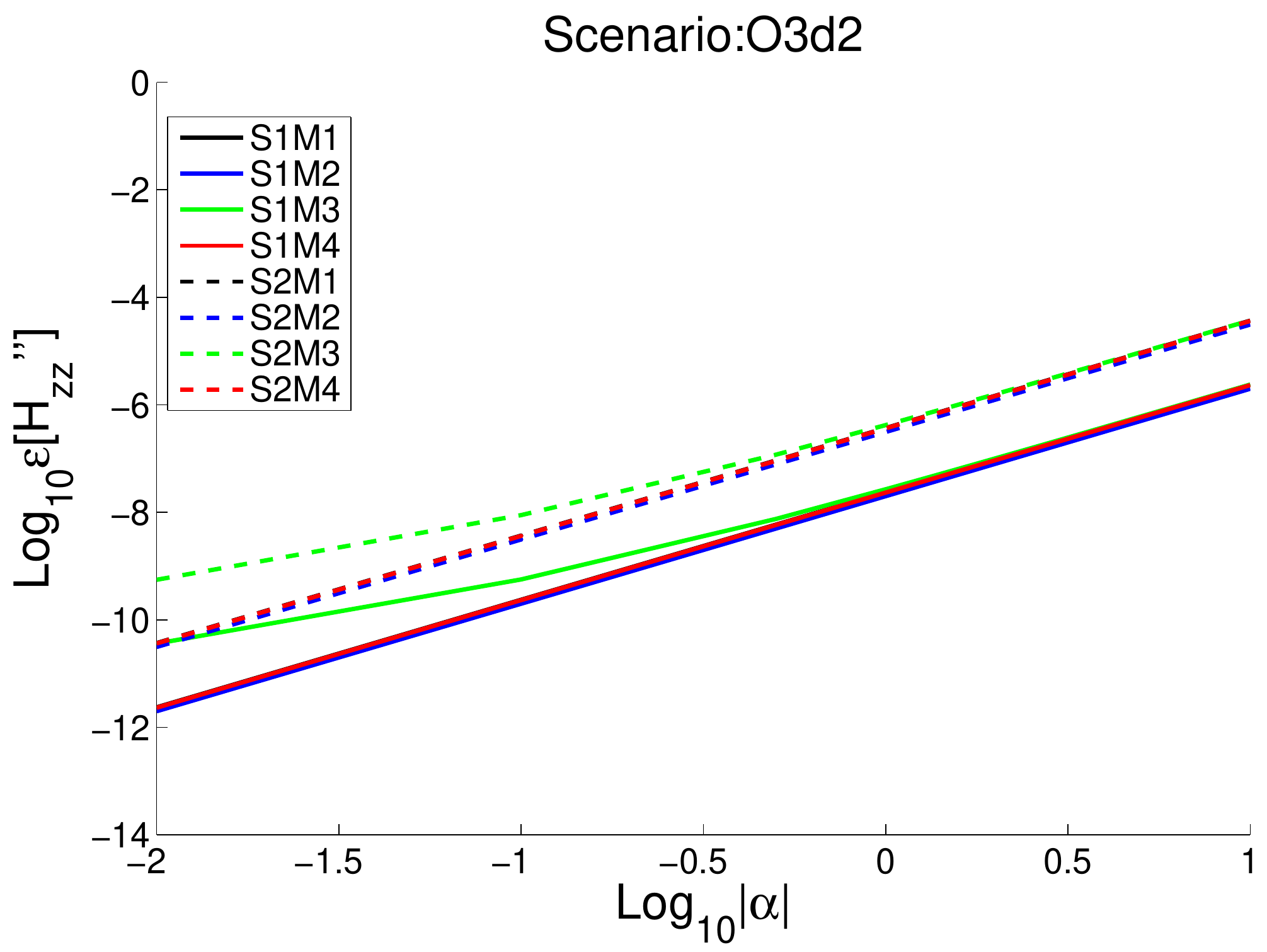}}
\caption{\small Relative error in computing $H_{zz}^{\prime \prime}$=Im[$H_{zz}$].}
\label{ImHzz}
\end{figure}

\begin{figure}[H]
\centering
\subfloat[\label{ReHxzO1D1}]{\includegraphics[width=3.25in]{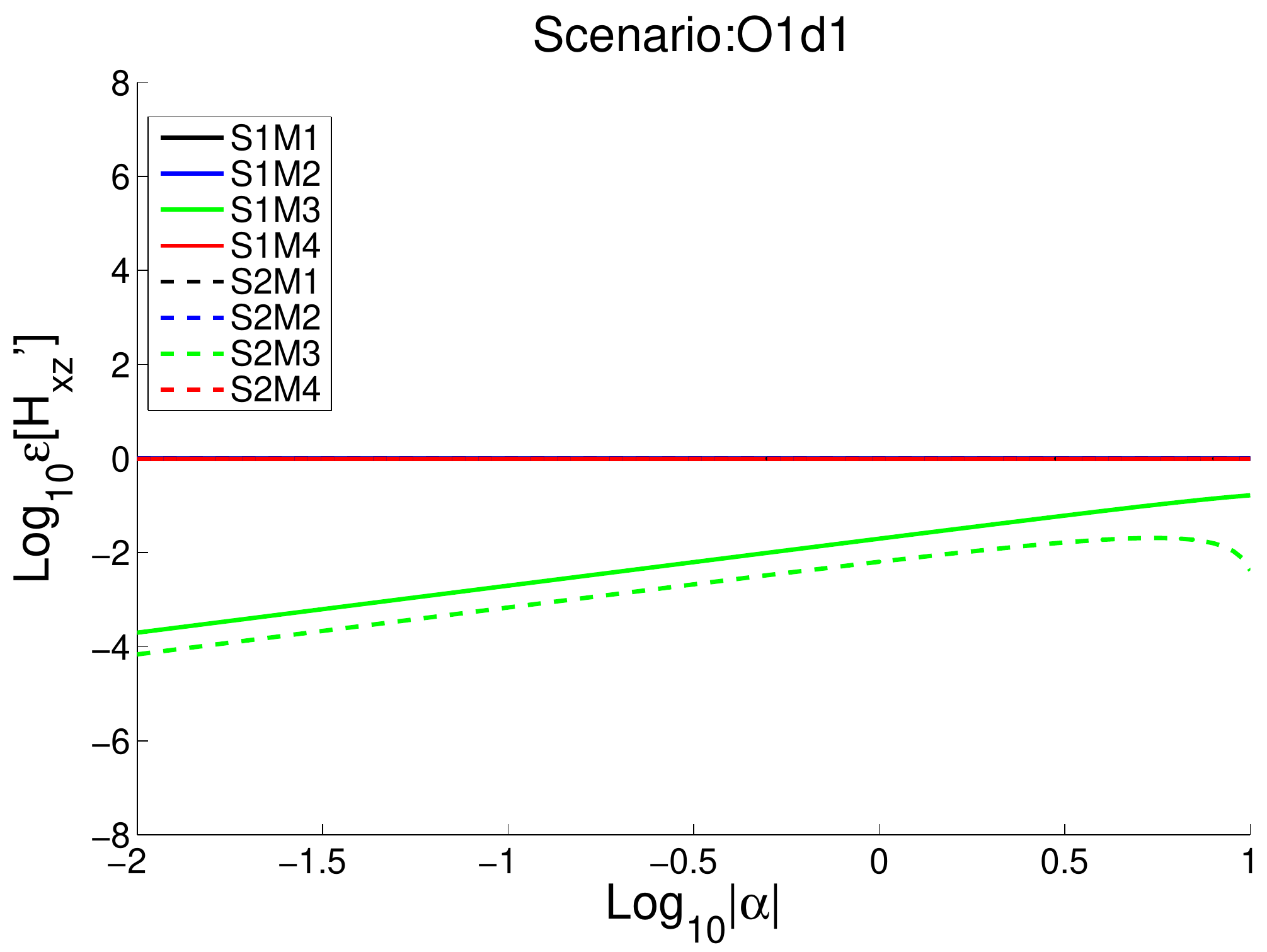}}
\subfloat[\label{ReHxzO1D2}]{\includegraphics[width=3.25in]{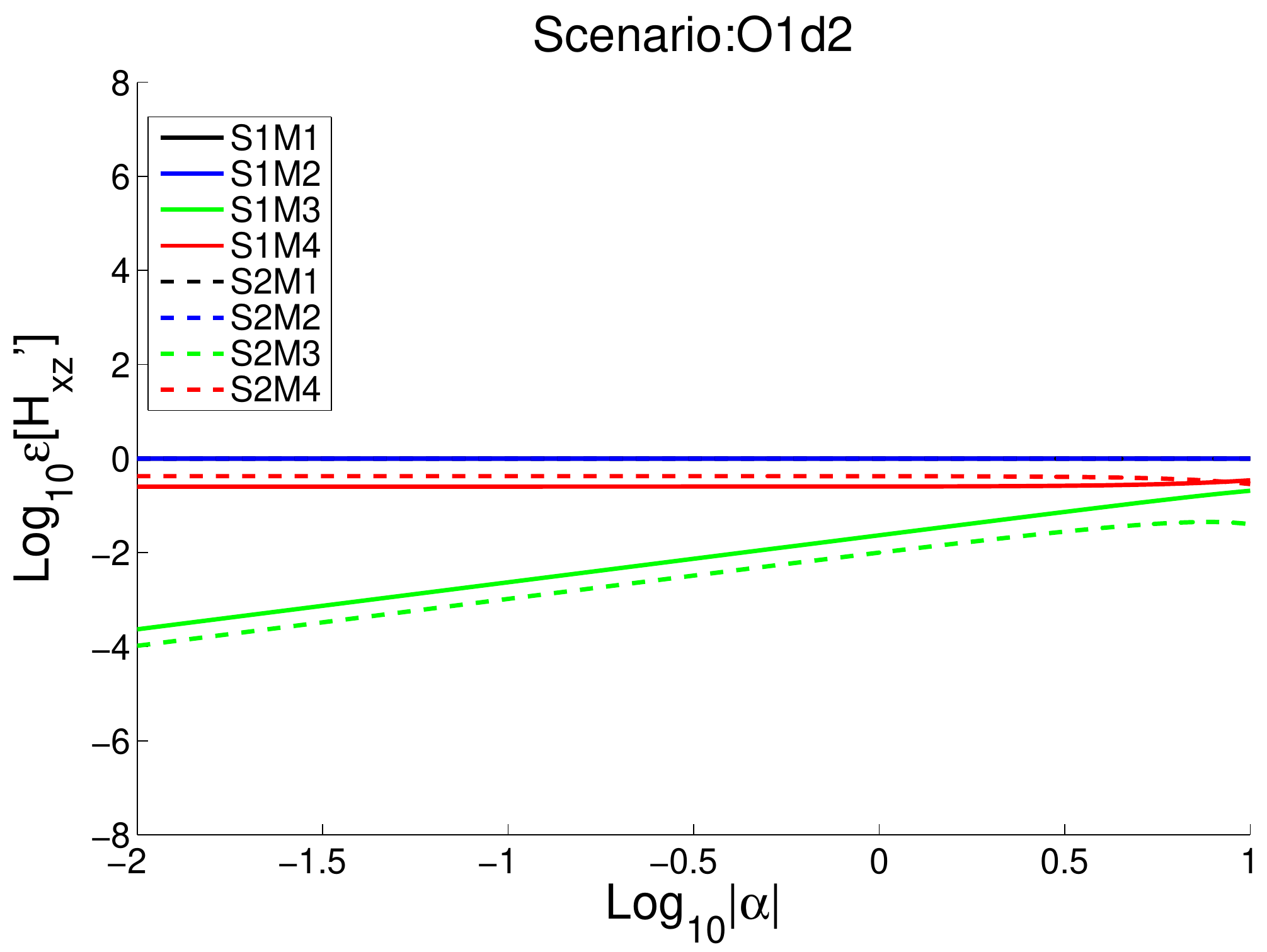}}

\subfloat[\label{ReHxzO2D1}]{\includegraphics[width=3.25in]{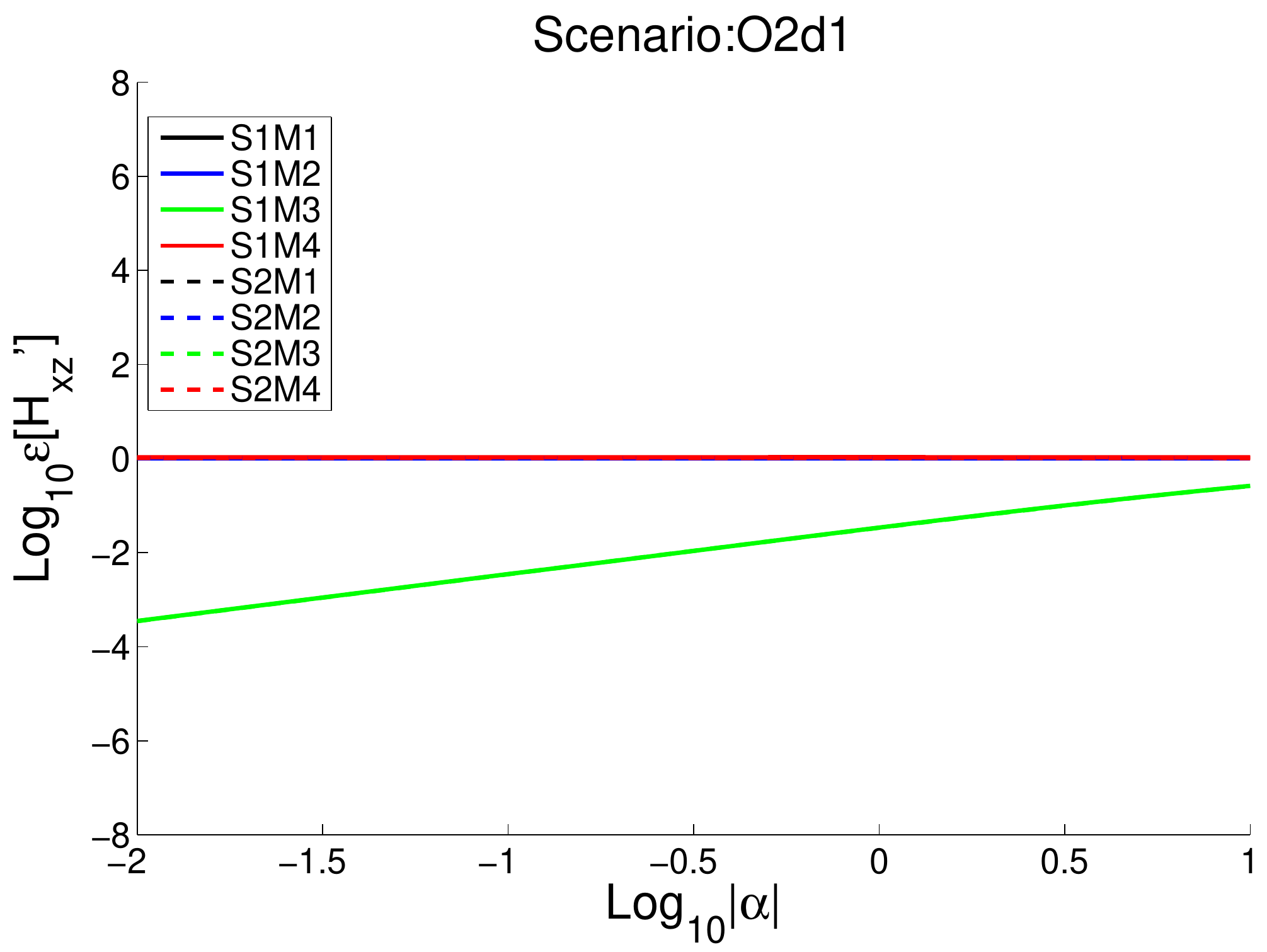}}
\subfloat[\label{ReHxzO2D2}]{\includegraphics[width=3.25in]{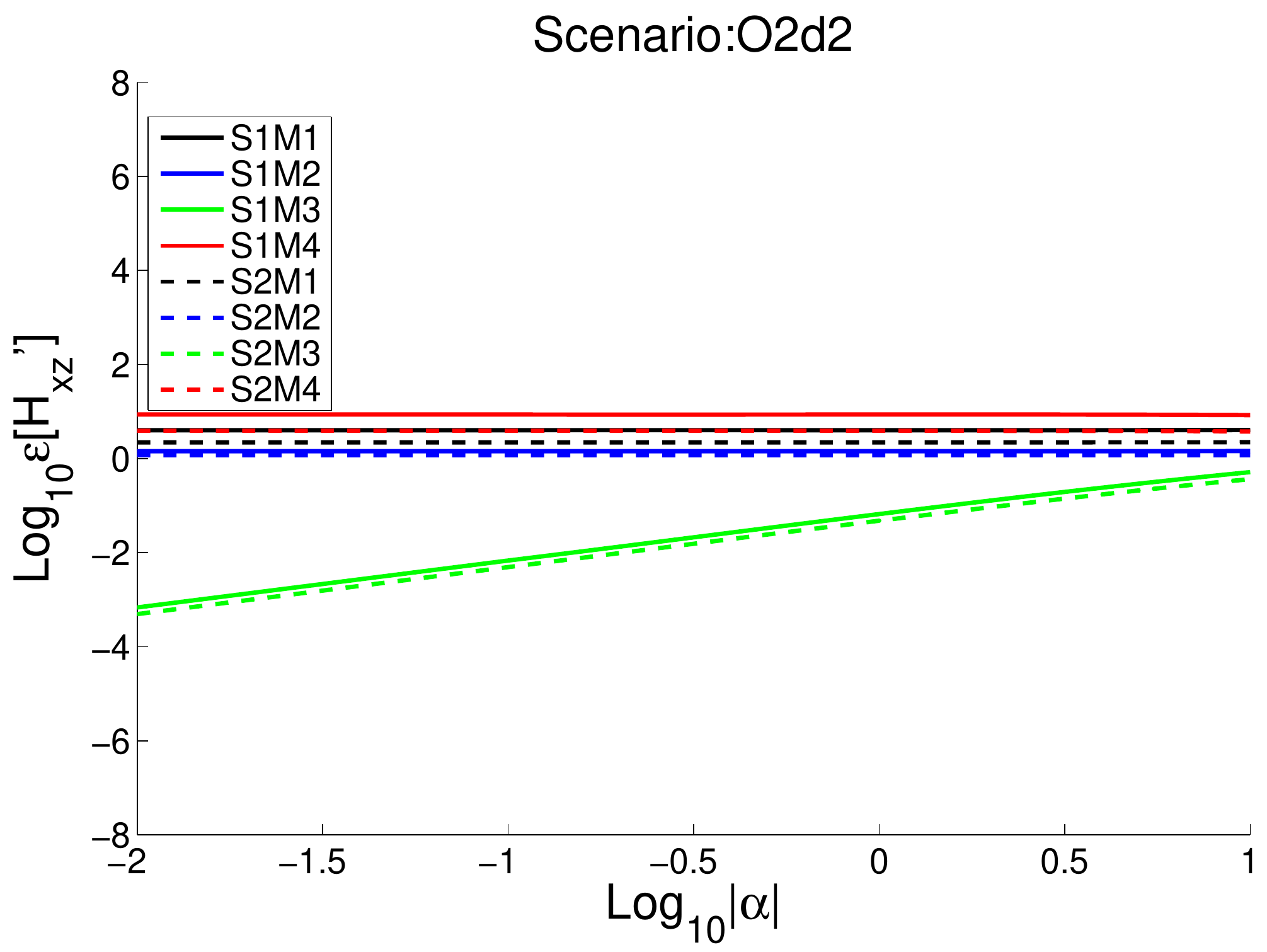}}

\subfloat[\label{ReHxzO3D1}]{\includegraphics[width=3.25in]{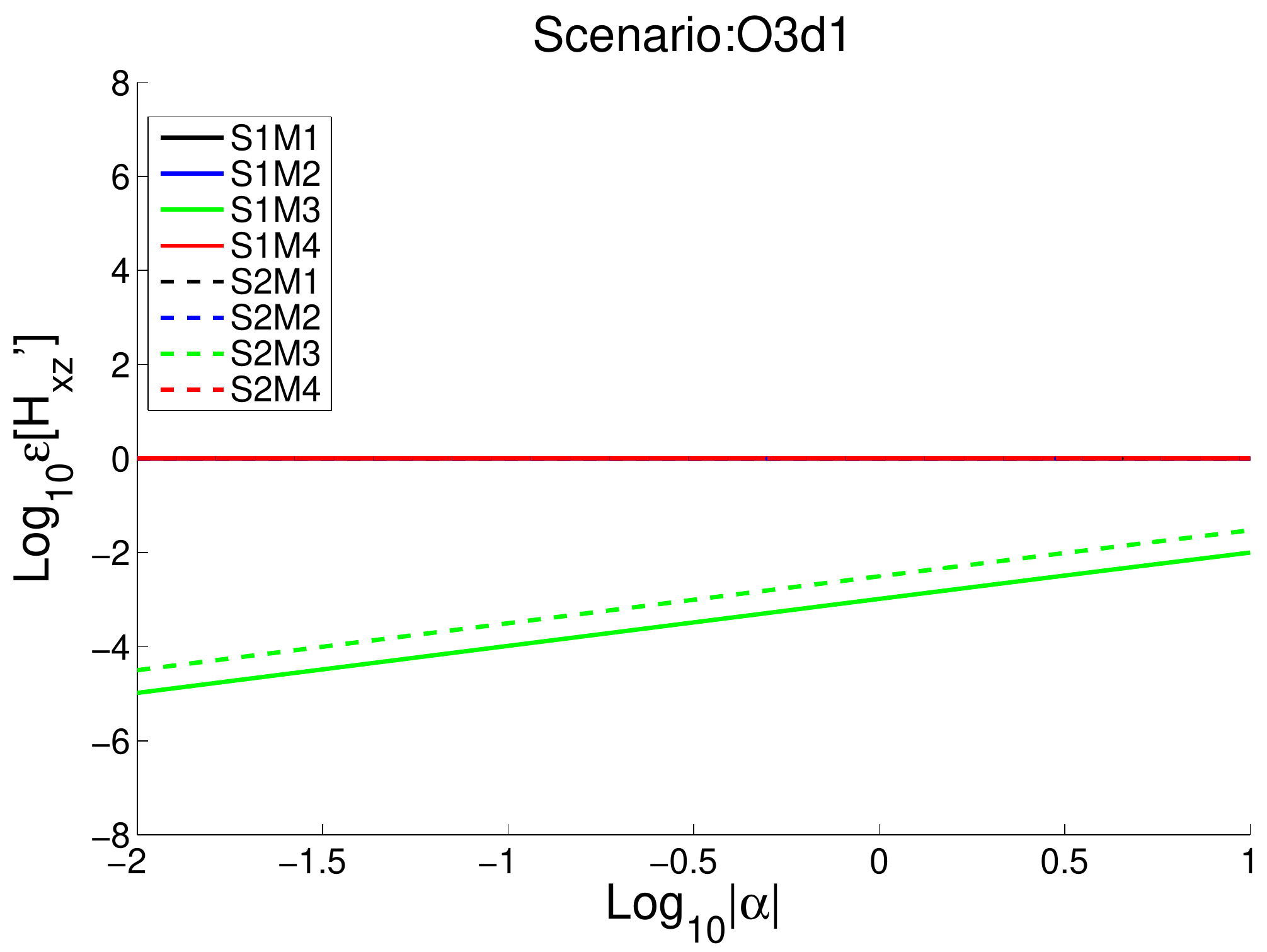}}
\subfloat[\label{ReHxzO3D2}]{\includegraphics[width=3.25in]{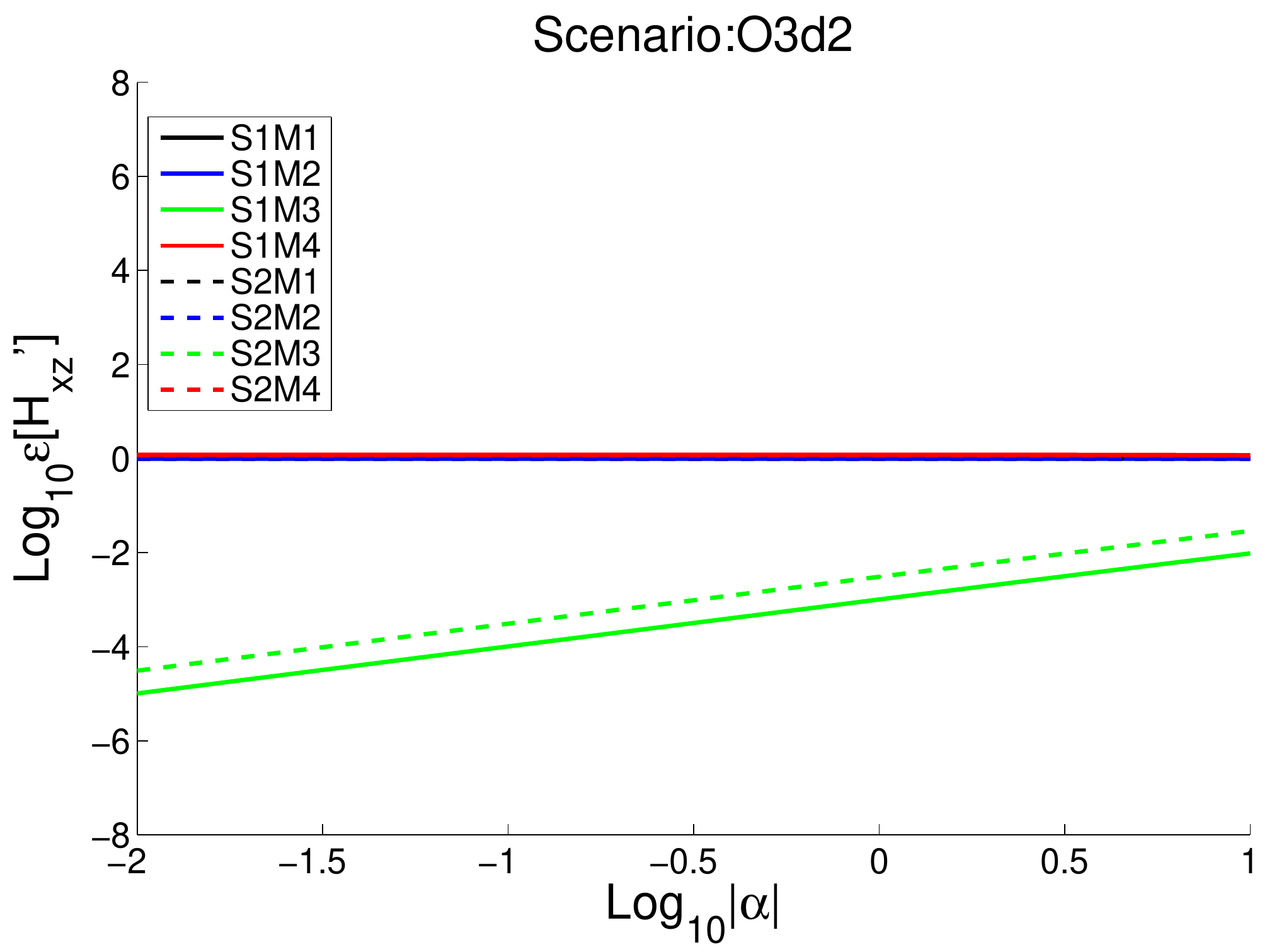}}
\caption{\small Relative error in computing $H_{xz}^{\prime}$=Re[$H_{xz}$].}
\label{ReHxz}
\end{figure}

\begin{figure}[H]
\centering
\subfloat[\label{ImHxzO1D1}]{\includegraphics[width=3.25in]{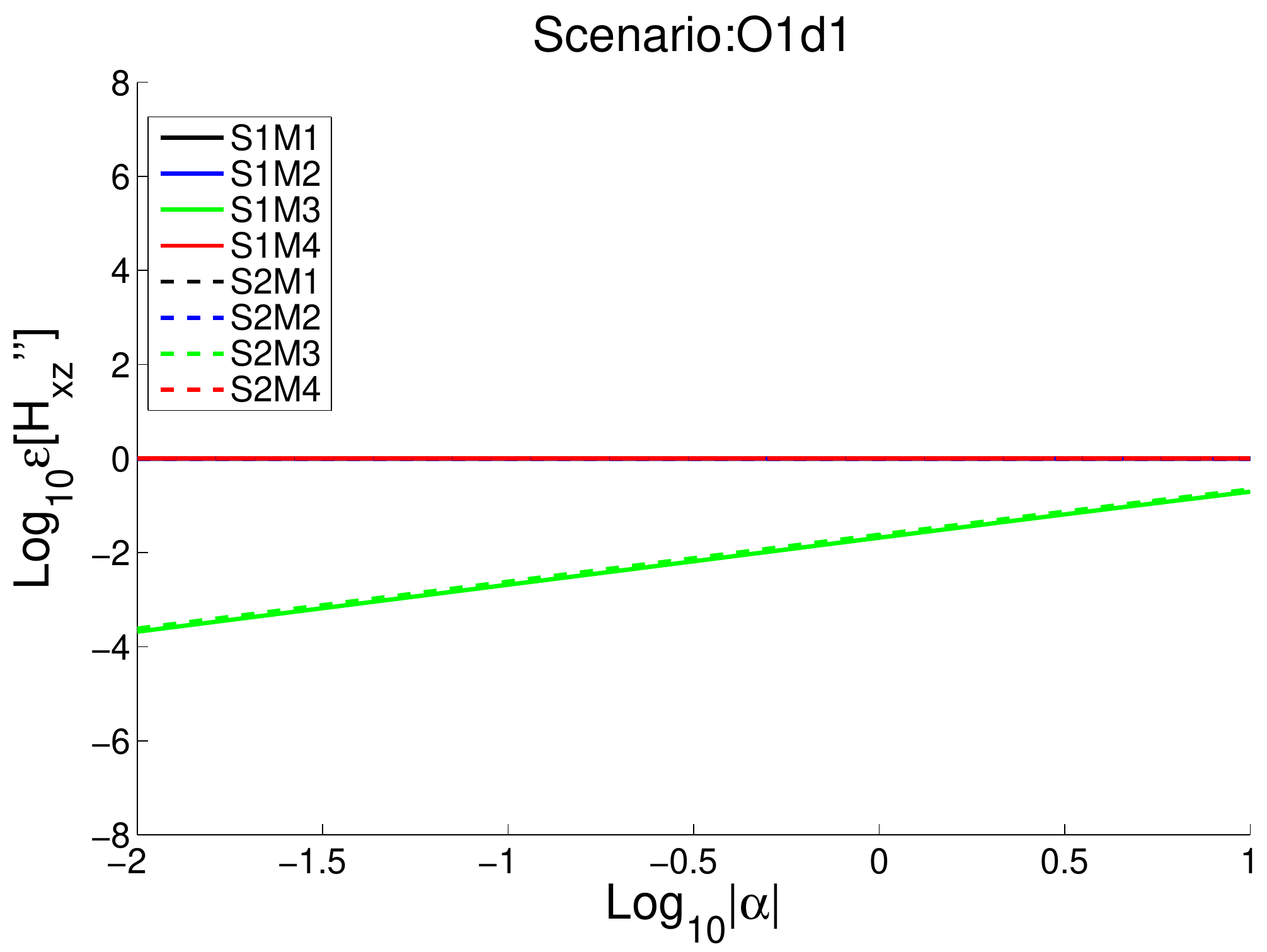}}
\subfloat[\label{ImHxzO1D2}]{\includegraphics[width=3.25in]{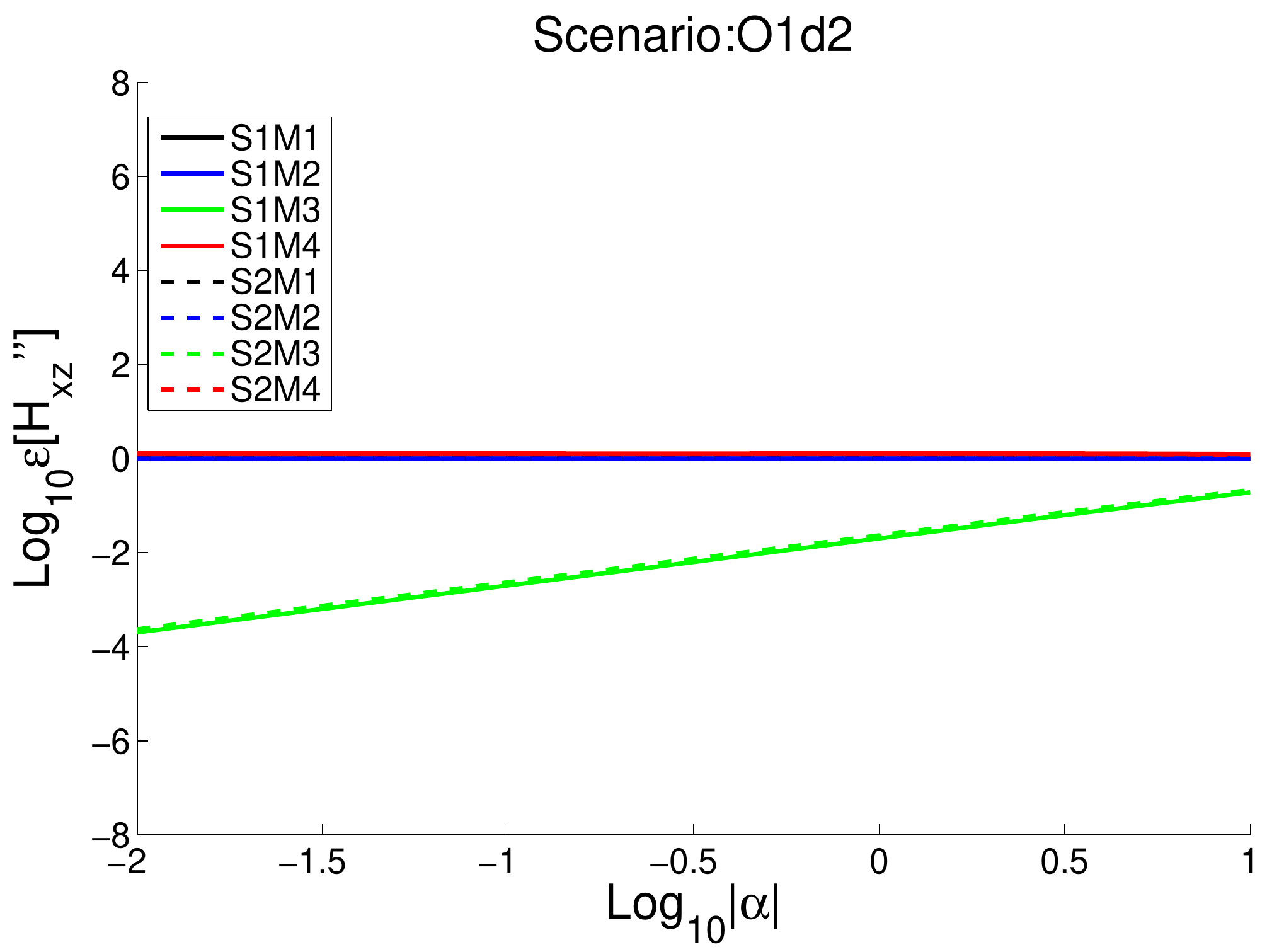}}

\subfloat[\label{ImHxzO2D1}]{\includegraphics[width=3.25in]{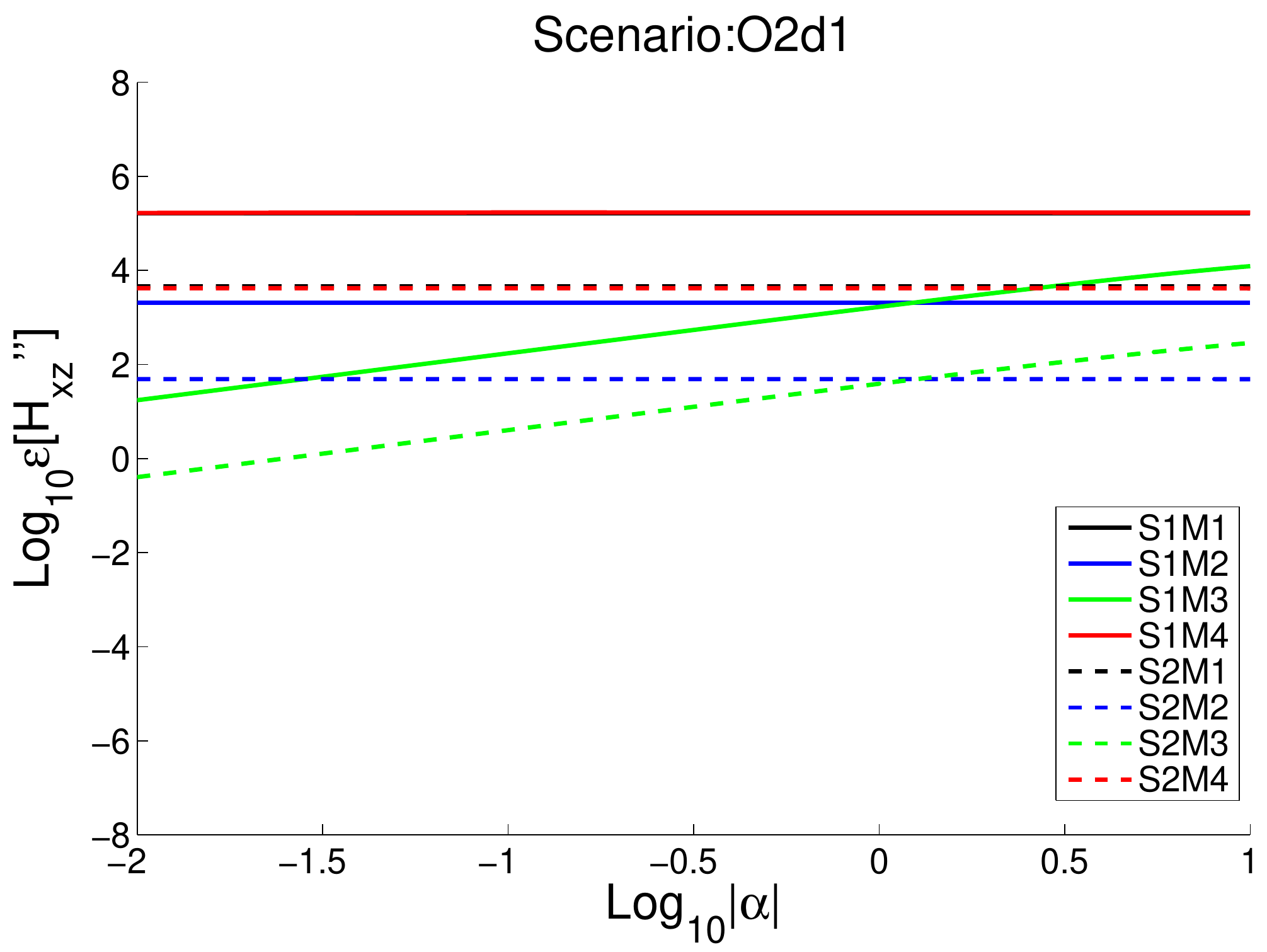}}
\subfloat[\label{ImHxzO2D2}]{\includegraphics[width=3.25in]{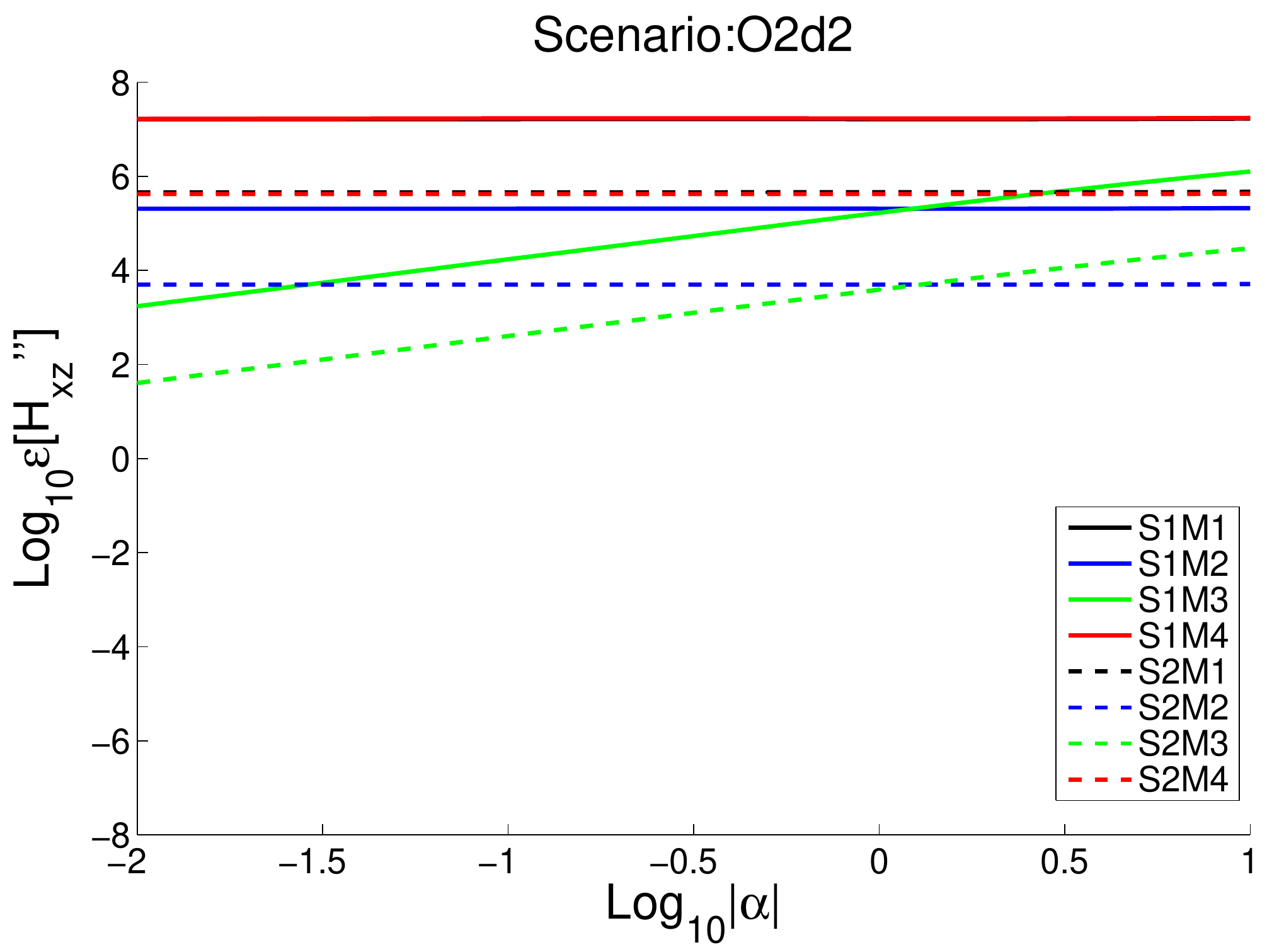}}

\subfloat[\label{ImHxzO3D1}]{\includegraphics[width=3.25in]{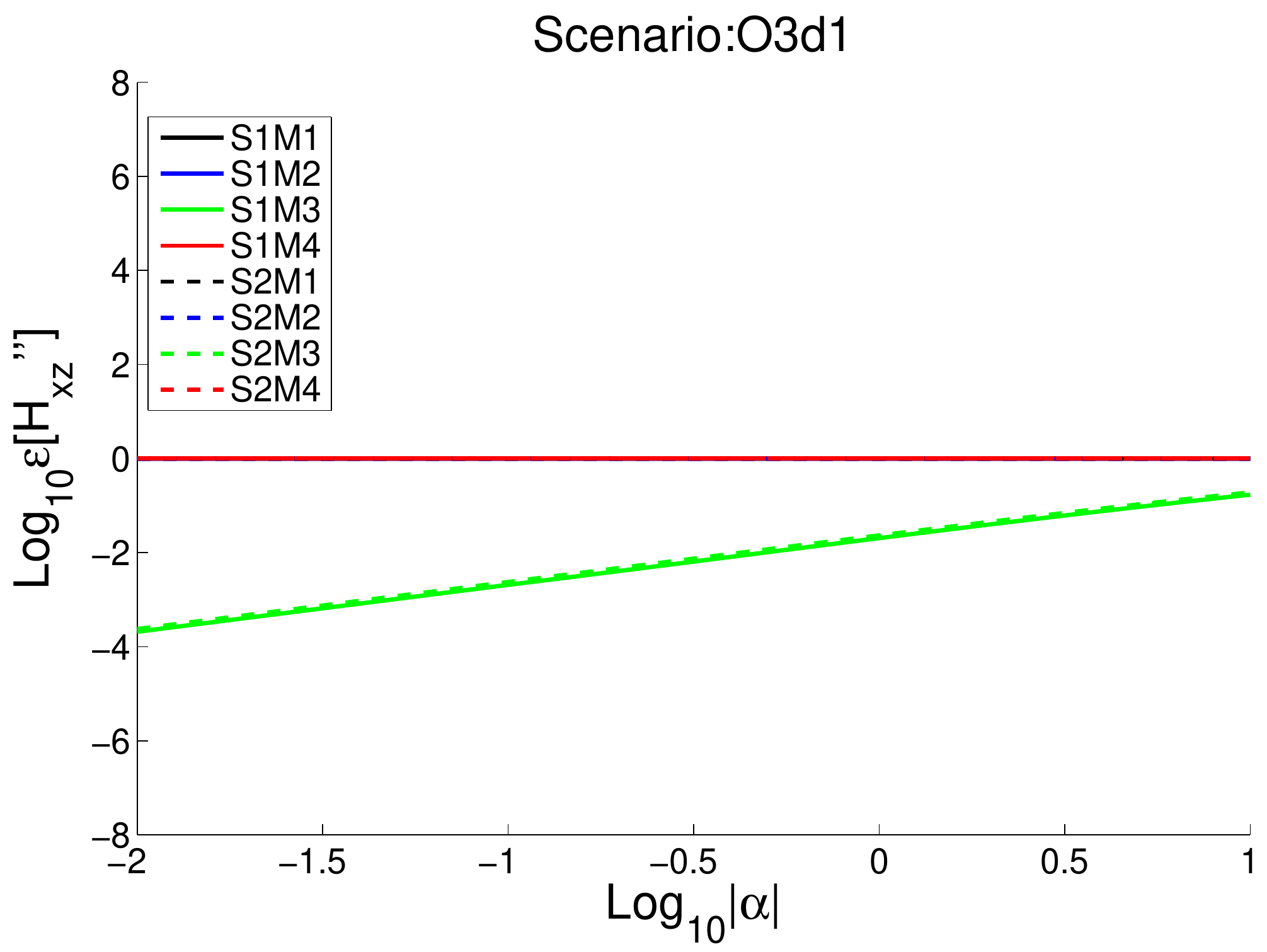}}
\subfloat[\label{ImHxzO3D2}]{\includegraphics[width=3.25in]{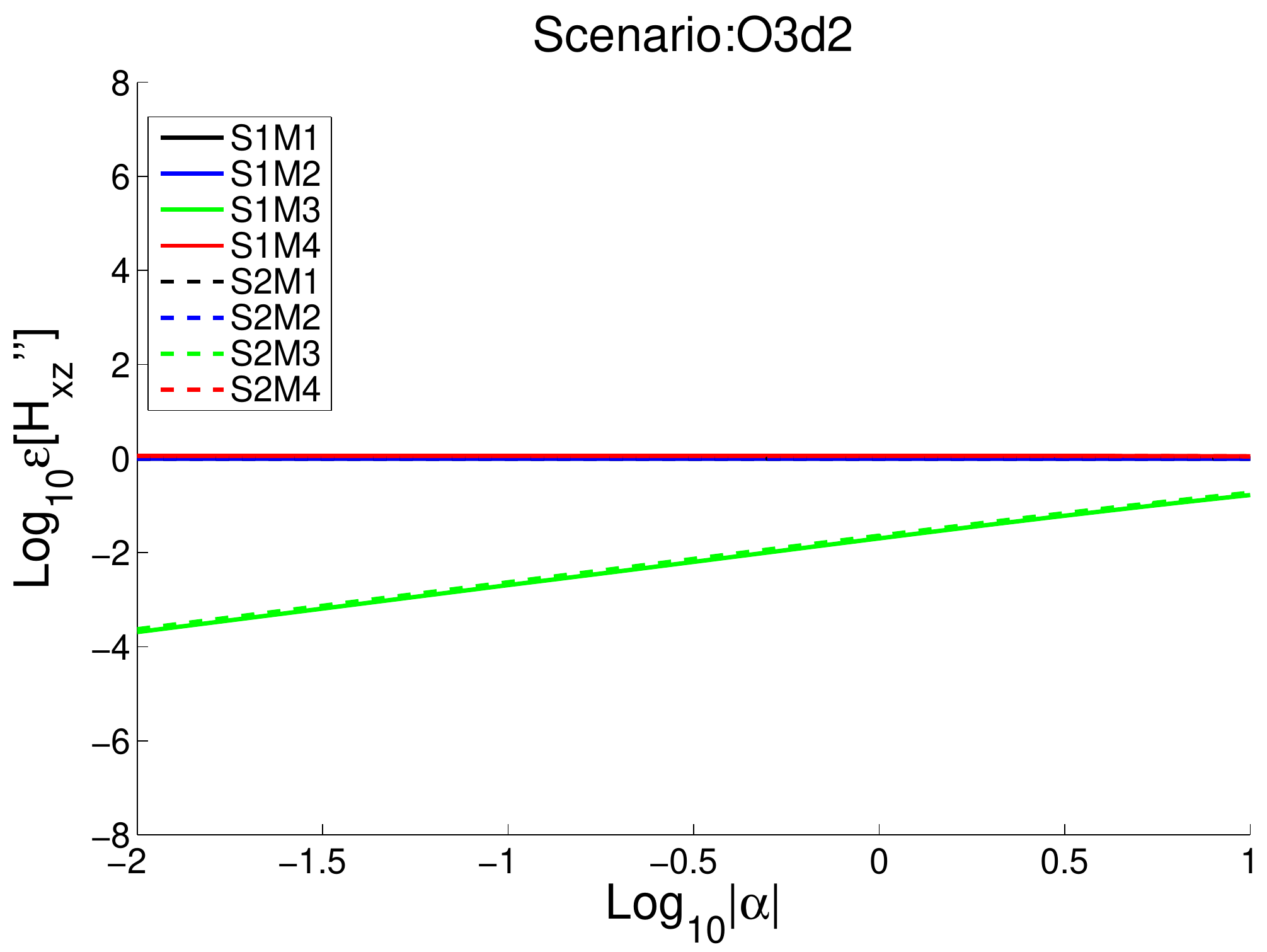}}
\caption{\small Relative error in computing $H_{xz}^{\prime \prime}$=Im[$H_{xz}$].}
\label{ImHxz}
\end{figure}

\begin{figure}[H]
\centering
\subfloat[\label{ReHzxO1D1}]{\includegraphics[width=3.25in]{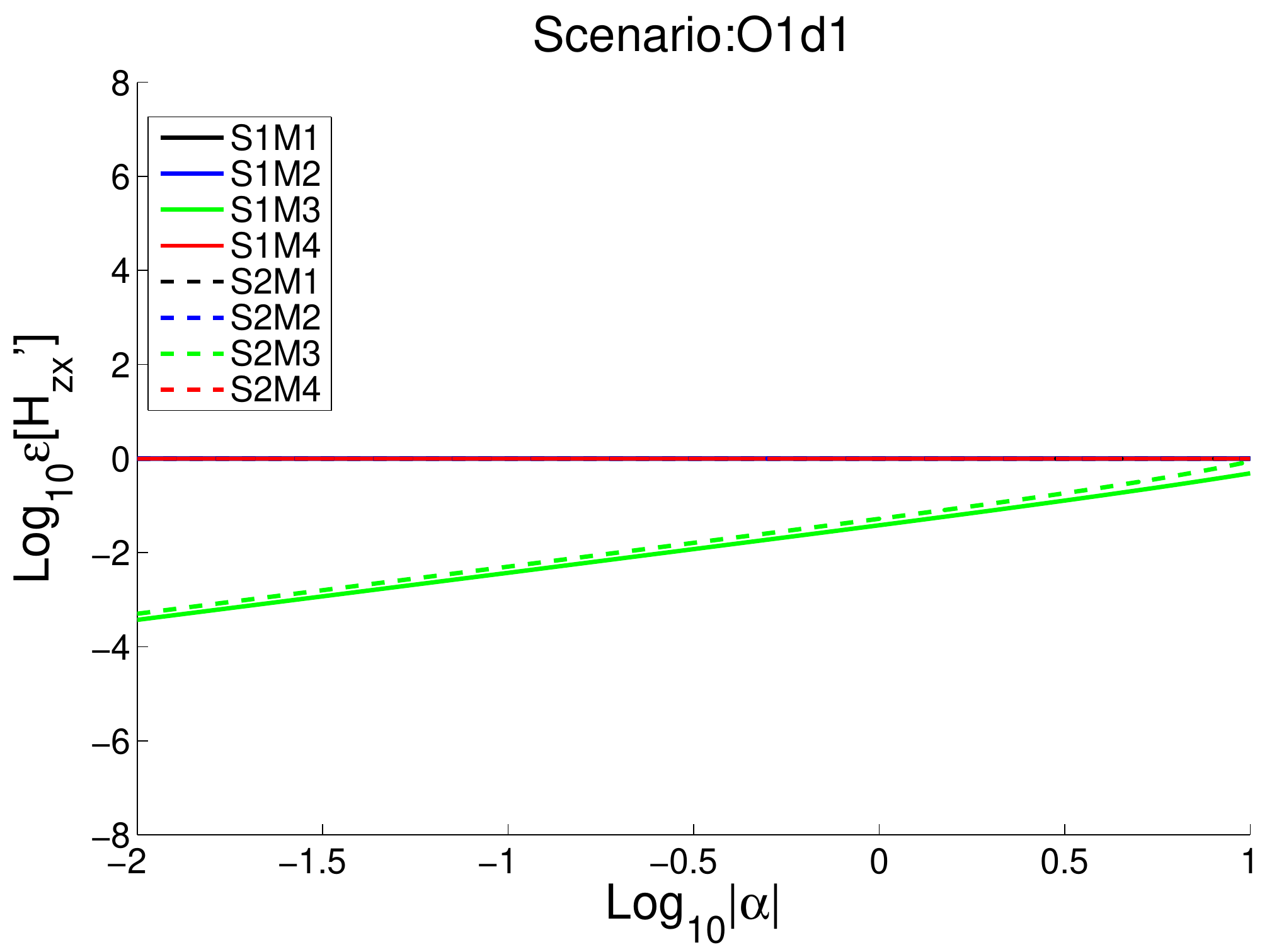}}
\subfloat[\label{ReHzxO1D2}]{\includegraphics[width=3.25in]{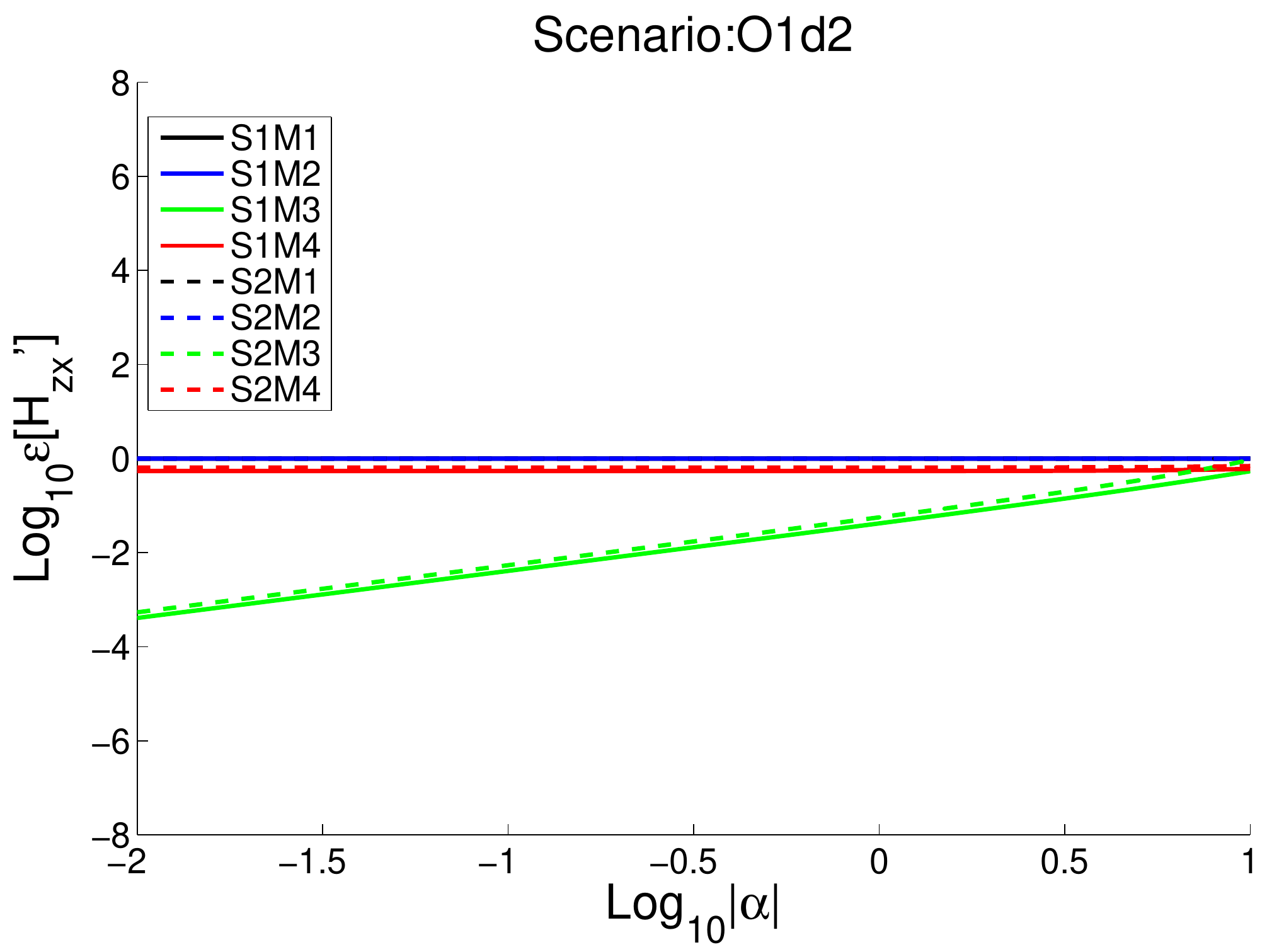}}

\subfloat[\label{ReHzxO2D1}]{\includegraphics[width=3.25in]{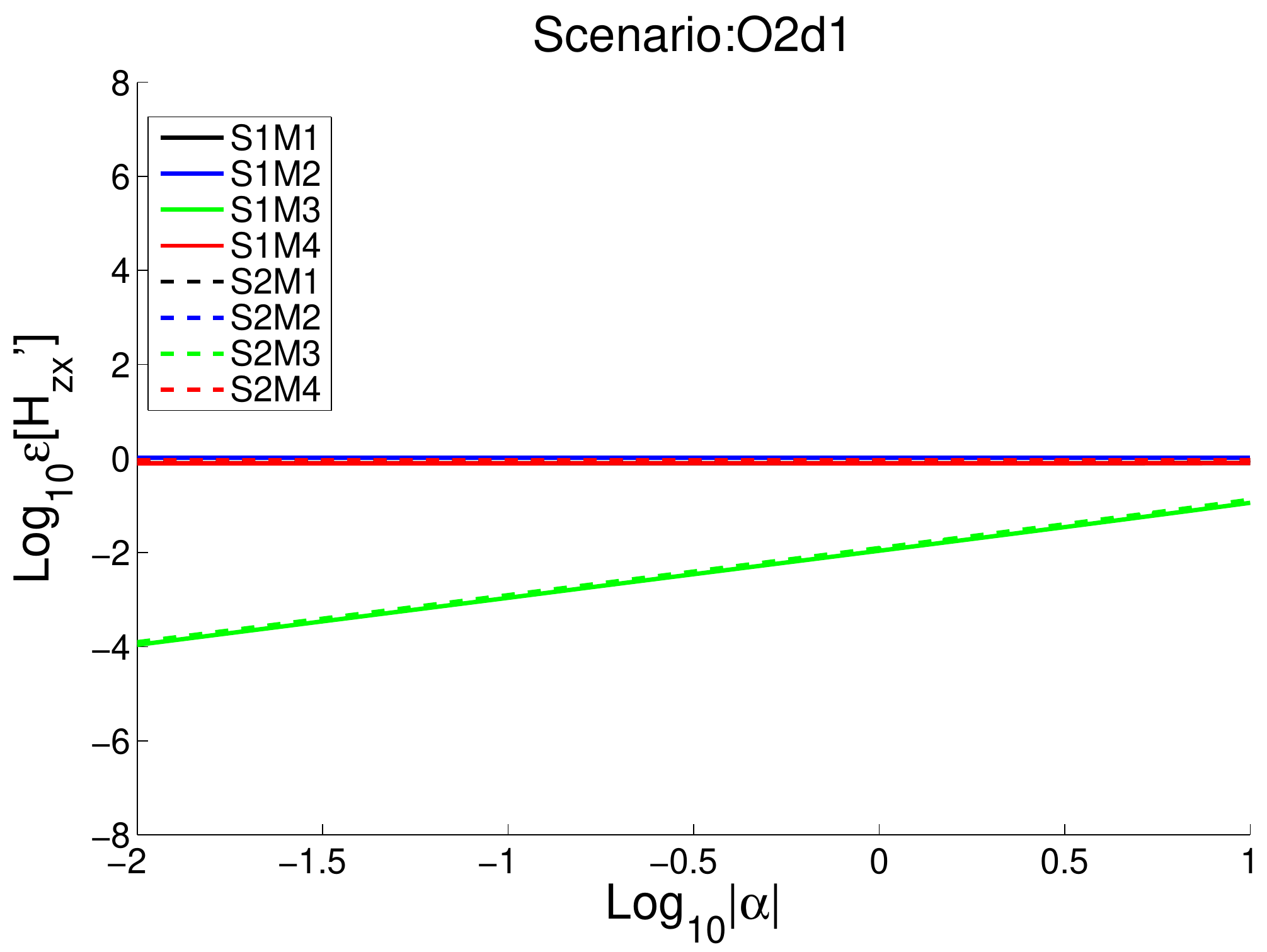}}
\subfloat[\label{ReHzxO2D2}]{\includegraphics[width=3.25in]{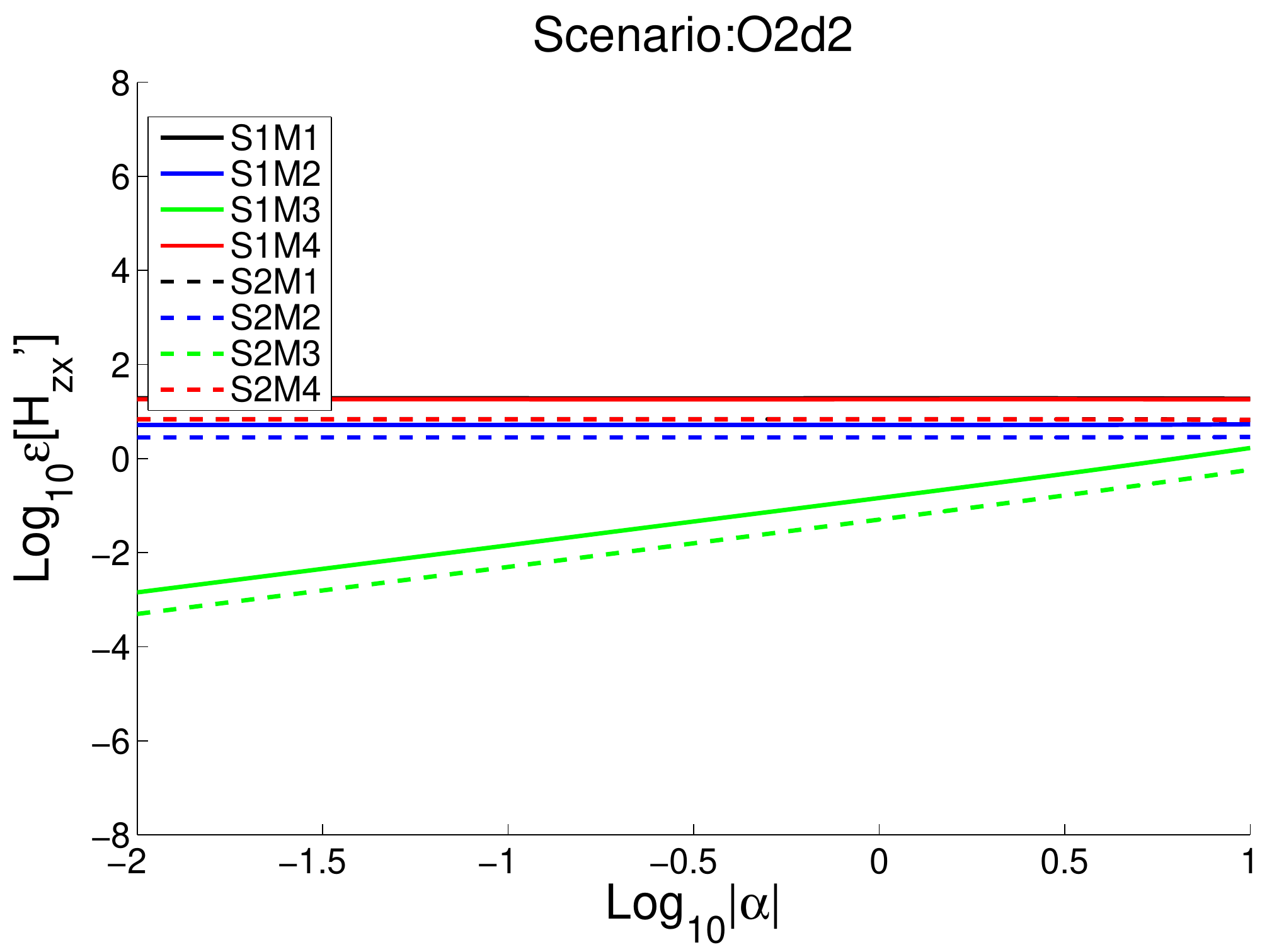}}

\subfloat[\label{ReHzxO3D1}]{\includegraphics[width=3.25in]{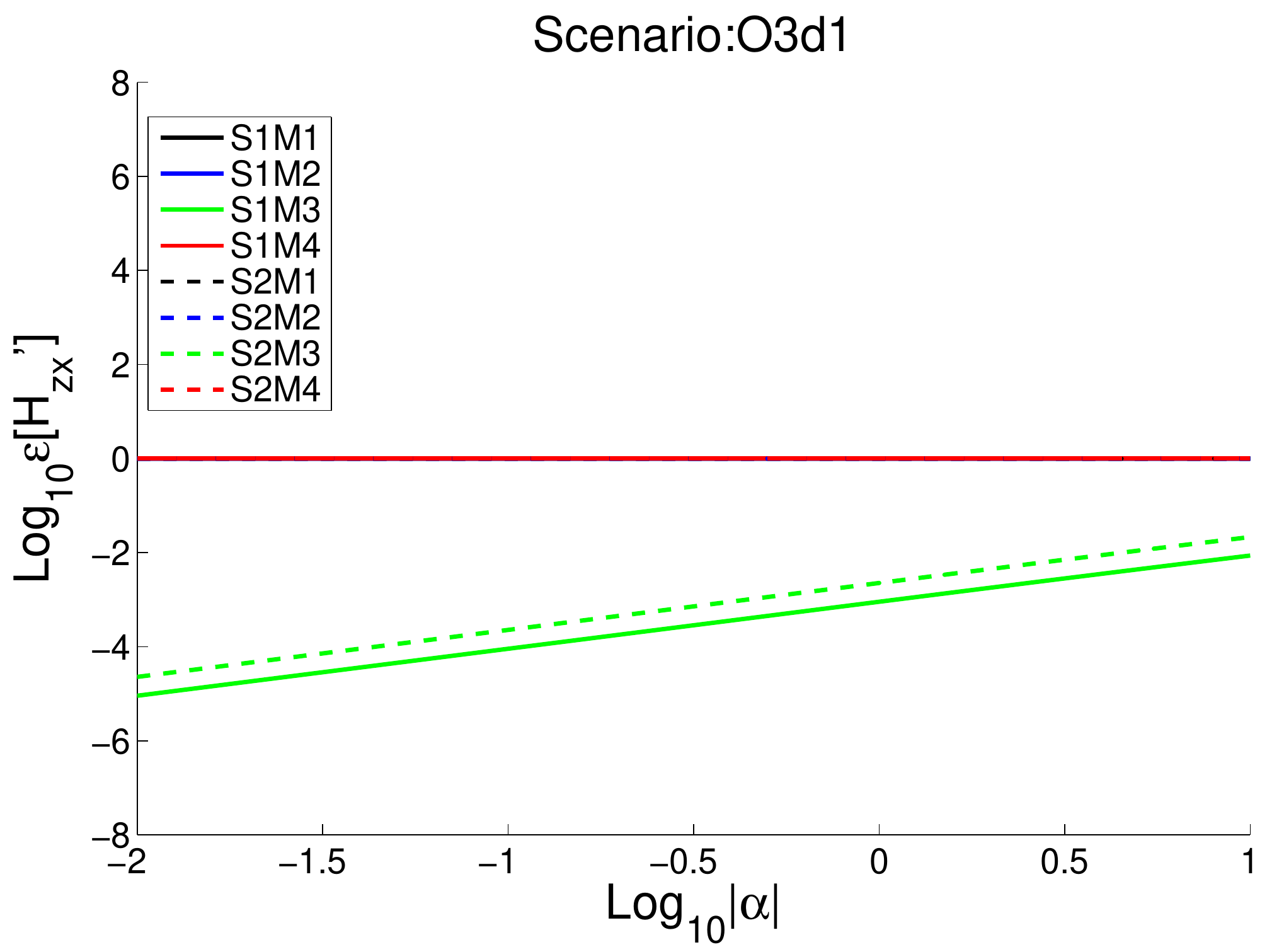}}
\subfloat[\label{ReHzxO3D2}]{\includegraphics[width=3.25in]{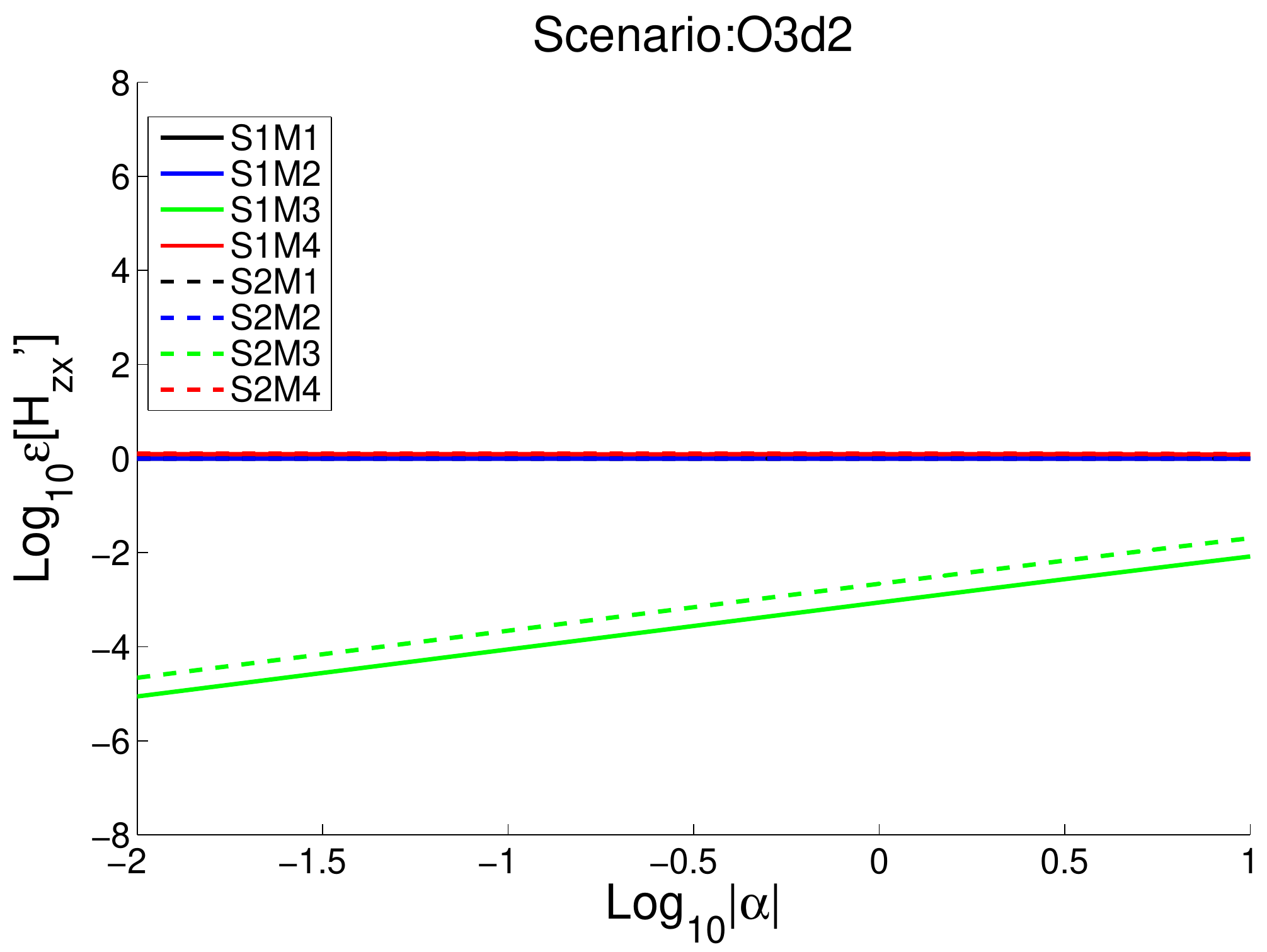}}
\caption{\small Relative error in computing $H_{zx}^{\prime}$=Re[$H_{zx}$].}
\label{ReHzx}
\end{figure}

\begin{figure}[H]
\centering
\subfloat[\label{ImHzxO1D1}]{\includegraphics[width=3.25in]{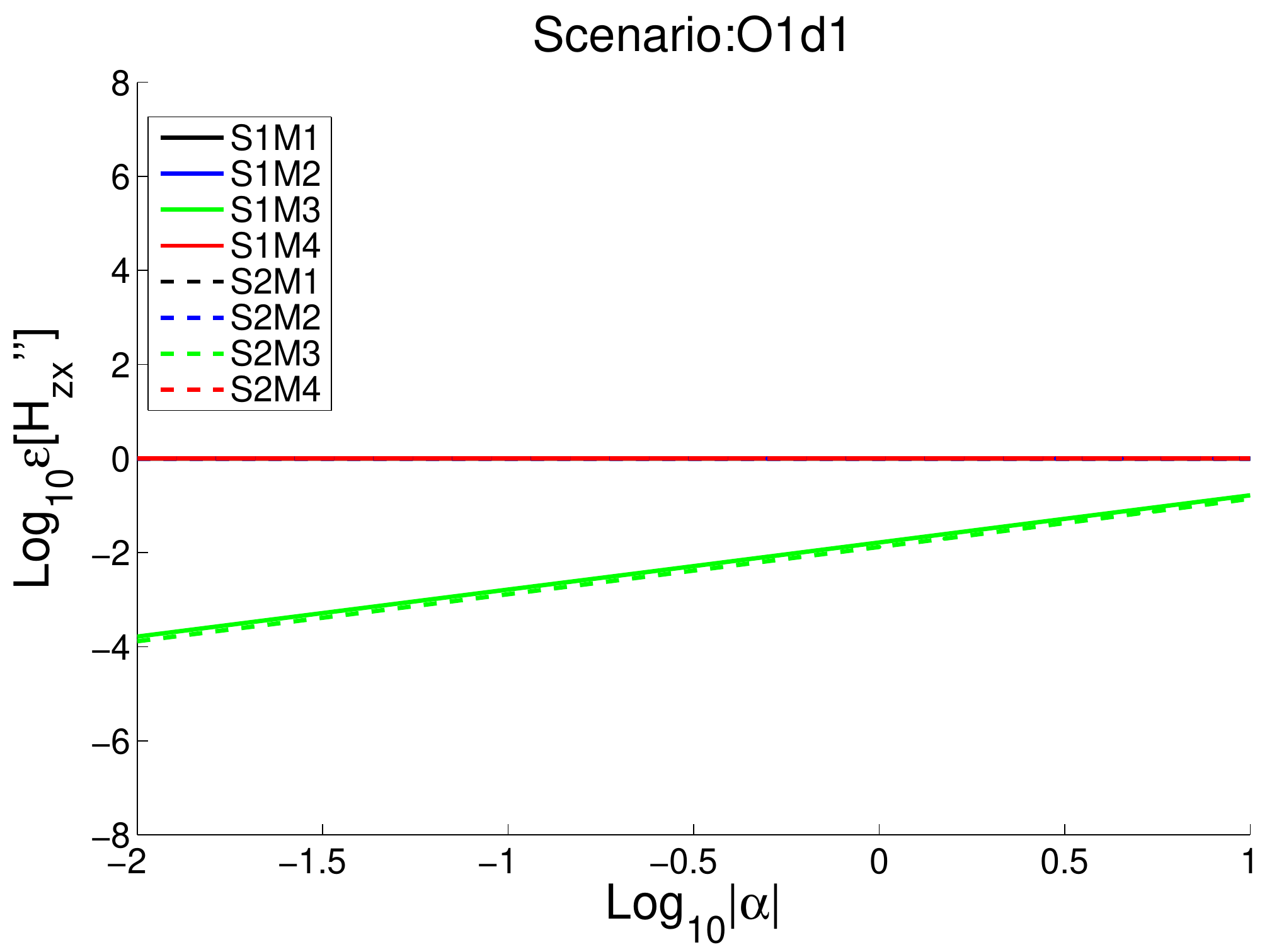}}
\subfloat[\label{ImHzxO1D2}]{\includegraphics[width=3.25in]{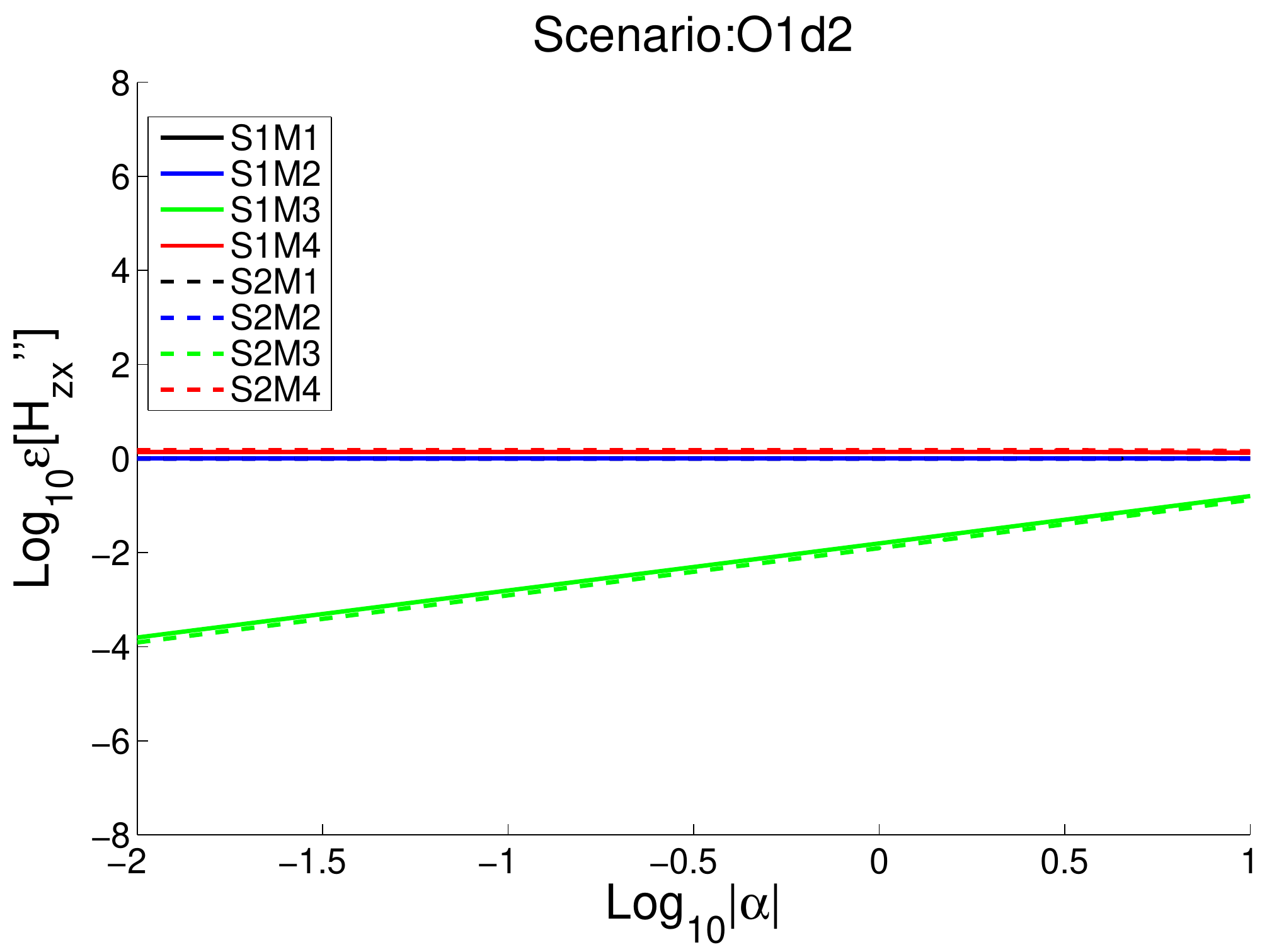}}

\subfloat[\label{ImHzxO2D1}]{\includegraphics[width=3.25in]{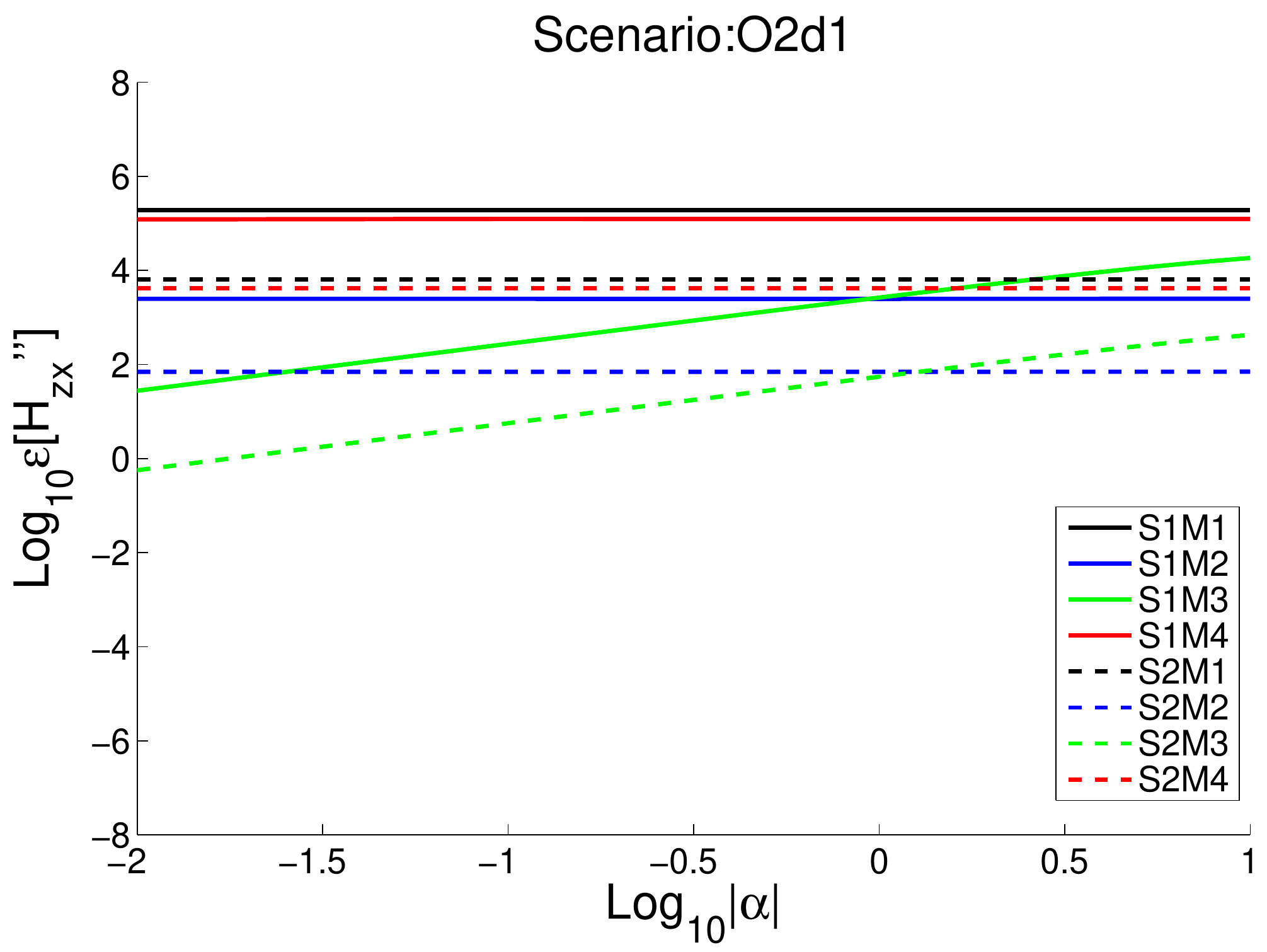}}
\subfloat[\label{ImHzxO2D2}]{\includegraphics[width=3.25in]{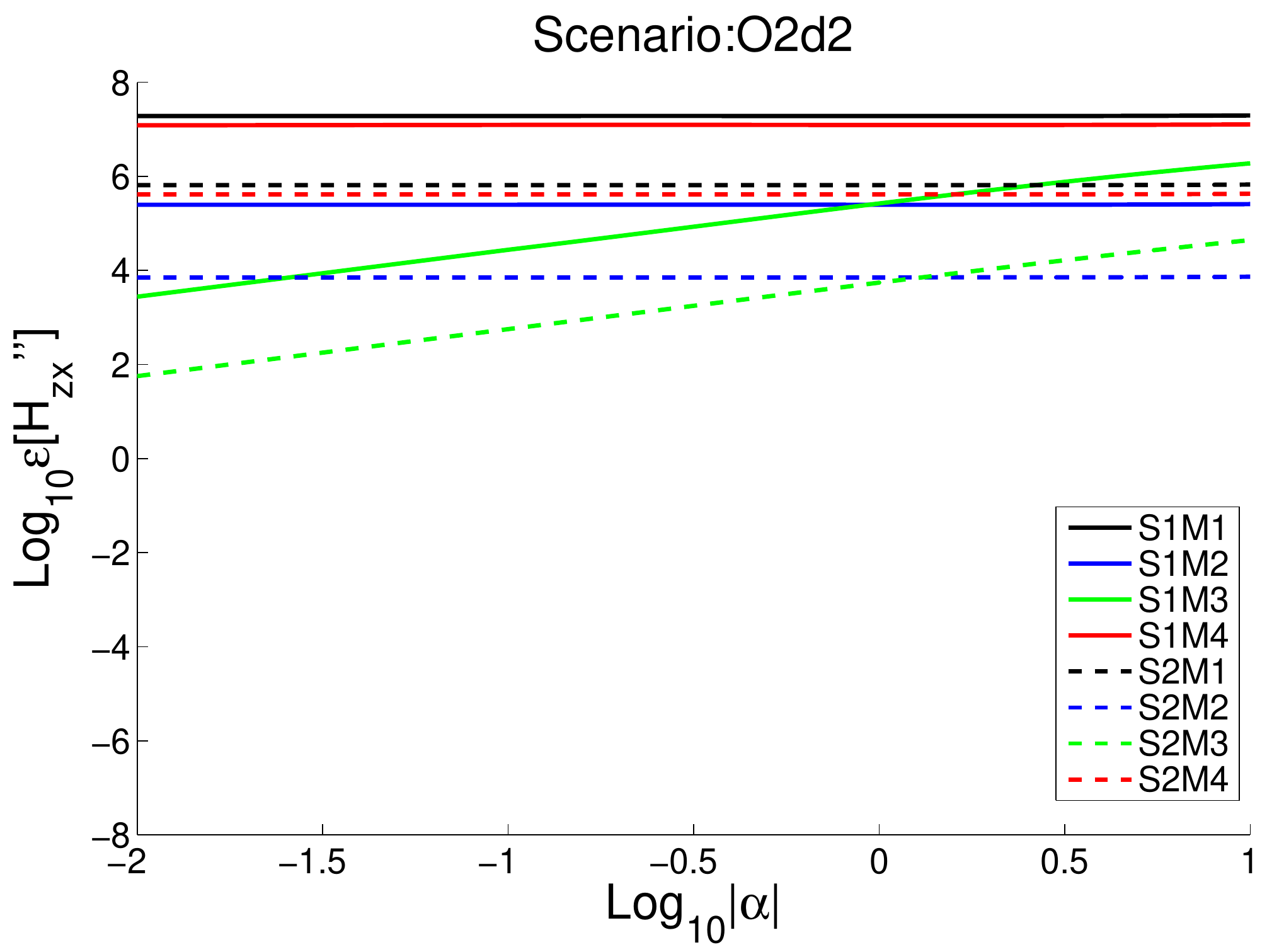}}

\subfloat[\label{ImHzxO3D1}]{\includegraphics[width=3.25in]{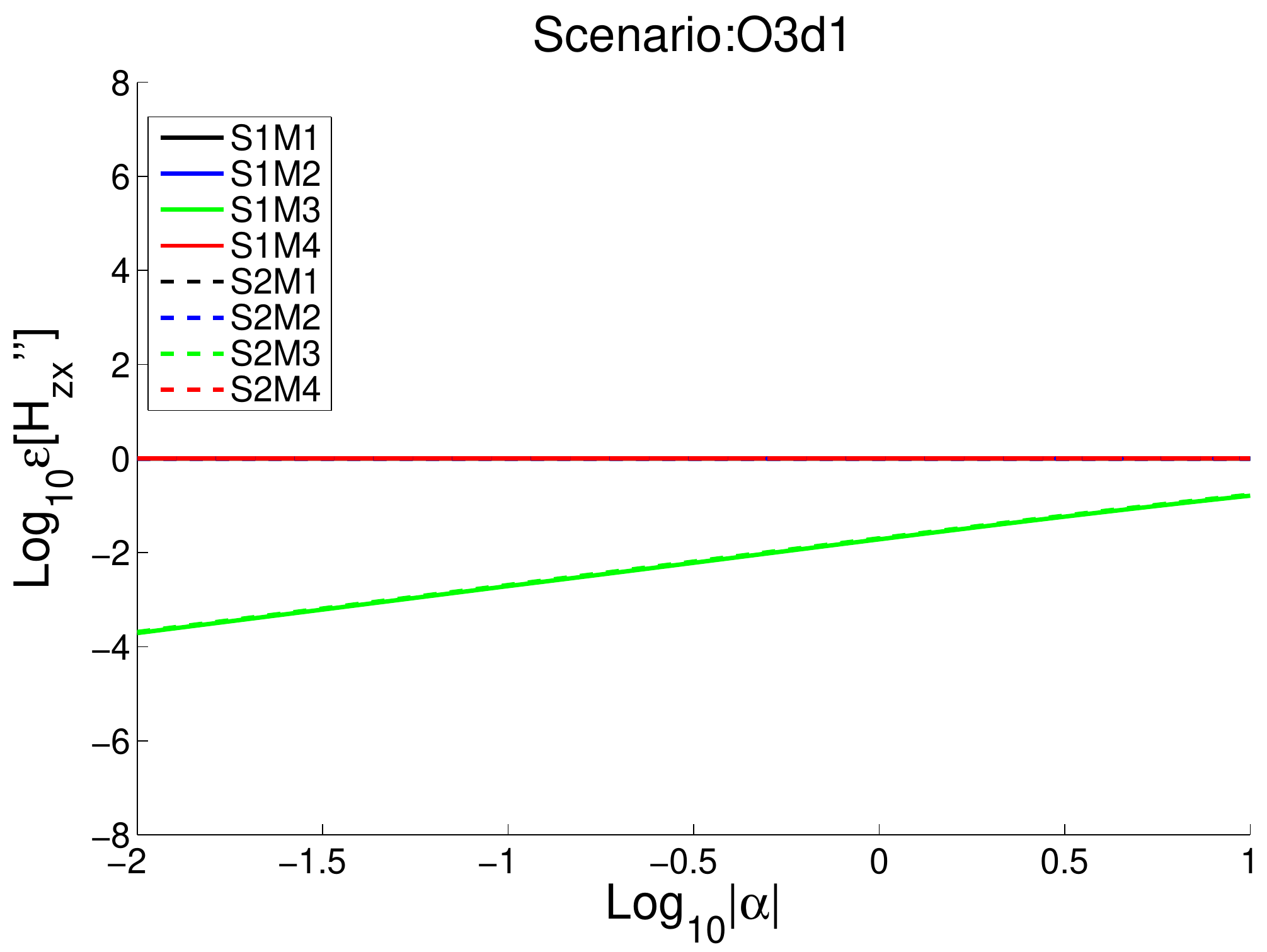}}
\subfloat[\label{ImHzxO3D2}]{\includegraphics[width=3.25in]{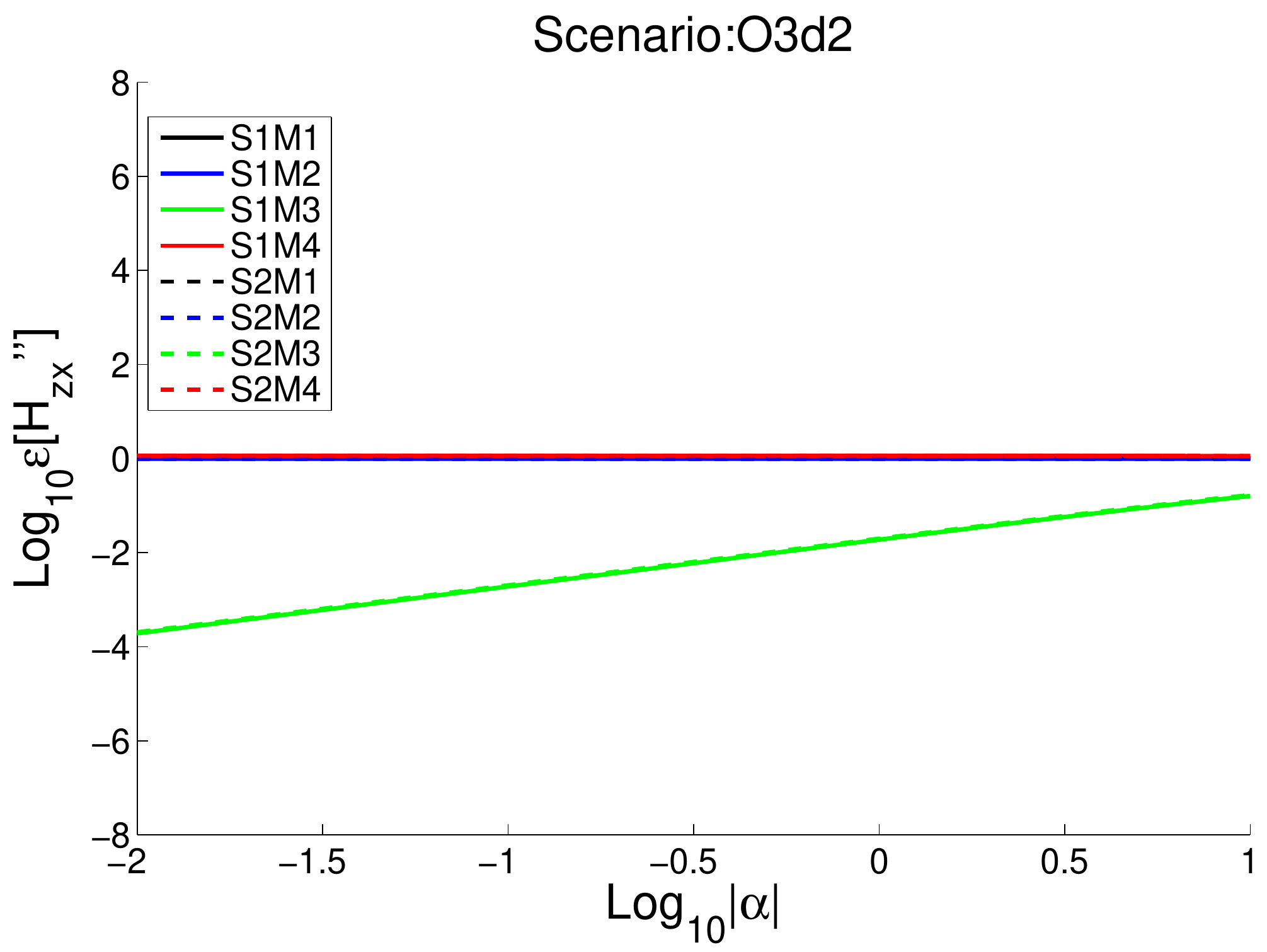}}
\caption{\small Relative error in computing $H_{zx}^{\prime \prime}$=Im[$H_{zx}$].}
\label{ImHzx}
\end{figure}

Figures \ref{ReHxx}-\ref{ImHzx} display errors (respectively) for the following field components: Re$[\mathcal{H}_{xx}]$, Im$[\mathcal{H}_{xx}]$, Re$[\mathcal{H}_{yy}]$, Im$[\mathcal{H}_{yy}]$, Re$[\mathcal{H}_{zz}]$, Im$[\mathcal{H}_{zz}]$, Re$[\mathcal{H}_{xz}]$, Im$[\mathcal{H}_{xz}]$, Re$[\mathcal{H}_{zx}]$, and Im$[\mathcal{H}_{zx}]$. Let $\Delta_{\epsilon}$ denote the rate of change of error ($\mathrm{Log}_{10}\epsilon$) versus $\mathrm{Log}_{10}|\alpha|$: A slope of two (one) indicates quadratic (linear) error variation versus effective tilt. We remark that in some plots, one or more material/tool-spacing scenario curves show strange ``dips" in the error levels (Figs. \ref{ReHxxO1D1}-\ref{ReHxxO1D2}, \ref{ImHyyO1D1}, \ref{ReHzzO2D2}, \ref{ReHzzO3D2}, and \ref{ReHxzO1D1}). Given how small the error dips are (typically $\leq 1$ DoP), we ascribe the dips to a combination of machine-dependent computation and problem geometry-dependent numerical cancellation (effective tilt, material profile, $d$, and sensor characteristics). There are also some ``kinks" in the error behavior, at very small tilt, that can be observed in one or more material/tool-spacing scenario curves within each of Figures \ref{ReHxxO2D1}-\ref{ReHxxO3D2}, \ref{ImHxxO2D1}, \ref{ImHxxO3D1}, \ref{ReHyyO2D1}-\ref{ReHyyO3D2}, \ref{ImHyyO1D1}, \ref{ImHyyO2D1}, \ref{ImHyyO3D1}, \ref{ReHzzO2D1}-\ref{ReHzzO3D2}, \ref{ImHzzO2D1}, and \ref{ImHzzO3D1}. Despite these two sporadically-occurring characteristics in the error curves (typically only 1-2 curves in each stated figure has these properties), the overall error trends are well preserved, and it is this we summarize in the following observations:
\begin{enumerate}
\item A two order of magnitude increase in $d$ (from 2mm [``d1" plots] to 200mm [``d2" plots]) produces low error variation (typically 0-1 DoP error increase). 
\item Error is typically 1-3 DoP higher when the transmitter and receiver are in different layers (``O2" plots), versus when they are in the same layer (``O1" and ``O3" plots). By contrast, the error is approximately equal if the transmitter and receiver are both either above or below the interface.  
\item Transmitter-receiver spacing (``S1" vs. ``S2" curves) has little effect on error levels (typically 0-1 DoP difference). 
\item The M3 scenario curves typically show greatest error (versus M1, M2, and M4 curves) in the co-polarized results, but the lowest error in the cross-polarized plots. There does not appear to be any obvious, systematic trend in error variation between the M1, M2, and M4 cases across the studied sensor/environment parameter permutations. 
\item The M1, M2, and M4 curves show quadratic error variation in the co-polarized results (their cross-polarized errors are intolerably high), while the M3 curves predominantly show instead linear error variation for both co-polarized and cross-polarized results. For Figures \ref{ImHxxO2D1}-\ref{ImHxxO2D2}, \ref{ImHyyO2D1}-\ref{ImHyyO2D2}, and \ref{ImHzzO2D1}-\ref{ImHzzO2D2} the M3 curves, interestingly, show quadratic error too however. 
\item The cross-pol response errors not only are much higher than their co-pol response error counterparts, but the errors vary more versus $d$, $L_s$, and sensor position $D$ too. 
\end{enumerate}

\newpage

\newpage

\newpage

\newpage

\newpage

\newpage

\newpage

\newpage

\newpage

\newpage
\newpage

\section{\label{Val}Application to Triaxial Induction Sensor Responses}

\begin{figure}[H]
\centering
\subfloat[\label{Fig2a}]{\includegraphics[width=2.75in]{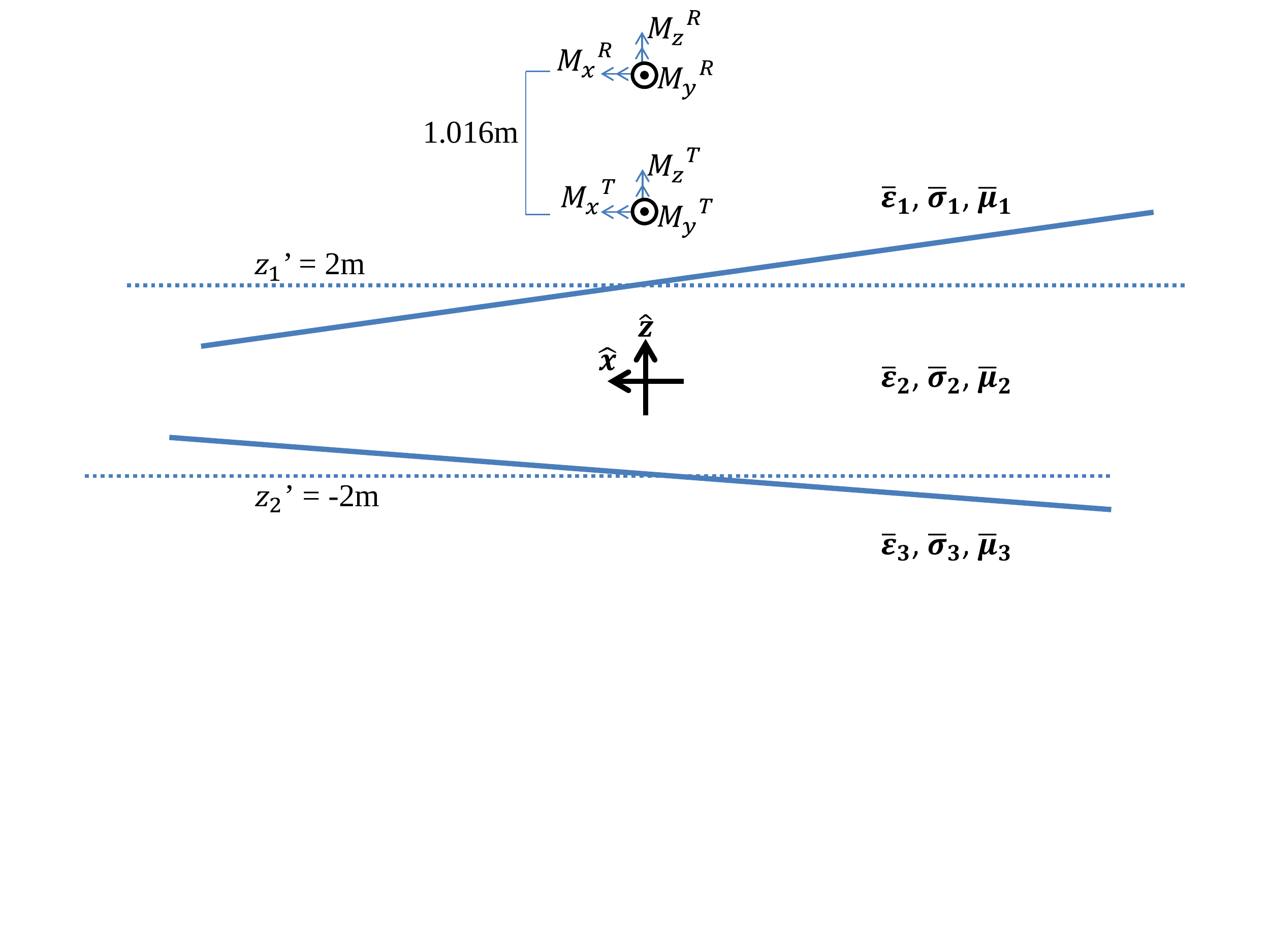}}
\subfloat[\label{Fig2b}]{\includegraphics[width=3.85in]{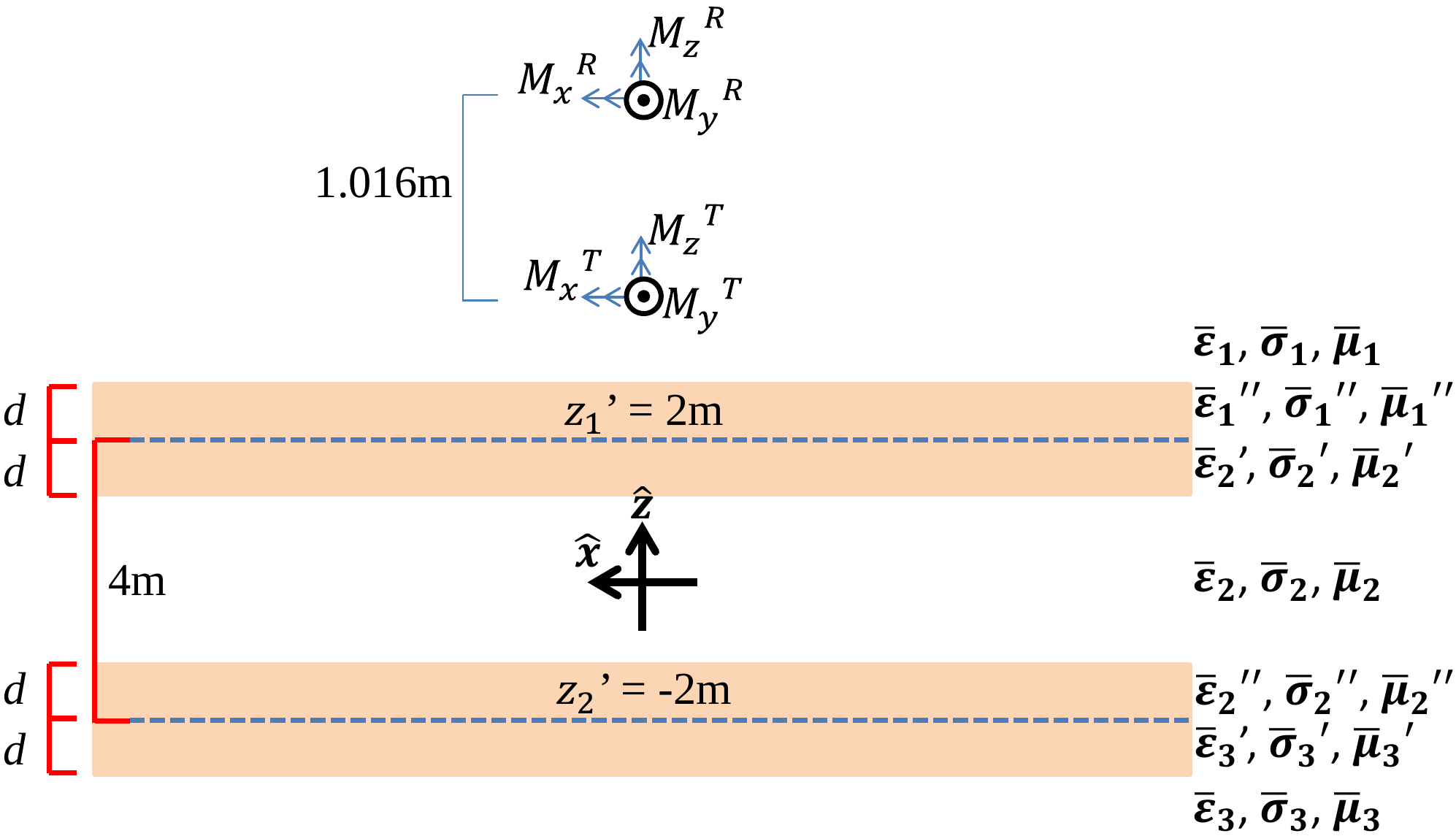}}
\caption{\small Original geometry (Fig. \ref{Fig2a}) and transformed, approximately equivalent geometry (Fig. \ref{Fig2b}) employed below. For clarity of illustration, the layers are shown tilted within the $xz$ plane ($\beta'=0^{\circ}$).}
\label{Fig2}
\end{figure}

Now we perform case studies, to illustrate the algorithm's flexibility in modeling media of diverse anisotropy and loss, involving twenty eight variations of a three-layered medium (seven tilt orientations, four central bed conductivity tensors), where the two interfaces exhibit (effective) relative tilting and are each flattened by two coating slabs $d$=2mm thick (i.e., one coating slab immediately above, and one coating slab immediately below, each flattened interface). Figure \ref{Fig2} depicts the well-logging scenario simulated: A vertically-oriented triaxial induction tool~\cite{wei,sofia}, that operates at $f$=100kHz, is transverse-centered at $(x,y)=(0,0)$, possesses three (co-located) electrically small loop antenna transmitters (modeled as unit-amplitude Hertzian [equivalent] magnetic current dipoles) directed along $x$ ($M_x^T$), $y$ ($M_y^T$), and $z$ ($M_z^T$), and possesses three (co-located) loop antenna receivers $\{M_x^R,M_y^R,M_z^R\}$ situated $L_s$=1.016m (forty inches) above the transmitters.\footnote{Note 1: ``T" superscript here denotes ``transmitter", \emph{not} non-Hermitian transpose.} 

All four material scenarios share the following common geological formation features prior to inserting the interface-flattening slabs: Top interface depth $z_1'=2$m and bottom interface depth $z_2'=-2$m,\footnote{For the equivalent problem with flattened interfaces, this choice of $z_1'$ and $z_2'$ results in the coating layer just above the bottom interface, and coating layer just below the top interface, having their respective coating layer/central formation layer junctions spaced 3.996m apart.} all three layers possess isotropic relative dielectric constant (i.e., excluding conductivity) $\epsilon_{r1}=\epsilon_{r2}=\epsilon_{r3}=1$ and isotropic relative magnetic permeability $\mu_{r1}=\mu_{r2}=\mu_{r3}=1$, layer one (top layer) possesses isotropic electric conductivity $\sigma_1=50$mS/m, layer three (bottom layer) possesses isotropic electric conductivity $\sigma_3=20$mS/m, the polar tilt angles of the two interfaces are equal to $\alpha_2'=-\alpha_1'=\alpha' \geq 0$, and the azimuth tilting orientation of the two interfaces are equal to $\beta_1'=\beta_2'=\beta'$. In other words, the \emph{relative} tilting between the two interfaces is $2\alpha'$ degrees while we choose to tilt both interfaces with identical azimuth orientation. This choice is made to isolate and better understand effects of relative polar tilting, without the added, confounding effect of relative azimuth deviation between the interfaces.

In the Figures below, the labels T1, T2, etc. in the legend represent induction log signature curves corresponding to different interface tilting scenarios, namely: $\{\alpha'=0^{\circ},\beta'=0^{\circ}\}$ (T1, solid black curve), $\{\alpha'=1^{\circ},\beta'=0^{\circ}\}$ (T2, solid blue), $\{\alpha'=1^{\circ},\beta'=45^{\circ}\}$ (T3, solid green), $\{\alpha'=1^{\circ},\beta'=90^{\circ}\}$ (T4, solid red), $\{\alpha'=3^{\circ},\beta'=0^{\circ}\}$ (T5, dash-dot blue), $\{\alpha'=3^{\circ},\beta'=45^{\circ}\}$ (T6, dash-dot green), $\{\alpha'=3^{\circ},\beta'=90^{\circ}\}$ (T7, dash-dot red). We chose the azimuth tilt orientations $\beta'=0^{\circ}$ and $\beta'=90^{\circ}$ specifically due to our having examined already, in the previous section, the error of co-polarized fields components when they are oriented either within or orthogonal to the plane of interface tilting. Moreover for a given $\alpha'$, the $\beta'=45^{\circ}$ curves exhibit results intermediate to the $\beta'=0^{\circ}$ and $\beta'=90^{\circ}$ results, suggesting confidence in the $\beta'=45^{\circ}$ results too. 

(Material) Scenario 1: The highly resistive central layer (mimicking a hydrocarbon-bearing reservoir) possesses isotropic electric conductivity $\sigma_2=5$mS/m. Scenario 2: The central layer's conductivity is characterized by a non-deviated uniaxial conductivity tensor $\boldsymbol{\bar{\sigma}}_2=\sigma_{h2}(\bold{\hat{x}}\bold{\hat{x}}+\bold{\hat{y}}\bold{\hat{y}})+\sigma_{v2}\bold{\hat{z}}\bold{\hat{z}}$ with $\sigma_{h2}=$5mS/m and $\sigma_{v2}=$1mS/m. Scenario 3: Same as Scenario 2, except the symmetric, non-diagonal conductivity tensor writes as~\cite{anderson1}
\begin{equation}
\boldsymbol{\bar{\sigma}}_2=
\begin{bmatrix}
\sigma_{xx2}&\sigma_{xy2}&\sigma_{xz2}\\\sigma_{xy2}&\sigma_{yy2}&\sigma_{yz2}\\\sigma_{xz2}&\sigma_{yz2}&\sigma_{zz2}
\end{bmatrix}
\end{equation}
\begin{align}
\sigma_{xx2}&= \sigma_{h2} + (\sigma_{v2}-\sigma_{h2})(\sin{\alpha_2}\cos{\beta_2})^2\\
\sigma_{xy2}&= (\sigma_{v2}-\sigma_{h2})(\sin{\alpha_2})^2\sin{\beta_2}\cos{\beta_2} \\
\sigma_{xz2}&= (\sigma_{v2}-\sigma_{h2})\sin{\alpha_2}\cos{\alpha_2}\cos{\beta_2} \\
\sigma_{yy2}&= \sigma_{h2} + (\sigma_{v2}-\sigma_{h2})(\sin{\alpha_2}\sin{\beta_2})^2 \\
\sigma_{yz2}&= (\sigma_{v2}-\sigma_{h2})\sin{\alpha_2}\cos{\alpha_2}\sin{\beta_2} \\
\sigma_{zz2}&= \sigma_{v2} - (\sigma_{v2}-\sigma_{h2})(\sin{\alpha_2})^2 
\end{align}
with tensor dip ($\alpha_2$) and strike ($\beta_2$) angles equal to $\alpha_2=30^{\circ}$ and $\beta_2=0^{\circ}$ (compared to Scenario 2, where $\alpha_2=\beta_2=0^{\circ}$).\footnote{Note: These conductivity tensor dip and strike angles, completely unrelated to the interface tilting angles $\{\alpha_m',\beta_m'\}$, describe the \emph{material tensor's} polar and azimuthal tilting (resp.) but follow a different convention~\cite{anderson1}.} Scenario 4: The central layer has full (albeit non-deviated) biaxial anisotropy characterized by the conductivity tensor $\boldsymbol{\bar{\sigma}}_2=\sigma_{x2}\bold{\hat{x}}\bold{\hat{x}}+\sigma_{y2}\bold{\hat{y}}\bold{\hat{y}}+\sigma_{z2}\bold{\hat{z}}\bold{\hat{z}}$, where $\{\sigma_{x2}=5,\sigma_{y2}=20,\sigma_{z2}=1\}$[mS/m].

\begin{figure}[H]
\centering
\subfloat[\label{ReHxx1}]{\includegraphics[width=3.25in]{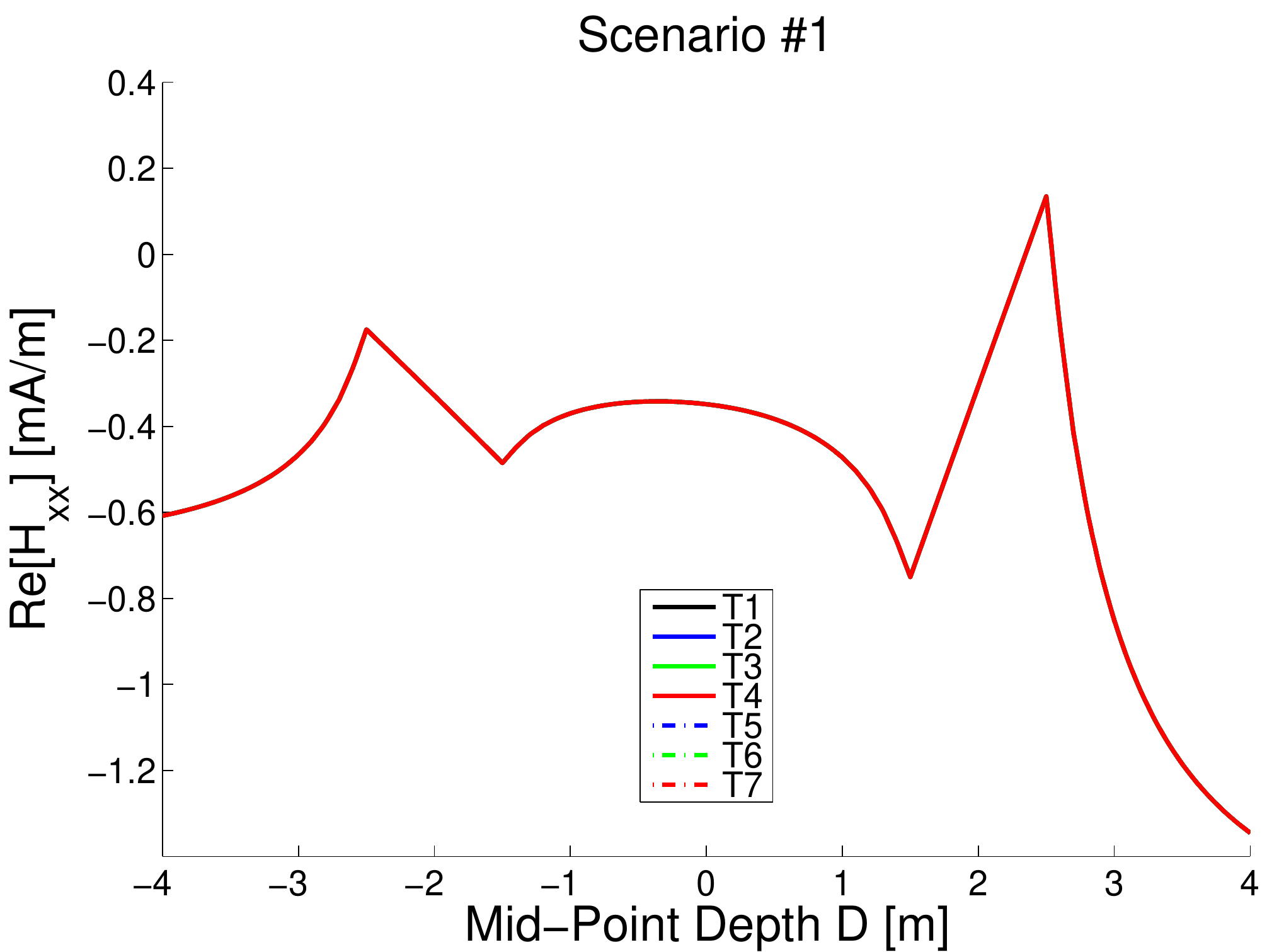}}
\subfloat[\label{ImHxx1}]{\includegraphics[width=3.25in]{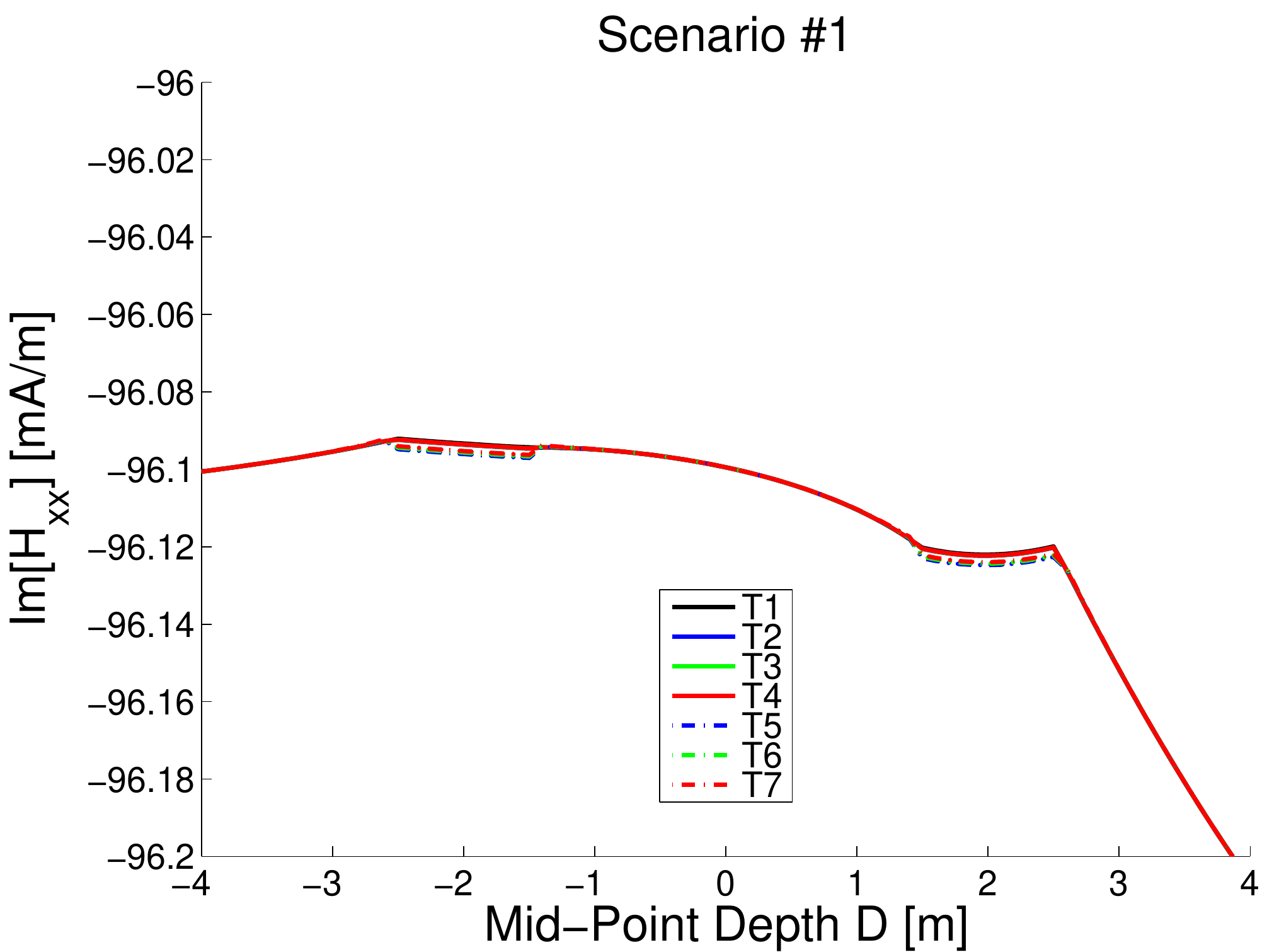}}

\subfloat[\label{ReHyy1}]{\includegraphics[width=3.25in]{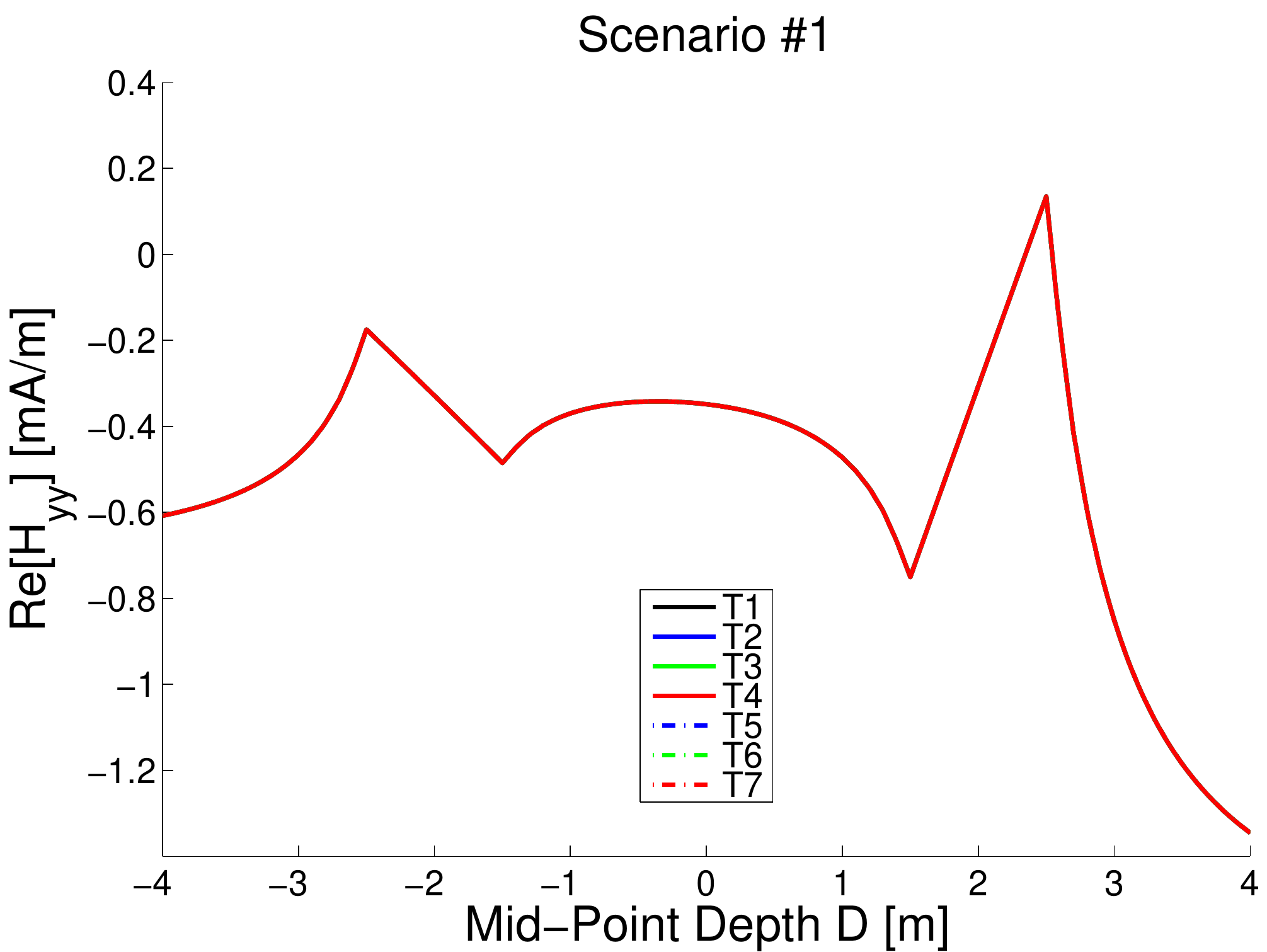}}
\subfloat[\label{ImHyy1}]{\includegraphics[width=3.25in]{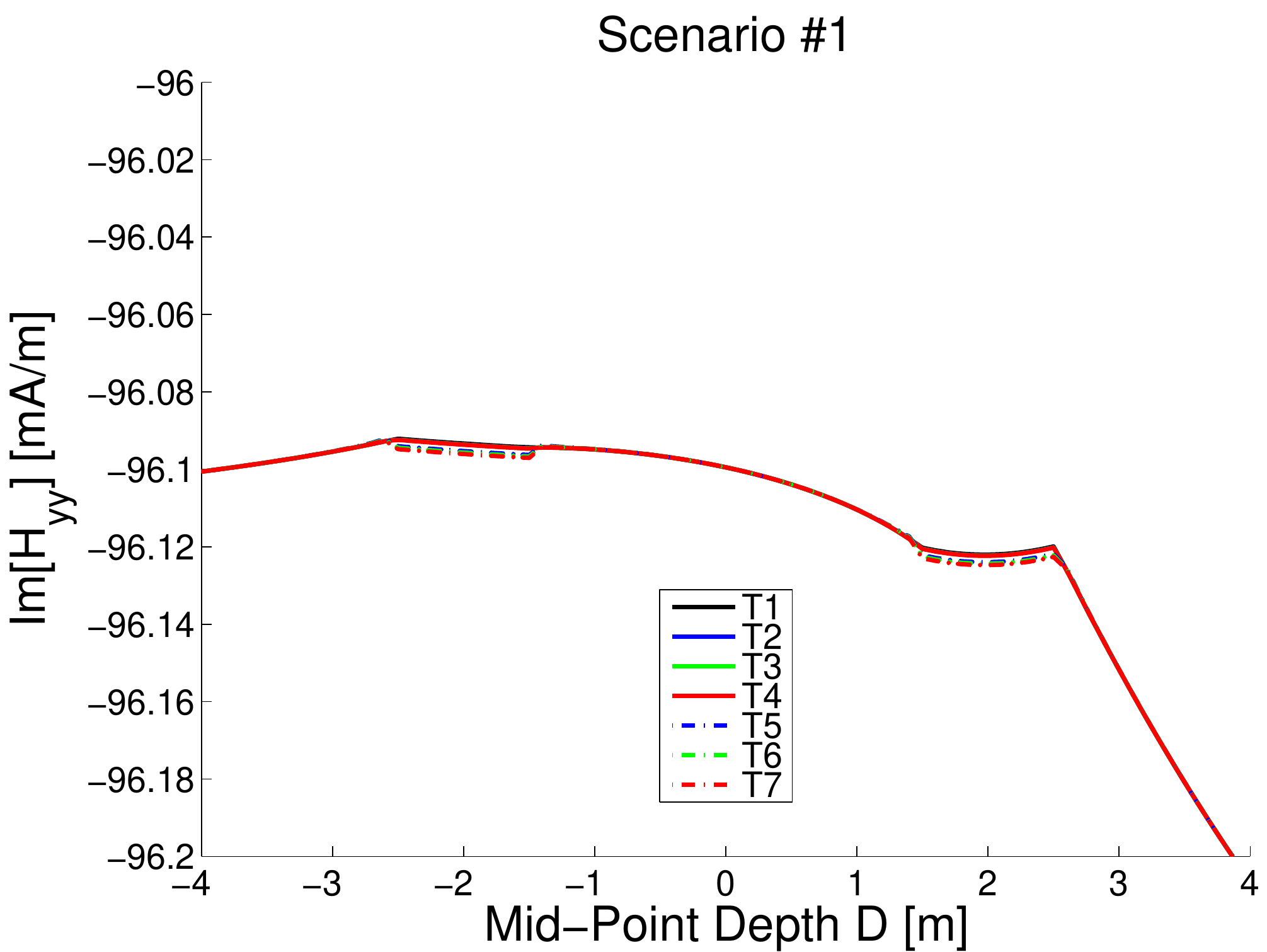}}

\subfloat[\label{ReHzz1}]{\includegraphics[width=3.25in]{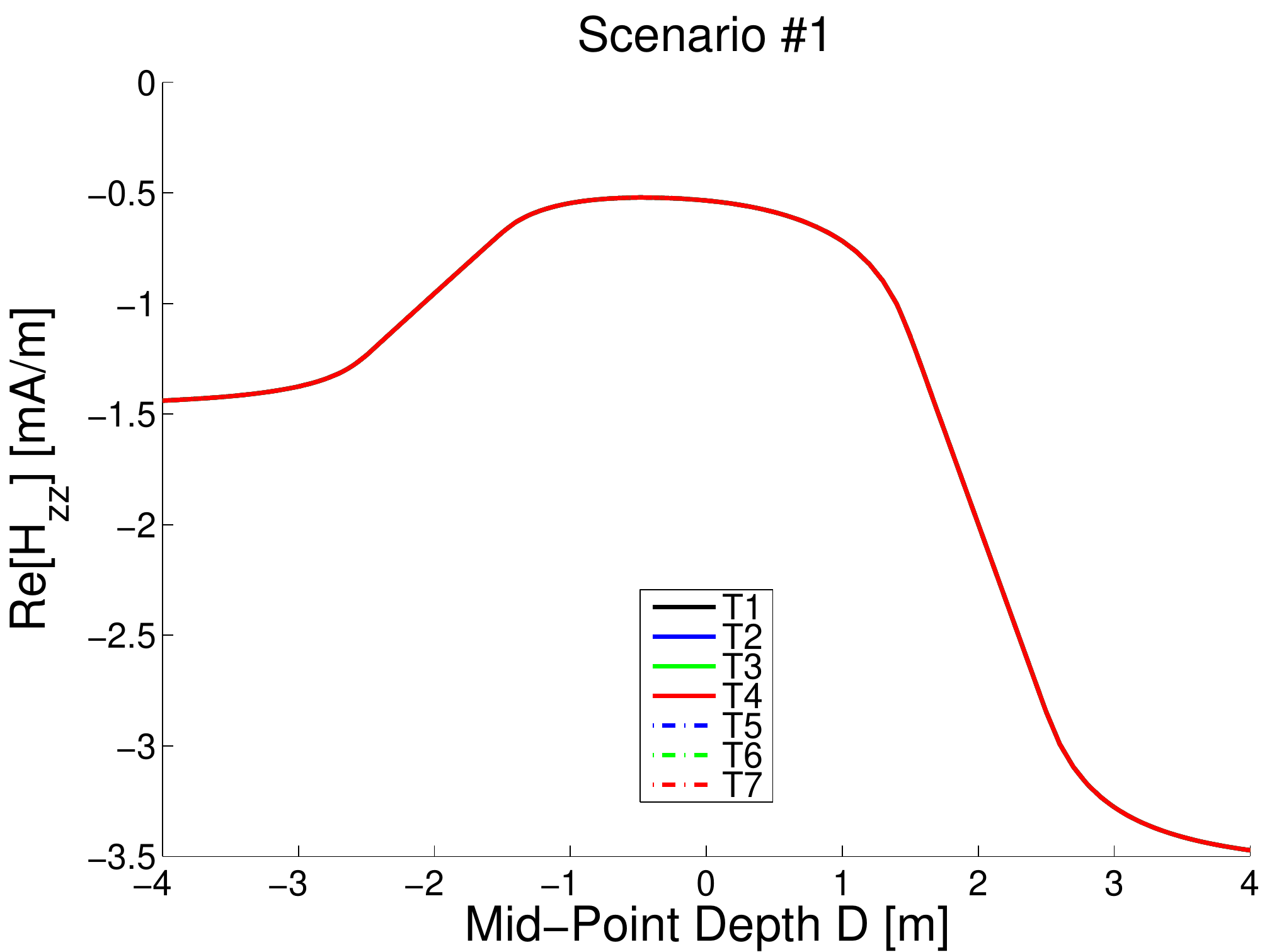}}
\subfloat[\label{ImHzz1}]{\includegraphics[width=3.25in]{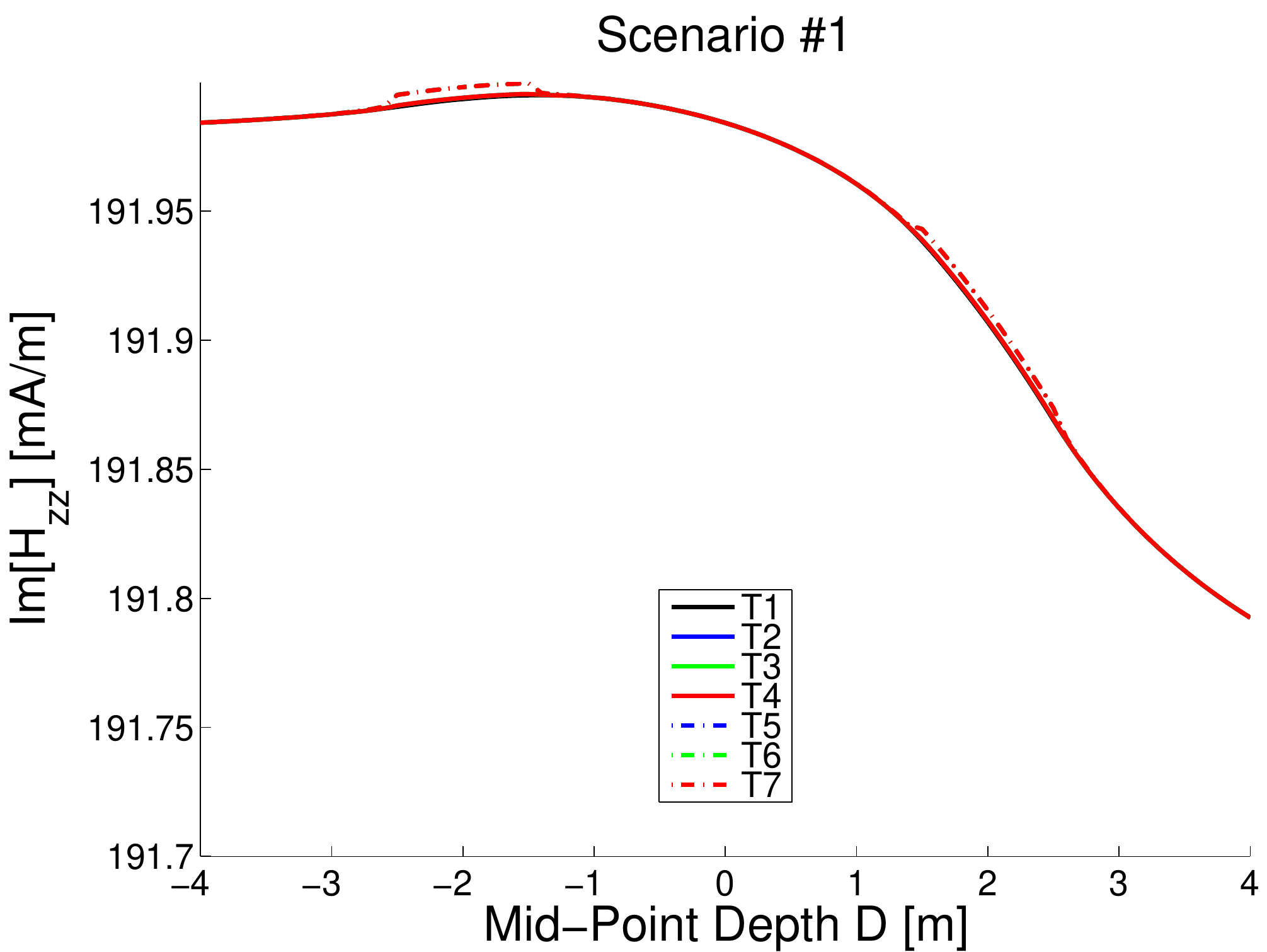}}
\caption{\small Co-polarized, complex-valued received magnetic fields: Material Scenario 1.}
\label{H1}
\end{figure}

Now we discuss the co-polarized results; note that as the cross-polarized results cannot be reliably modeled nearly as well as the co-pol fields (see previous section), we omit them. Observing Figure \ref{H1}: 
\begin{enumerate}
\item The real part of all three co-polarized measurements has no visibly noticeable sensitivity to interface tilting, even when the sensor is near bedding junctions. The imaginary part of these measurements \emph{does}, by contrast, exhibit tilting sensitivity. 
\item The imaginary parts' sensitivity to tilting depends on the sensor position, with sensitivity being largest when the sensor is near the interfaces ($\pm 2$m). This is expected, since the conductive formation exponentially attenuates fields scattered and propagating away from the interfaces.  
\item The 4m bed thickness and sensor frequency (100kHz) preclude observation of inter-junction coupling effects due to wave ``multi-bounce" within the slab layers~\cite{chew}[Ch. 2].
\item At a fixed polar interface tilting (T2, T3, and T4 [$\pm 1^{\circ}$] versus T5, T6, and T7 [$\pm 3^{\circ}$]), Im$[H_{z'z'}]$ shows no visible sensitivity to the interface's azimuth orientation. This lack of azimuthal sensitivity is quite sensible, given the azimuthal symmetry of this $z$-transmit/$z$-receive measurement~\cite{zhdanov}. 
\item In contrast to Im$[H_{z'z'}]$, Im$[H_{x'x'}]$ and Im$[H_{y'y'}]$ do show azimuthal sensitivity. For fixed polar tilt, both measurements show greatest excursion (i.e., away from the black, zero-tilt curve) when the interfaces' common azimuth tilt orientation is aligned with the transmitter and receiver orientation ($0^{\circ}$ [blue curves] and $90^{\circ}$ [red curves] for Im$[H_{x'x'}]$ and Im$[H_{y'y'}]$, respectively). On the other hand, minimal excursion occurs when the azimuth tilt orientation is orthogonal to the transmitter and receiver orientation ($90^{\circ}$ and $0^{\circ}$ for Im$[H_{x'x'}]$ and Im$[H_{y'y'}]$, respectively).  
\item Tilting, irrespective of the polar orientation's sign (e.g., $1^{\circ}$ versus $-1^{\circ}$ polar tilt), results in the responses Im$[H_{x'x'}]$ and Im$[H_{y'y'}]$ having downward excursions. Indeed for a fixed azimuth tilt (curve color), observe the sensor response near the two (effectively) oppositely-tilted interfaces for both the solid and dash-dot curves. 
\item In contrast to Im$[H_{x'x'}]$ and Im$[H_{y'y'}]$ exhibiting downward excursions, Im$[H_{z'z'}]$ always shows an \emph{upward} excursion irrespective of polar tilt sign. 
\end{enumerate}
The above observations also visibly manifest for the three remaining cases involving anisotropic media. We find, for the small tilting range explored here at least (zero to six degrees of relative tilt), that the anisotropy primarily serves to alter ``baseline" (zero-tilt) sensor responses which are then perturbed by the effect of tilt. 

\newpage

\newpage
\begin{figure}[H]
\centering
\subfloat[\label{ReHxx2}]{\includegraphics[width=3.25in]{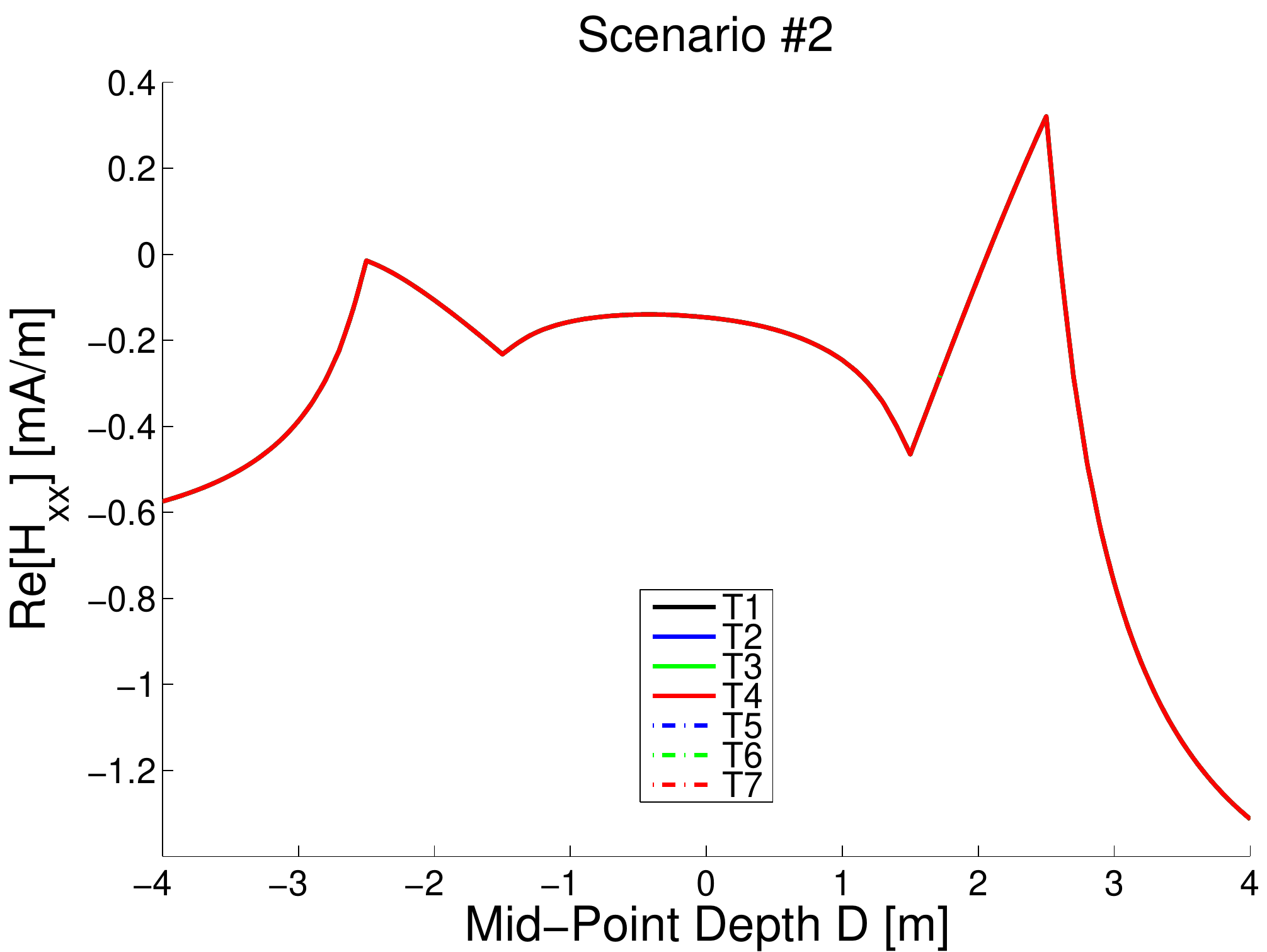}}
\subfloat[\label{ImHxx2}]{\includegraphics[width=3.25in]{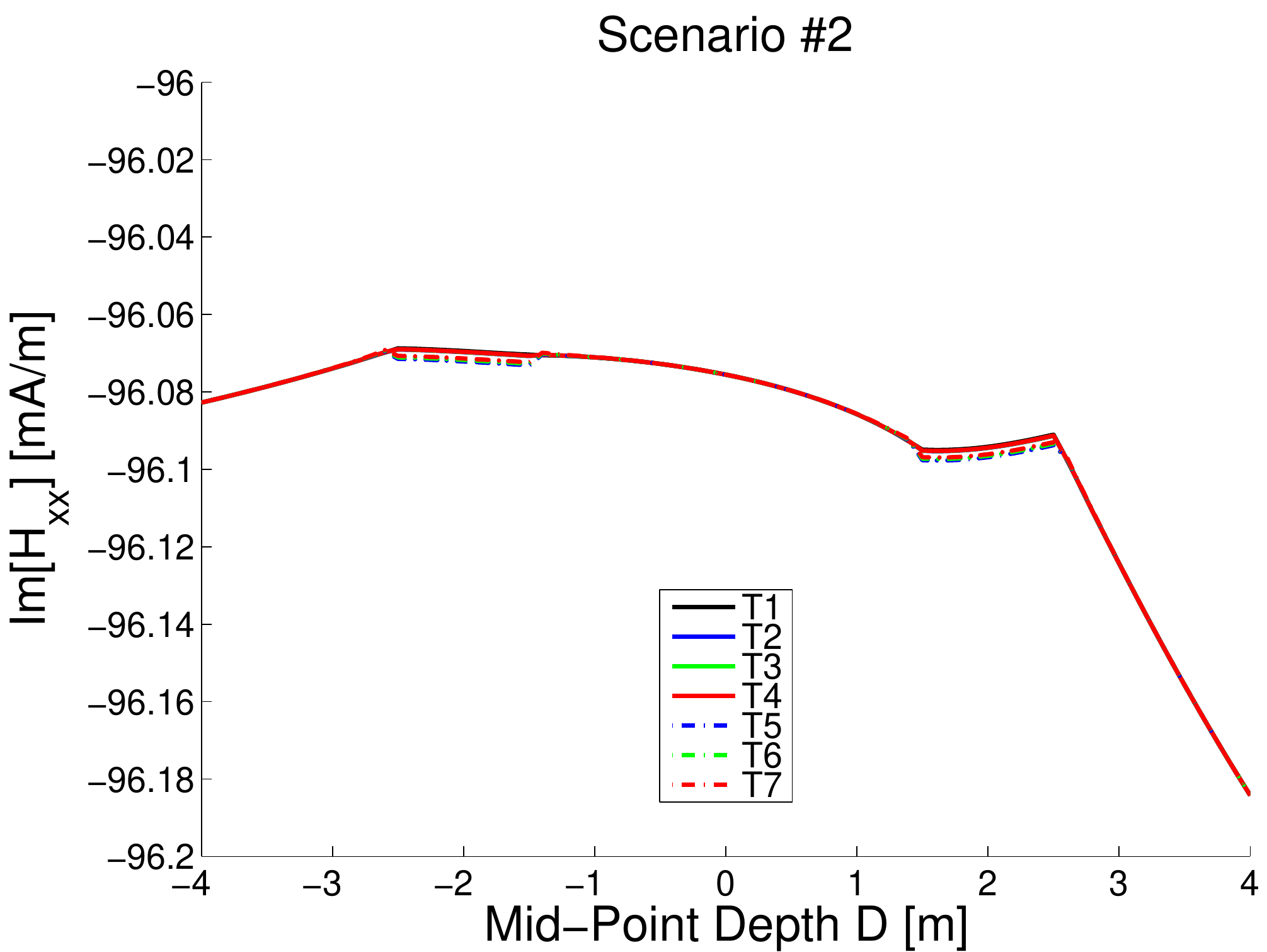}}

\subfloat[\label{ReHyy2}]{\includegraphics[width=3.25in]{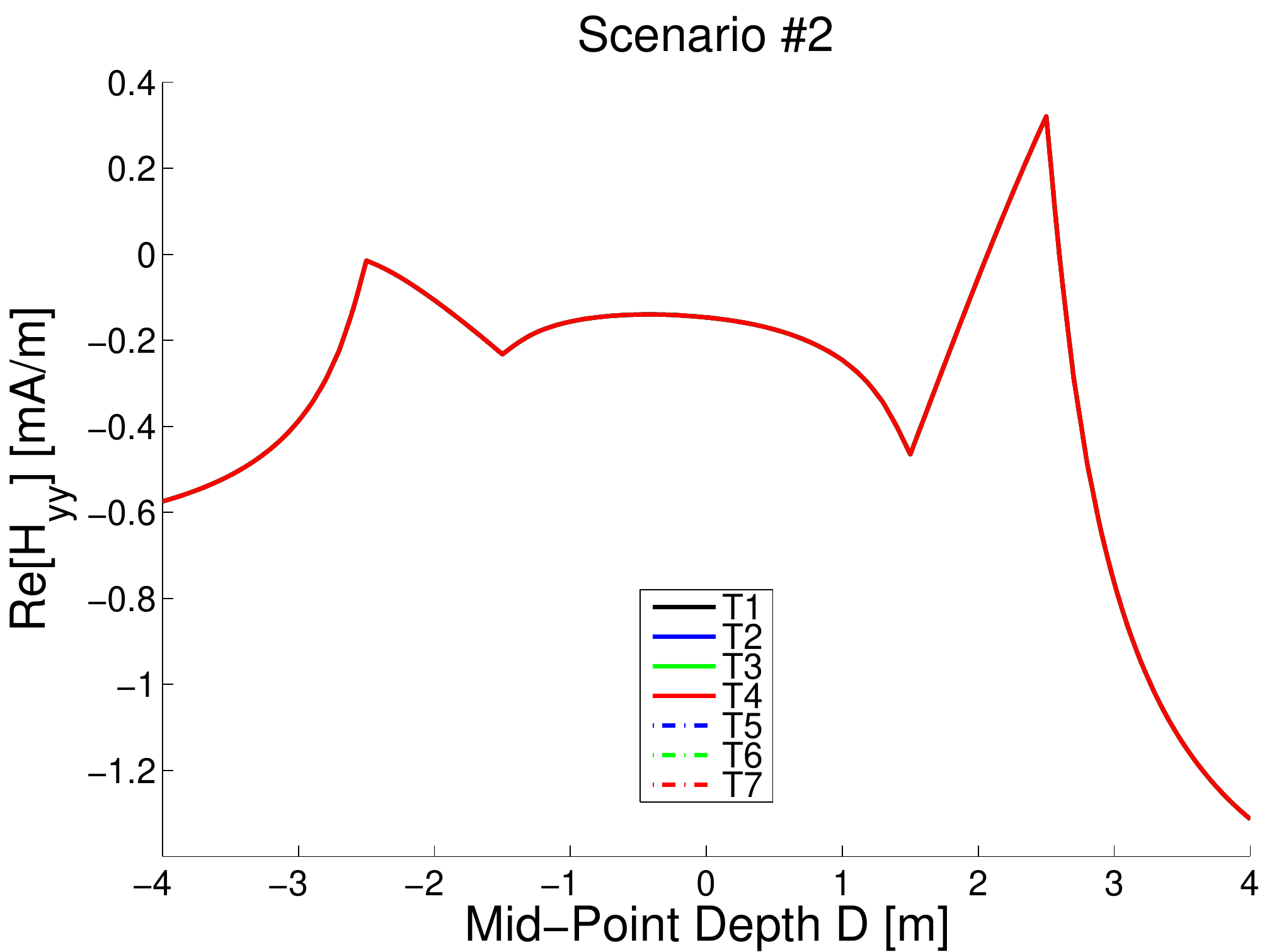}}
\subfloat[\label{ImHyy2}]{\includegraphics[width=3.25in]{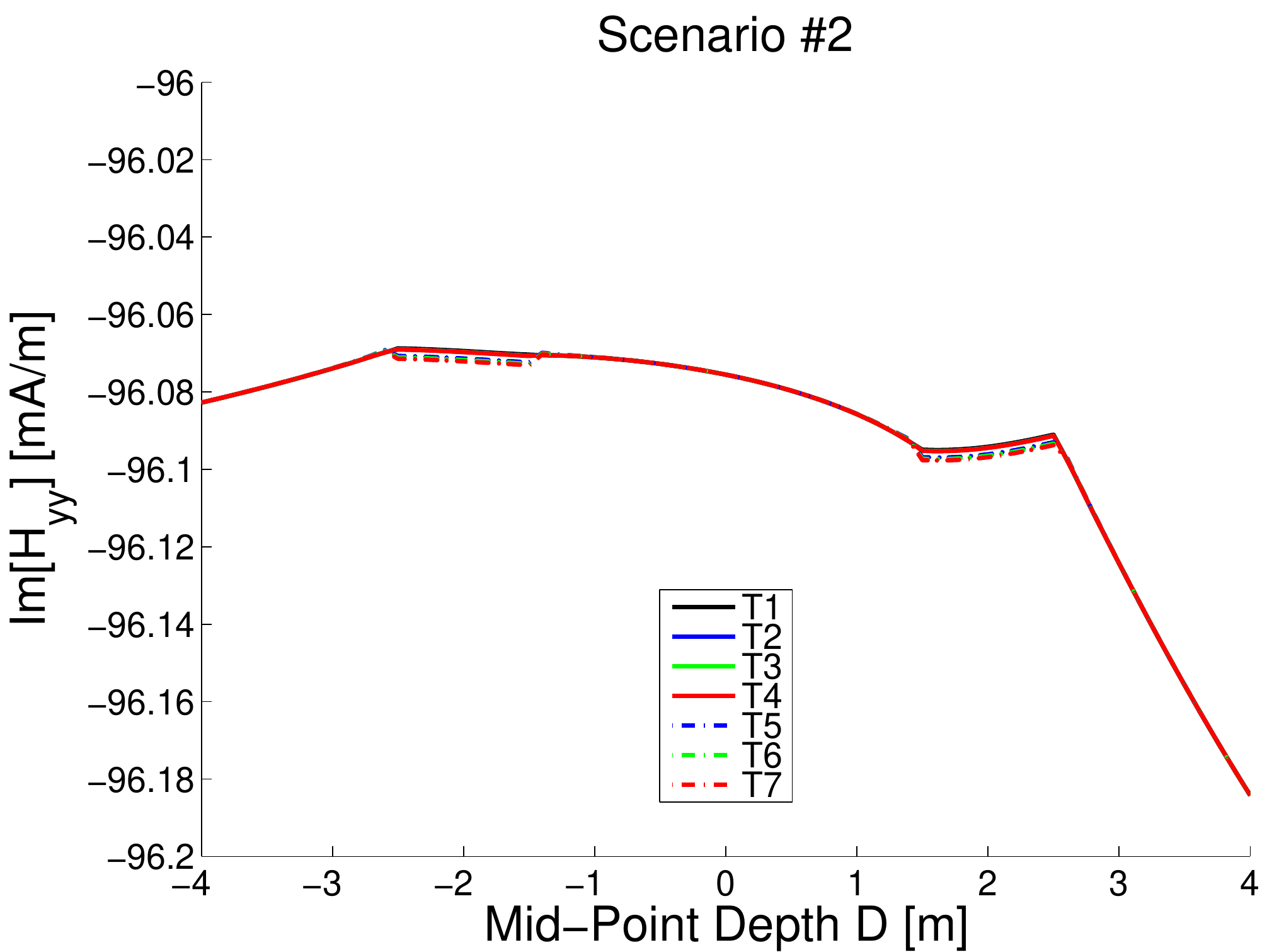}}

\subfloat[\label{ReHzz2}]{\includegraphics[width=3.25in]{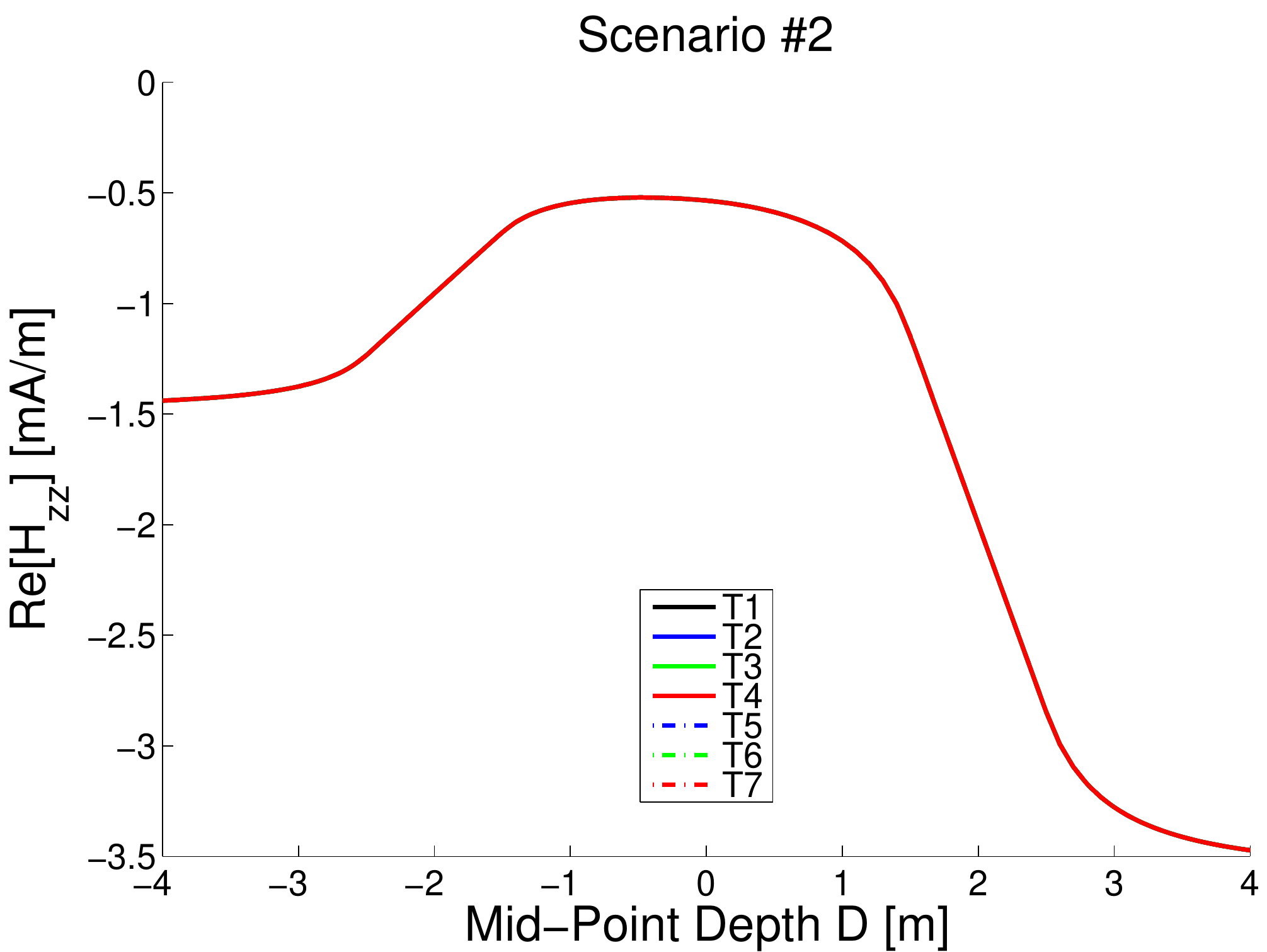}}
\subfloat[\label{ImHzz2}]{\includegraphics[width=3.25in]{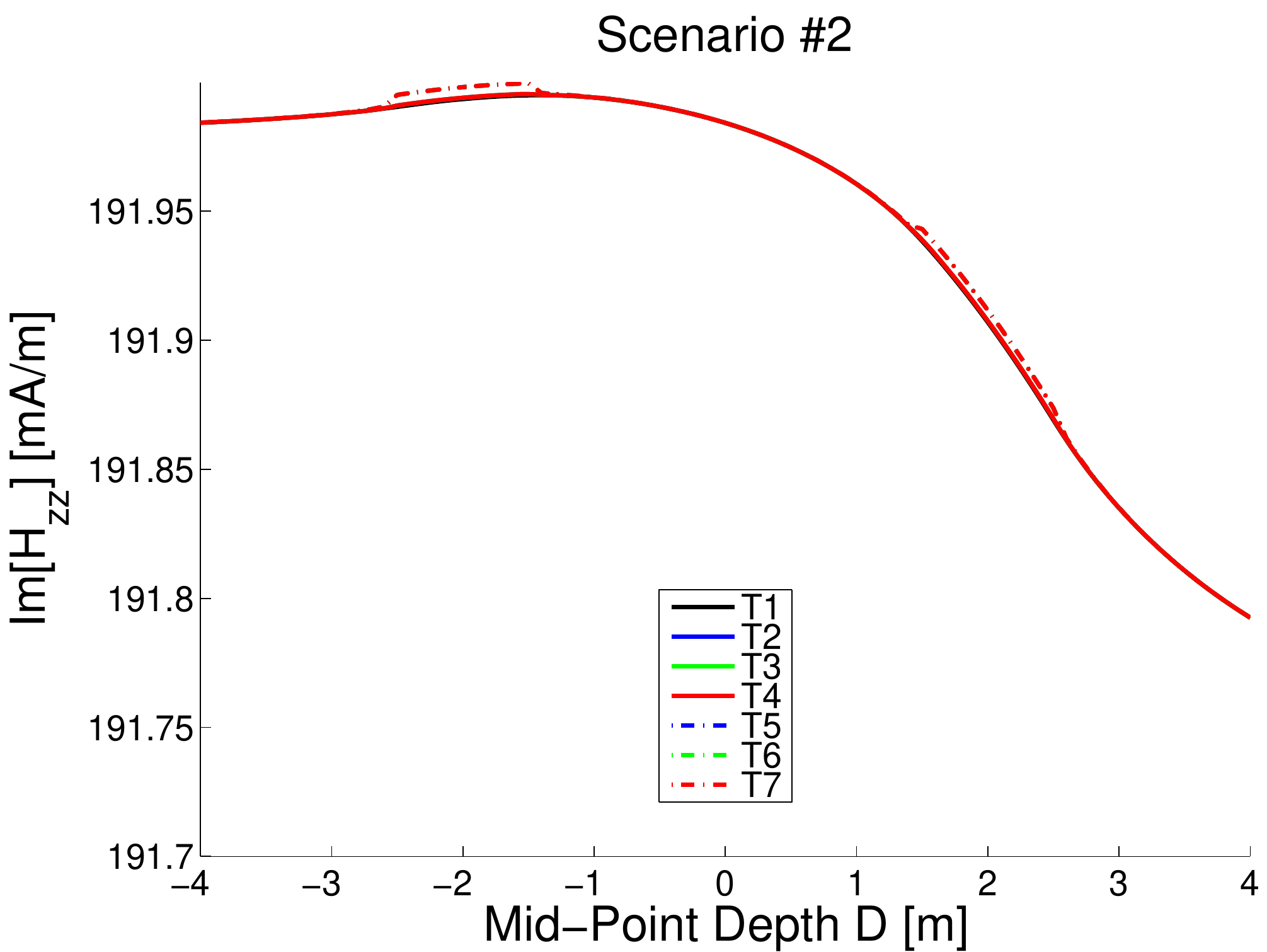}}
\caption{\small Co-polarized, complex-valued received magnetic fields: Material Scenario 2.}
\label{H2}
\end{figure}

\newpage
\begin{figure}[H]
\centering
\subfloat[\label{ReHxx3}]{\includegraphics[width=3.25in]{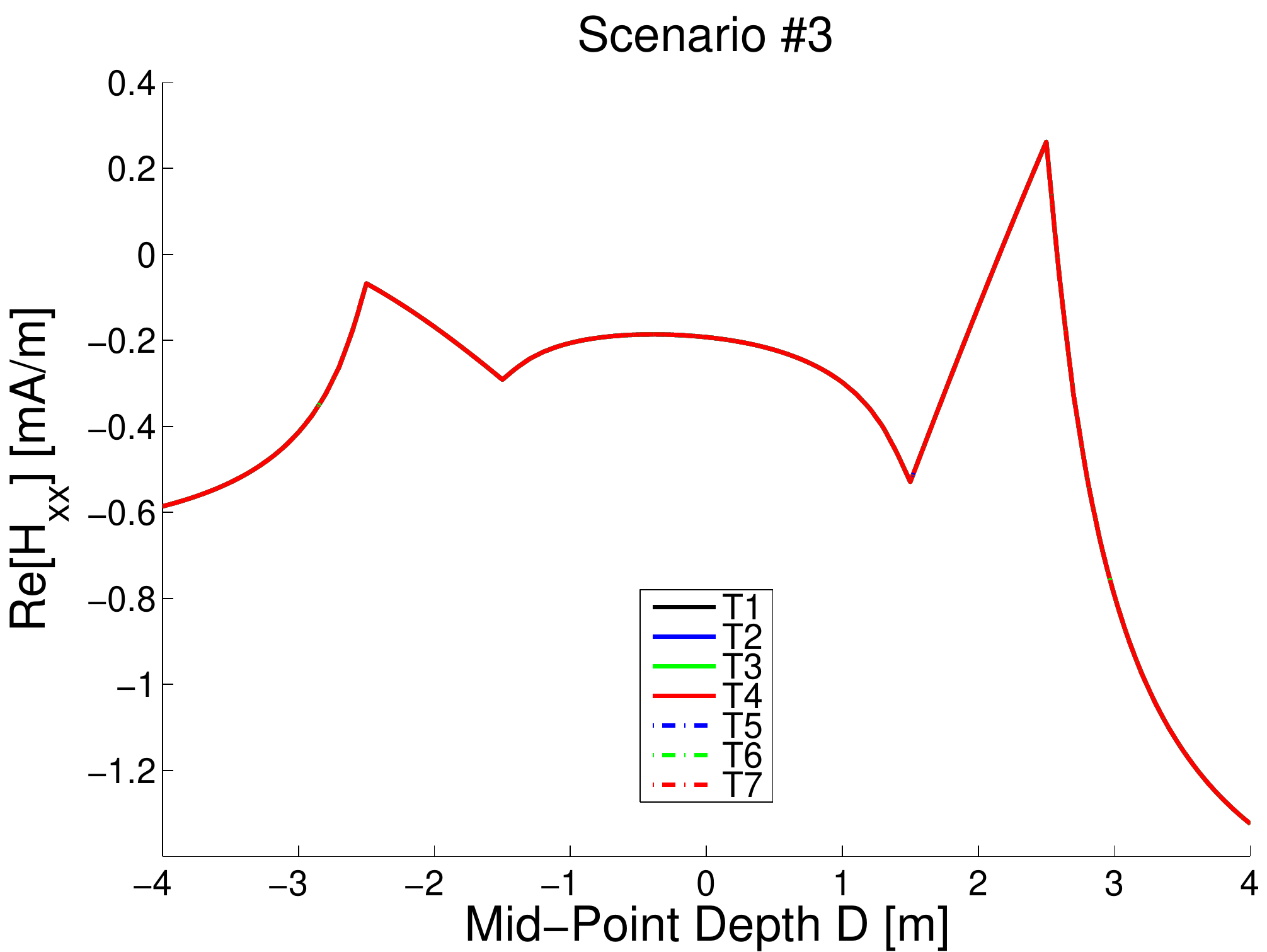}}
\subfloat[\label{ImHxx3}]{\includegraphics[width=3.25in]{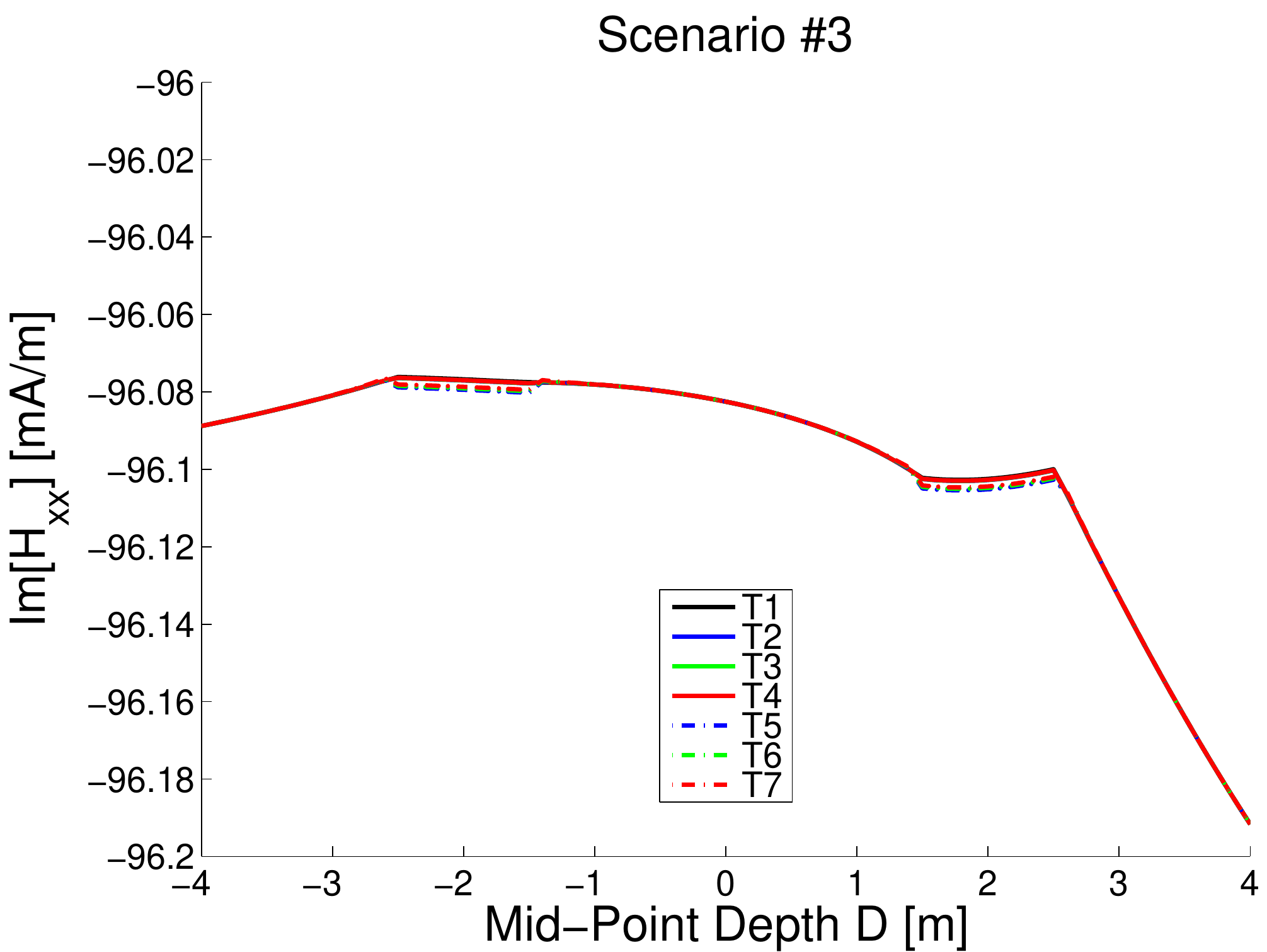}}

\subfloat[\label{ReHyy3}]{\includegraphics[width=3.25in]{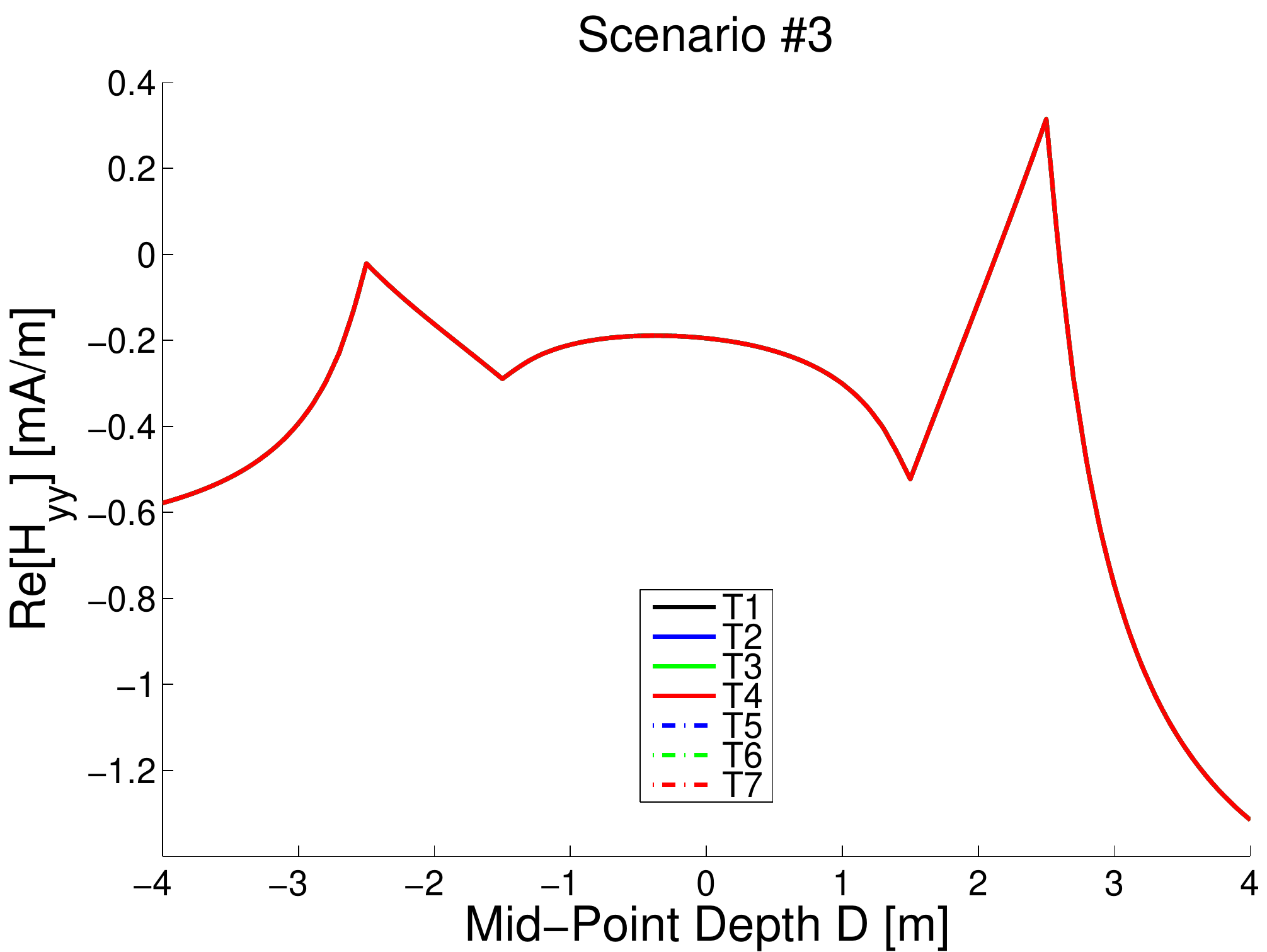}}
\subfloat[\label{ImHyy3}]{\includegraphics[width=3.25in]{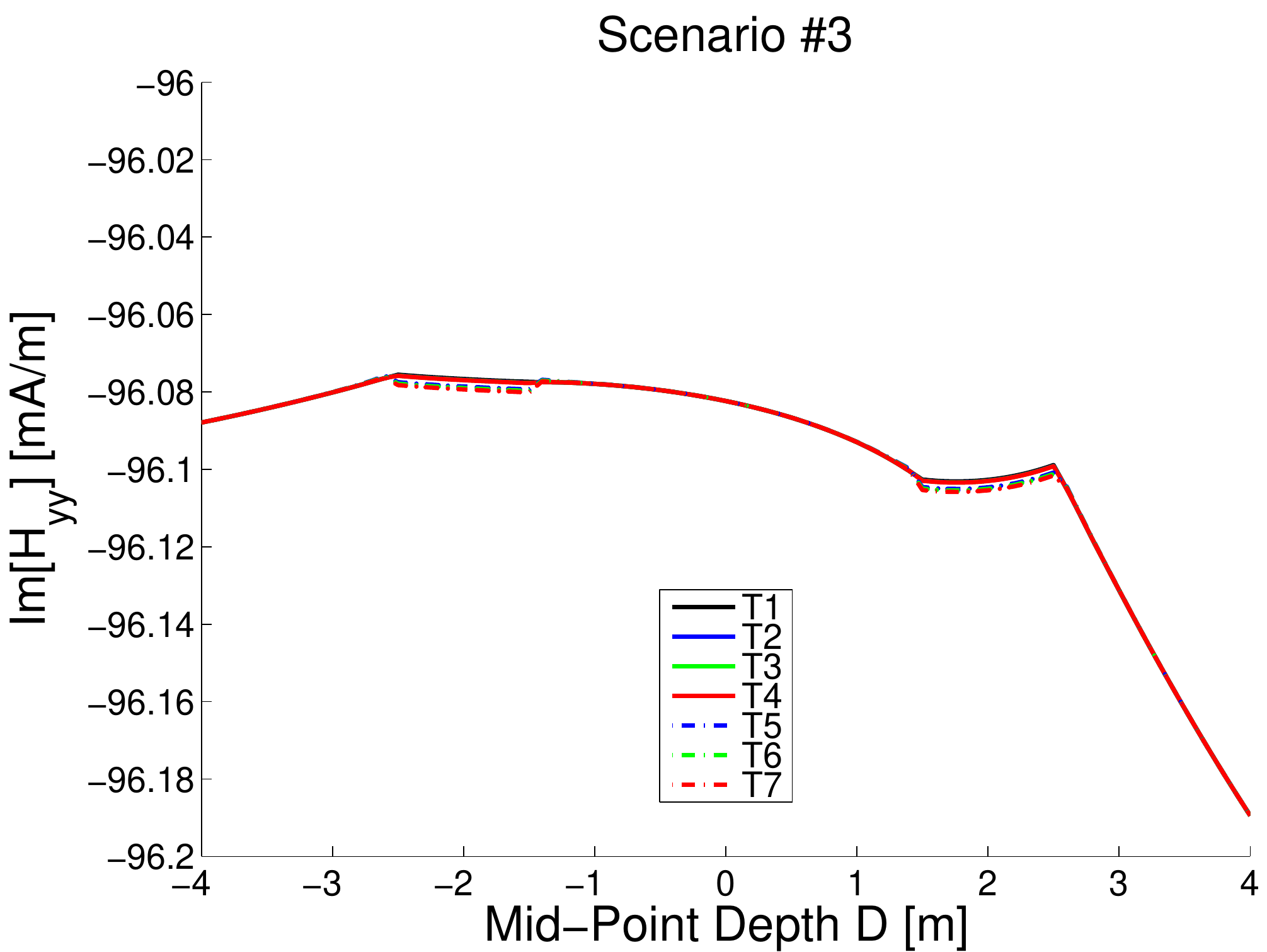}}

\subfloat[\label{ReHzz3}]{\includegraphics[width=3.25in]{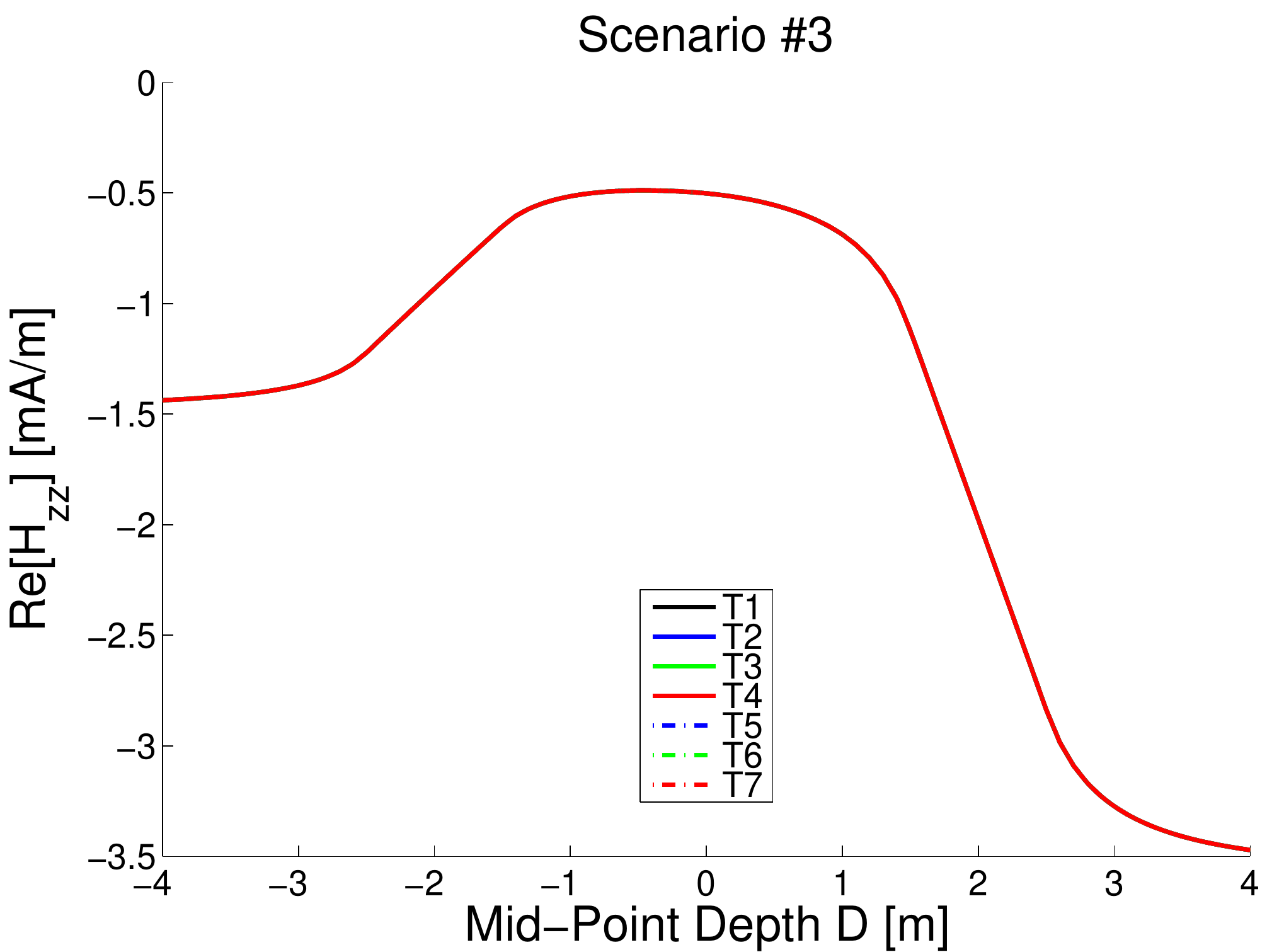}}
\subfloat[\label{ImHzz3}]{\includegraphics[width=3.25in]{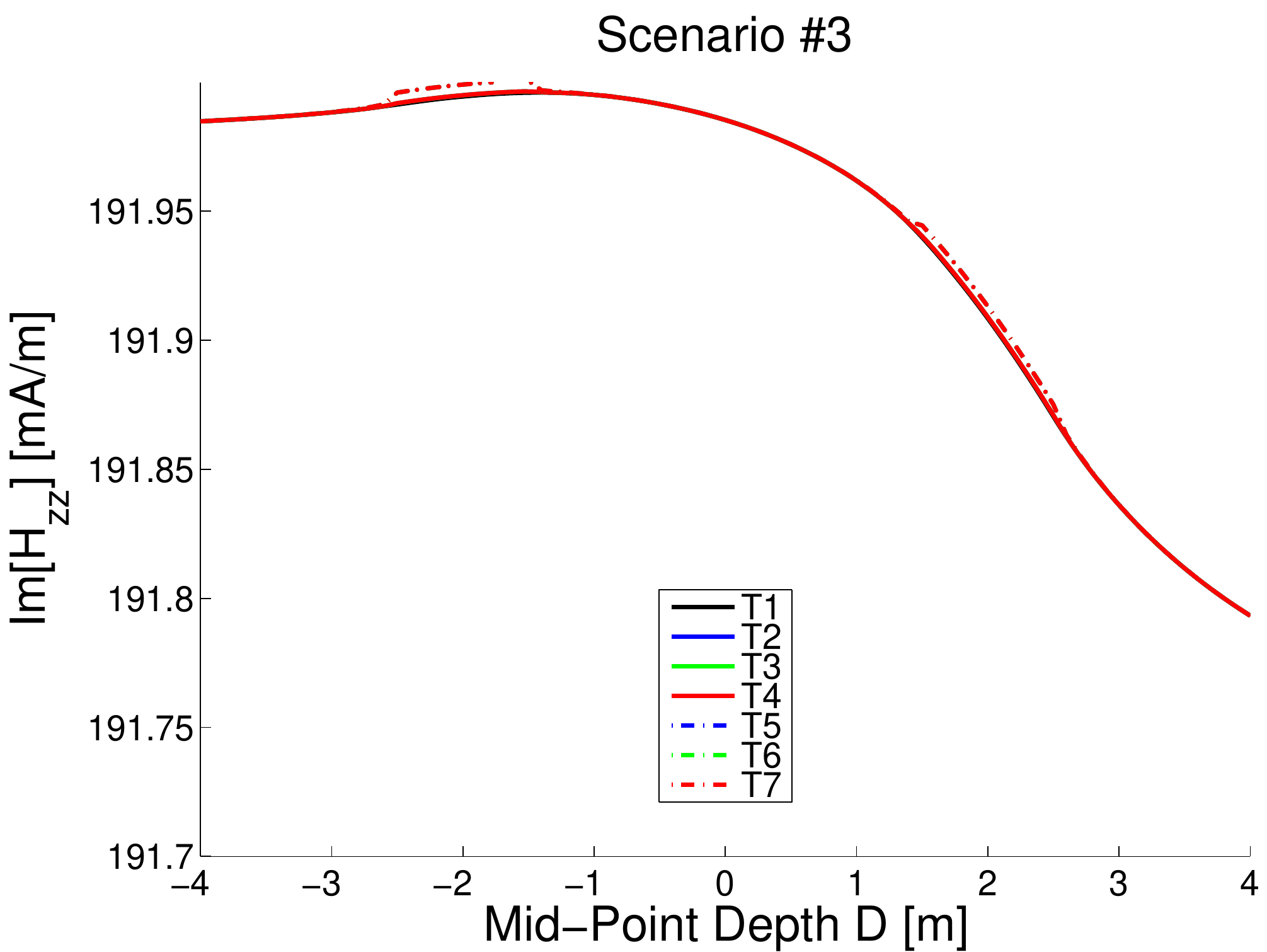}}
\caption{\small Co-polarized, complex-valued received magnetic fields: Material Scenario 3.}
\label{H3}
\end{figure}

\newpage
\begin{figure}[H]
\centering
\subfloat[\label{ReHxx4}]{\includegraphics[width=3.25in]{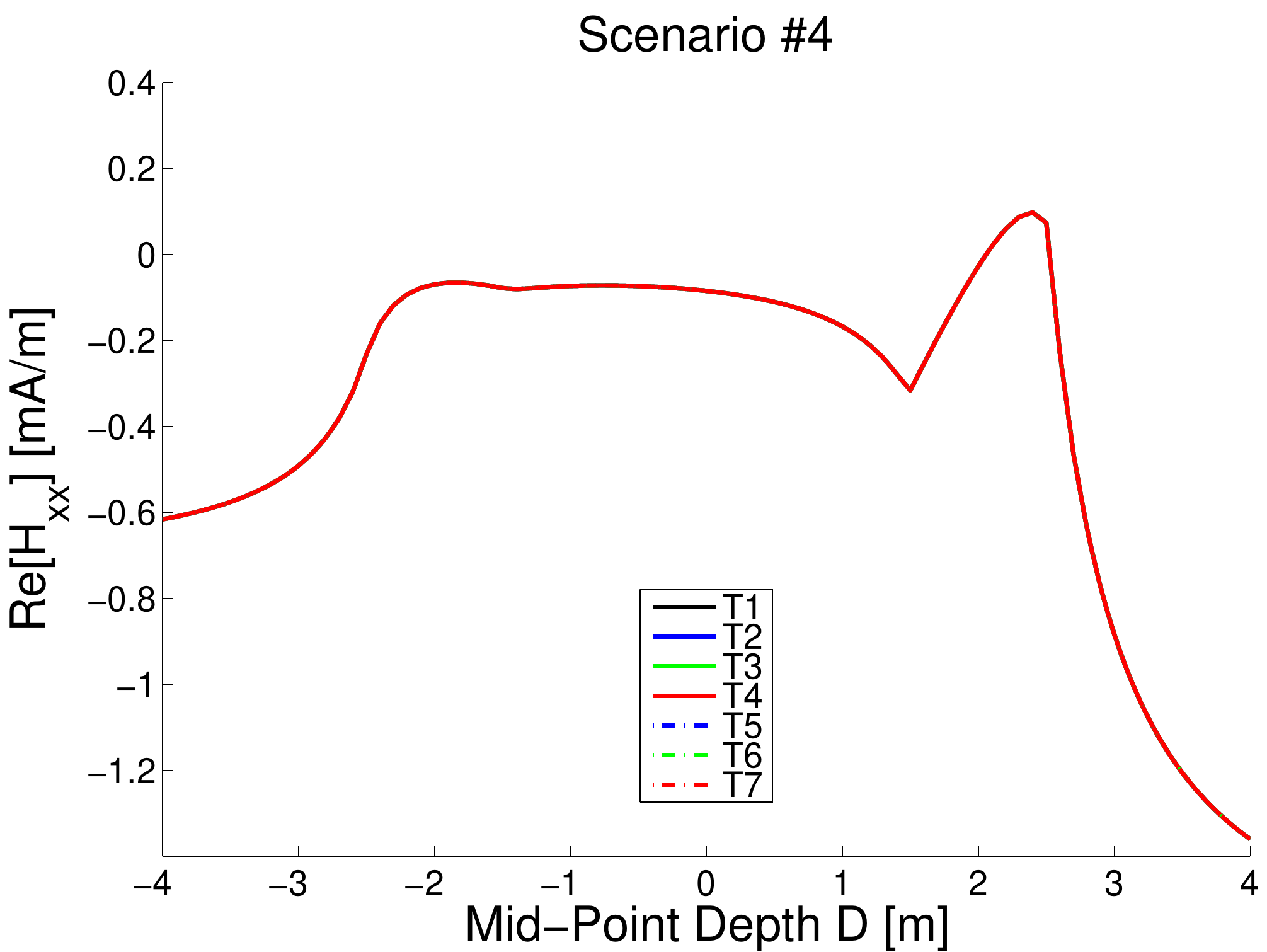}}
\subfloat[\label{ImHxx4}]{\includegraphics[width=3.25in]{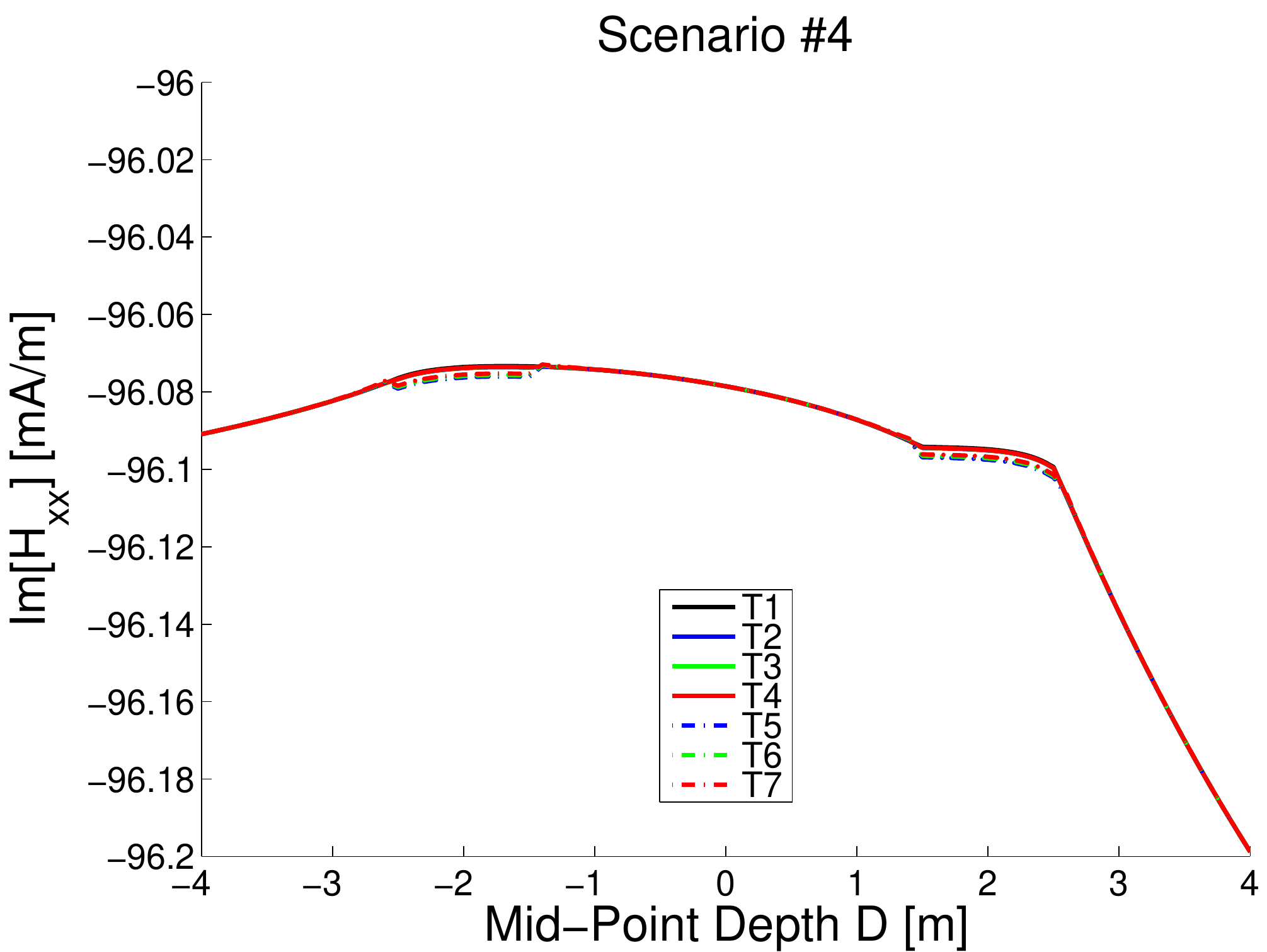}}

\subfloat[\label{ReHyy4}]{\includegraphics[width=3.25in]{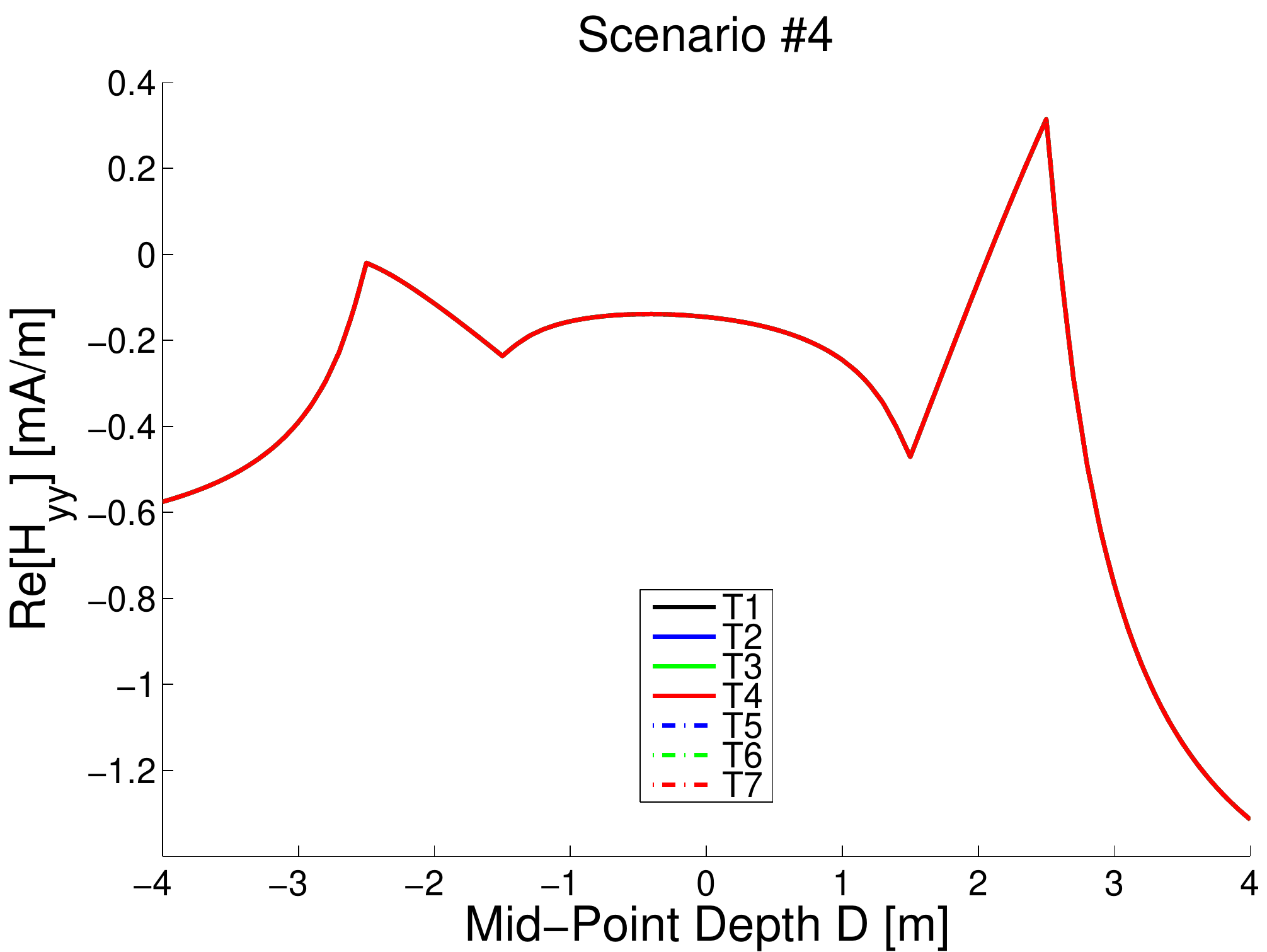}}
\subfloat[\label{ImHyy4}]{\includegraphics[width=3.25in]{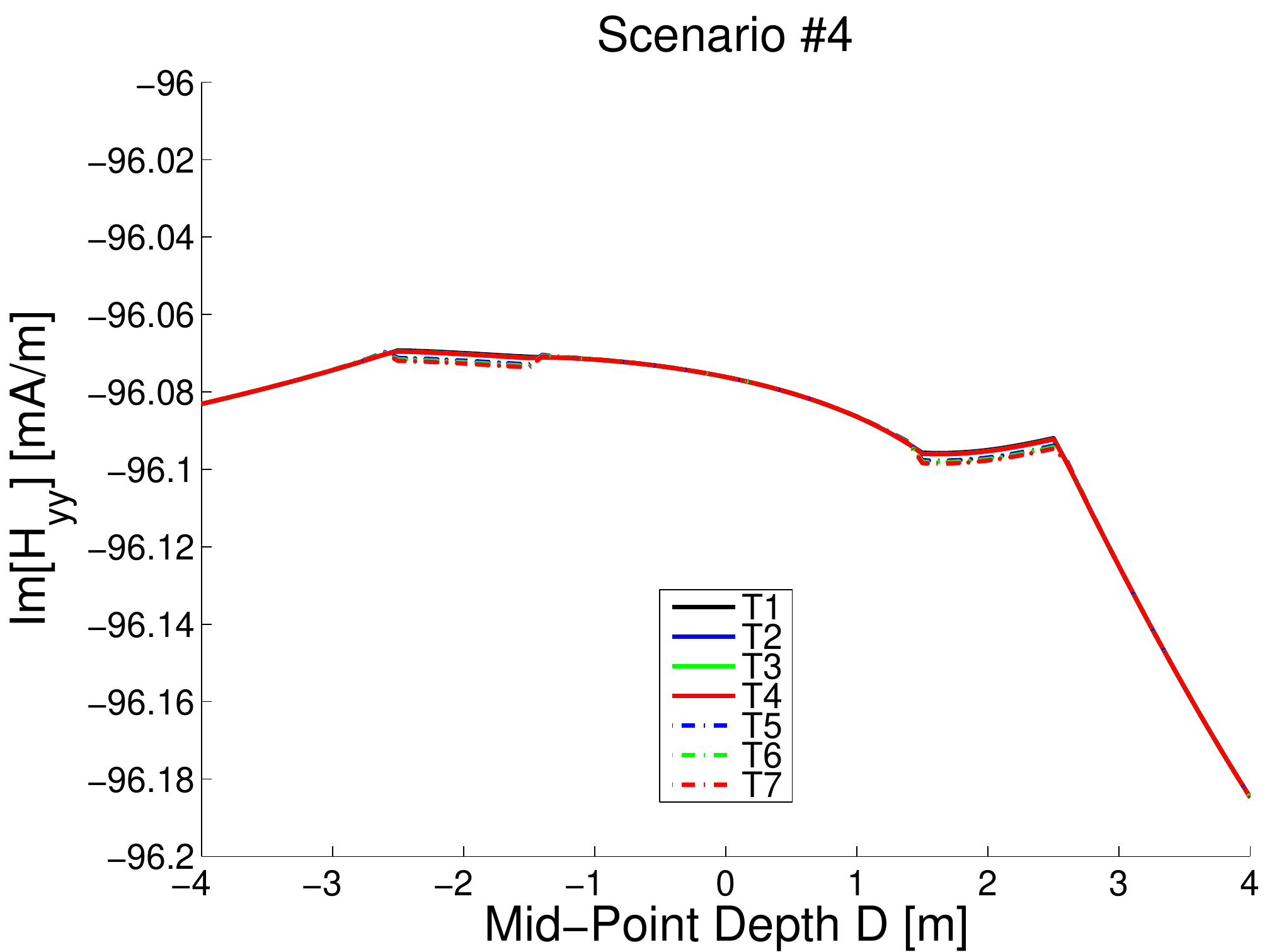}}

\subfloat[\label{ReHzz4}]{\includegraphics[width=3.25in]{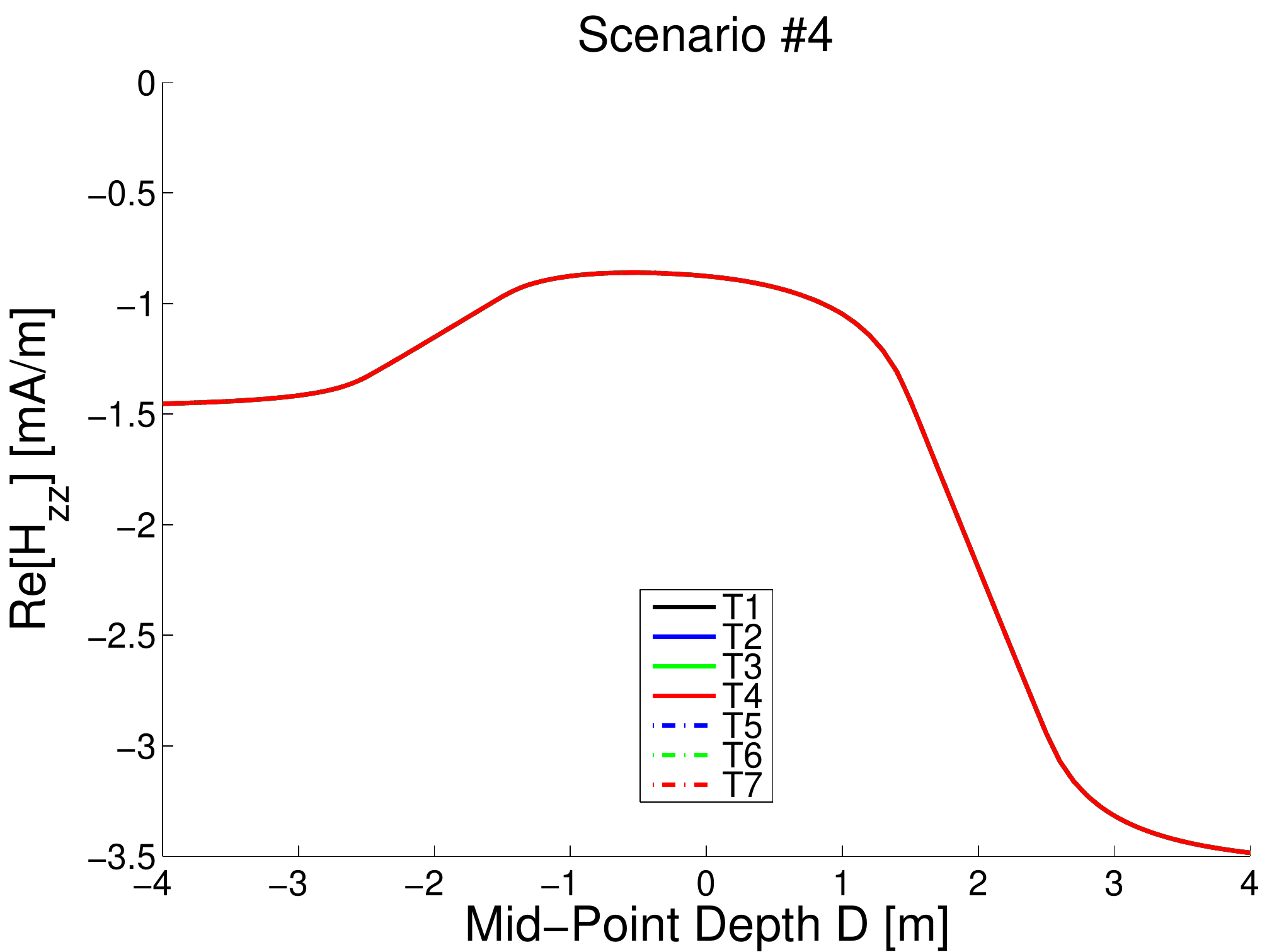}}
\subfloat[\label{ImHzz4}]{\includegraphics[width=3.25in]{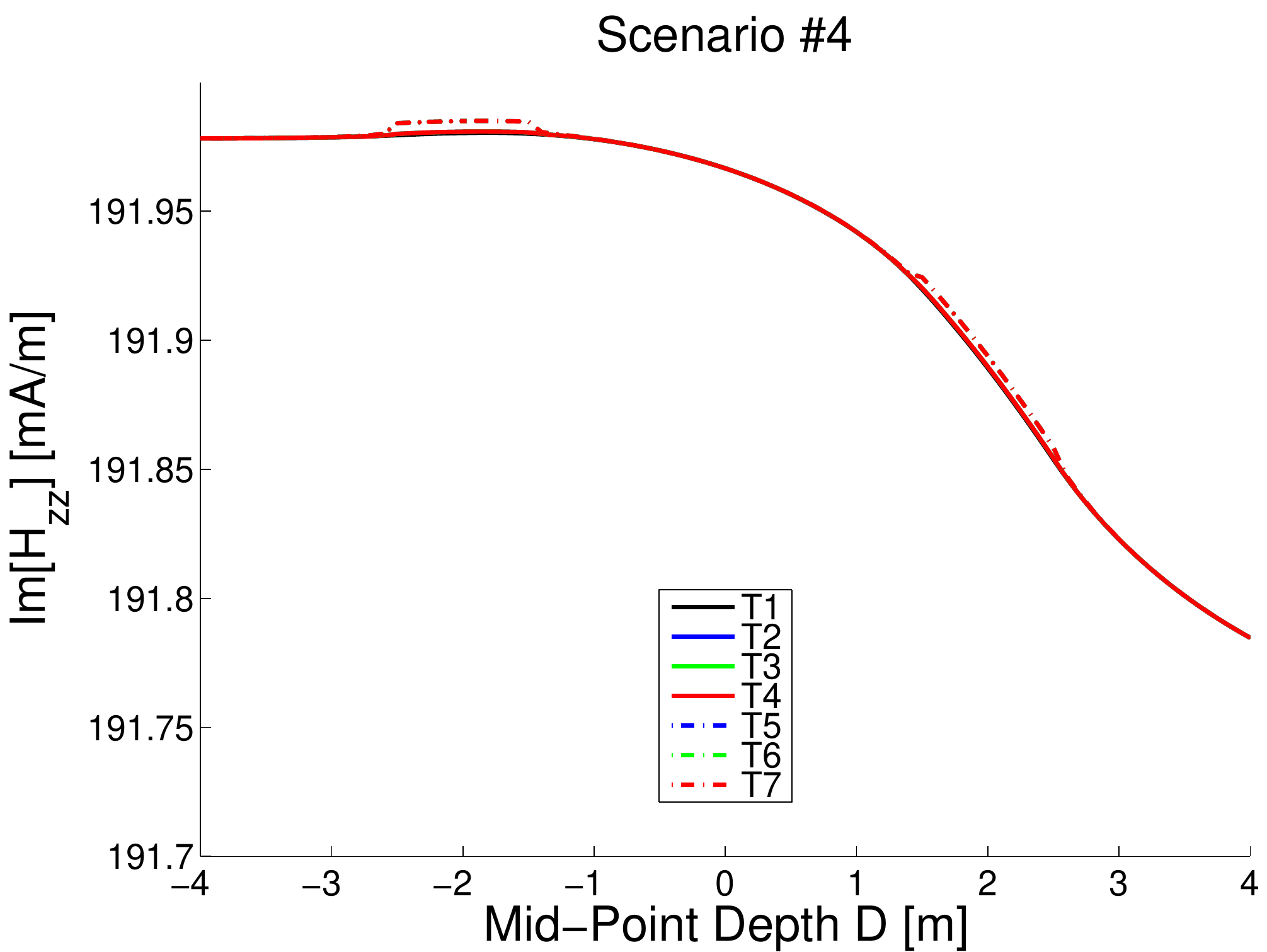}}
\caption{\small Co-polarized, complex-valued received magnetic fields: Material Scenario 4.}
\label{H4}
\end{figure}

\section{\label{Conc}Conclusion}

We have introduced and profiled (both quantitatively and qualitatively) a methodology, which augments robust full-wave numerical pseudo-analytical algorithms admitting formation layers of general anisotropy and loss, to incorporate the effects of planar interface tilting. Our proposed methodology, directed at extending the range of applicability of such eigenfunction expansion methods by relaxing the traditional constraint of parallelism between layers, consists of first defining a spatial coordinate transformation within a thin planar region surrounding each interface to be ``flattened". Such coordinate transformation locally distorts the EM field, effectively altering the local angle of incidence between EM waves and flat interfaces so as to mimic the presence of interfaces possessing effective, independently-defined tilt orientations. During the second stage of flattening the interfaces, the coordinate transformation is incorporated into the EM material properties of said planar regions via application of T.O. principles, which exploits the well-known ``duality" between spatial coordinate transformations and equivalent material properties ``implementing" these coordinate transformations in flat space. 

As the proposed methodology is not limited by the loss and anisotropy properties of any layer, some combination of beds such as non-deviated sand-shale laminates, cross-bedded clean-sand depositions, formations fractured by invasive drilling processes, and simpler isotropic conductive beds can all be included along with flexibly-defined (effective) bed junction tilting with respect to polar deviation magnitude as well as azimuth orientation. Exhibiting representative examples of the new methodology's said flexibilities, we applied it to demonstrating the expected qualitative properties of multi-component induction tool responses when planar bedding deviation is present. This being said, one should not confuse modeling flexibility and manifestation of expected qualitative trends with \emph{quantitative} accuracy. Indeed, although T.O. theory informs us that said interface-flattening media facilitate numerical prediction of tilted-interface effects via employing specially-designed coating slabs, the necessary \emph{orientation} of the slabs' truncation surfaces (i.e., planes parallel to the $xy$ plane) predicts that spurious scattering will arise. Particularly, our numerical error analysis demonstrated that artificial scattering typically scales quadratically with effective interface tilt. We found that one could nonetheless safely model tilting effects so long as the magnitude of each interface's polar deviation is kept small, which guided the choice of explored interface tilt ranges when examining qualitative trends of sensor responses produced by induction instruments within tilted geophysical layers.

\section{Acknowledgments}
NASA NSTRF and OSC support acknowledged.
\newpage
\bibliographystyle{model1-num-names}
\bibliography{reflist}
\end{document}